\documentclass[a4paper,11pt]{article}
\pdfoutput=1 

\usepackage{jcappub}

\usepackage{graphicx}
\usepackage{amsmath}
\usepackage{amssymb}
\usepackage{amsfonts}
\usepackage{mathrsfs}
\usepackage{lscape}
\usepackage{hyperref}
\usepackage{verbatim}
\usepackage{xcolor}
\usepackage{bm}
\usepackage{tabularx}
\usepackage{mathtools}
\usepackage{ulem}
\newcommand{\be}{\begin{equation}}
\newcommand{\ee}{\end{equation}}
\newcommand{\ba}{\begin{eqnarray}}
\newcommand{\ea}{\end{eqnarray}}
\newcommand{\nn}{\nonumber}

\def\bea{\begin{eqnarray}}
\def\eea{\end{eqnarray}}
\def\eqi{\begin{equation}}
\def\eqf{\end{equation}}
\def\eqia{\begin{eqnarray}}
\def\eqfa{\end{eqnarray}}
\def\lcdm{$\Lambda$CDM } 

\definecolor{darkgreen}{rgb}{0,0.6,0}

\newcommand{\vp}{\varphi}
\newcommand{\mc}{\mathcal}
\renewcommand{\(}{\left(}
\renewcommand{\)}{\right)}
\renewcommand{\[}{\left[}
\renewcommand{\]}{\right]}

\makeatother

\title{An effective fluid description of scalar-vector-tensor theories under the sub-horizon and quasi-static approximations}

\author[\star, 1]{Wilmar Cardona,\note{Corresponding author.}}
\author[\dagger, 2]{J. Bayron Orjuela-Quintana,}
\author[\ddagger, 2]{C\'esar A. Valenzuela-Toledo,}

\affiliation[1]{ICTP South American Institute for Fundamental Research \& Instituto de F\'isica Te\'orica, Universidade Estadual Paulista, 01140-070, S\~ao Paulo, Brazil}

\affiliation[2]{Departamento de F\'isica, Universidad del Valle, Ciudad Universitaria Mel\'endez, 760032, Cali, Colombia}

\emailAdd{$\star$ wilmar.cardona@unesp.br}
\emailAdd{\mbox{$\dagger$ john.orjuela@correounivalle.edu.co}}
\emailAdd{$\ddagger$ cesar.valenzuela@correounivalle.edu.co}

\date{\today}

\abstract{We consider scalar-vector-tensor (SVT) theories with second-order equations of motion and tensor propagation speed equivalent to the speed of light. Under the sub-horizon and the quasi-static approximations we find analytical formulae for an effective dark energy fluid, i.e., sound speed, anisotropic stress as well as energy density and pressure. We took advantage of our general, analytical fluid description and showed that it is possible to design SVT cosmological models which are degenerate with $\Lambda$CDM at the background level while having gravity strength $G_{\rm eff}<G_{\rm N}$ at late-times as well as non-vanishing dark energy perturbations. We implemented SVT designer models in the widely used Boltzmann solver \texttt{CLASS} thus making it possible to test SVT models against astrophysical observations. Our effective fluid approach to SVT models reveals non trivial behaviour in the sound speed and the anisotropic stress well worth an investigation in light of current discrepancies in cosmological parameters such as $H_0$ and $\sigma_8$.}

\begin{document}

\maketitle

\section{Introduction \label{Section:Introduction}}

Pretty interesting discrepancies in cosmological parameters are challenging the standard model of cosmology $\Lambda$CDM \cite{Abdalla:2022yfr,Planck:2018vyg}. Low red-shift measurements of pulsating (Cepheid variables) and exploding stars (Supernovae type Ia) allow a determination of the Hubble constant $H_0$ which disagrees ($\approx 5 \sigma$) with the value inferred by the Planck Collaboration in the context of the \lcdm model \cite{Riess:2019cxk,Freedman:2021ahq,Riess:2021jrx}. A similar situation involves the strength of matter clustering parameterized as $S_8$. The \lcdm value obtained from Cosmic Microwave Background (CMB) anisotropies differs ($2-3 \sigma$) when compared to $S_8$ values determined by probes such as weak lensing and galaxy clustering     \cite{DES:2022ign,Gatti:2021uwl,Zurcher:2021bjz,Huang:2021tvo,Kobayashi:2021oud,Loureiro:2021ruj,PhysRevLett.111.161301,PhysRevD.91.103508,Abbott:2017wau,DES:2021wwk,Blanchard:2021dwr,Heymans:2020gsg,Philcox:2021kcw,Joudaki:2016mvz}. Although Big Bang nucleosynthesis (BBN) is a key ingredient in the standard cosmological model, there is a factor of $\approx 3.5$ discrepancy between the theoretically expected abundance of lithium-7 and the observationally inferred abundance obtained from absorption spectroscopy of metal-poor stars in the galactic halo \cite{Fields:2011zzb,Cyburt:2015mya,Mathews:2019hbi,Sbordone:2010zi,Pitrou:2020etk,Iocco:2008va}. By measuring the sky-averaged 21 cm brightness temperature at red-shift $z \approx 20$, the EDGES experiment indicates a value for the baryon temperature cooler than expected in \lcdm \cite{Bowman:2018yin,PhysRevD.98.083525}. However, a recent analysis bears out earlier concerns and shows no evidence for non-standard cosmology \cite{Singh:2022ivh}. 

These exemplifying discordances could be due to unaccounted for systematic errors in analyses of current data sets, but thus far different analyses do not show a preference for this explanation. There exists also the very interesting possibility that the lack of agreement in some cosmological parameters measured by different experiments could also be  due to a misunderstanding of the underlying physics. In other words, data would be hinting at new physics not taken into consideration within the standard cosmological model \cite{PhysRevLett.124.161301,Poulin:2021bjr,Hill:2021yec}. For instance, a configuration of vector fields leading to an effective Early Dark Energy (EDE) fluid with equation of state $w_{\rm EDE}(z)$, sound speed $c_{s, \rm EDE}^2$, and non-vanishing anisotropic stress $\pi_{\rm EDE}(z)$ could simultaneously soften $H_0$ and $S_8$ tensions \cite{Sabla:2022xzj}.   

The concordance cosmological model \lcdm is not only a good fit for most astrophysical measurements. Being relatively simple (i.e., it is described by six cosmological parameters), the \lcdm model is preferred over its alternatives in analyses performing Bayesian model comparison \cite{Heavens:2017hkr}. However, we must bear in mind that, despite of its success explaining observations, \lcdm is just a pretty good Universe's phenomenological description resting upon elements yet to be understood, e.g., the cosmological constant \cite{Weinberg:1988cp,Carroll:2000fy}, the nature of dark matter \cite{Bertone:2016nfn,PhysRevLett.125.211101,PhysRevLett.126.041303,Carr:2020xqk,PhysRevD.104.043520,PhysRevLett.123.061302,Hertzberg:2019bvt,doi:10.1146/annurev.aa.32.090194.002531}, inflation \cite{Planck:2015sxf,Steinhardt:2002ih,Chowdhury:2019otk,Ijjas:2014nta,Ijjas:2013vea}. Unravelling the nature of dark matter as well as deciphering the reason why the Universe is speeding up from fundamental physics has motivated a considerable amount of research over the past decades.  

In order to address the problem of the Universe late-time accelerating expansion two paths are usually followed in alternative models to $\Lambda$CDM. On the one hand, it is expected that new kinds of matter with the right properties (e.g., a negative equation of state $w<-1/3$) will be discovered in laboratories. For instance, new particles in more complete theories of fundamental interactions could dominate the energy content at  late times, avoid fine-tuning issues and be the reason why the Universe is speeding up \cite{Ratra:1987rm,ArmendarizPicon:2000dh,Copeland:2006wr}. These exotic matter fields are collectively known as dark energy (DE). On the other hand, General Relativity (GR) might require modifications despite its success \cite{Clifton:2011jh,Li:2018tfg}, making plausible to consider modified gravity (MG) theories (for a review on DE and MG models see, for instance, \cite{Bamba:2012cp}). Nevertheless, thus far several tests performed up to extra-galactic and cosmological scales agree very well with GR \cite{PhysRevLett.116.221101,Collett:2018gpf,Ade:2015rim}.

A number of popular DE and MG models (e.g., $f(R)$, Brans-Dicke, kinetic gravity braiding, quintessence, K-essence, etc.) can be nicely encompassed in a unified framework put forward by G. W. Horndeski in 1974 \cite{Horndeski:1974wa}. The generalisation of covariant Galileons led to the rediscovery of Horndeski's theory which ever since got a lot of attention \cite{Deffayet:2011gz,Kobayashi:2011nu,Charmousis:2011bf}. The theory constitutes the most general Lorentz-invariant extension of GR in four dimensions considering non-minimal couplings between the metric tensor and a scalar field, restricting the equations of motion to being second order in the derivatives of the field functions. Although recent measurements of the propagation speed for gravitational waves severely reduced the Horndeski Lagrangian \cite{Amendola:2017orw,Copeland:2018yuh,Crisostomi:2017pjs,Creminelli:2017sry,Sakstein:2017xjx,Ezquiaga:2017ekz,Baker:2017hug,TheLIGOScientific:2017qsa,Kase:2018aps}, remaining degrees of freedom are well worth an investigation.  

Construction of cosmological models is as important as model testing because in this way we can decide about plausibility of different theories for describing nature. Testing cosmological models is not a trivial task and besides the huge effort dealing with astrophysical measurements (e.g., instruments design, data pipelines), it also requires development of software to compute theoretical predictions for a given model. Boltzmann solvers are codes widely used in cosmology nowadays and they have become fundamental for model testing. These codes allow to not only compute the background evolution of a cosmological model, they also solve the involved differential equations governing the evolution of perturbations. \texttt{CAMB} \cite{Lewis:1999bs} and \texttt{CLASS} \cite{Blas:2011rf} are two popular Boltzmann codes which focus on linear order perturbations\footnote{Non-linear corrections are usually taken into consideration via fitting functions such as HALOFIT \cite{Smith:2002dz}.} and compute observables such as the CMB angular power spectrum and the matter power spectrum, and thus make possible testing cosmological models against data sets. 

There are in the literature a number of works where \texttt{CLASS} and \texttt{CAMB} were modified to include cosmological models differing from $\Lambda$CDM. For instance, MG models have been implemented by using functions parameterizing deviations from GR (i.e., $\mu(a,k)$, $\gamma(a,k)$) in \texttt{MGCAMB} \cite{Zhao:2008bn,Hojjati:2011ix,Sakr:2021ylx}; by solving the full system of differential equations for a given model as in \texttt{FRCAMB} \cite{He:2012wq,Xu:2015usa}; by exploiting an Effective Field Theory approach as in \texttt{EFTCAMB} \cite{Gubitosi:2012hu,Hu:2013twa}; by using to good advantage a gauge invariant formalism in the framework of an equation of state (EoS) approach for perturbations in \texttt{CLASS} \cite{Battye:2015hza,Battye:2017ysh}. A remarkable step for the investigation of alternative cosmological models was the development of \texttt{Hi-CLASS} \cite{Zumalacarregui:2016pph} which implements Horndeski theories and, without using neither sub-horizon nor quasi-static approximations, solves differential equations for linear order perturbations. This was indeed not a trivial task due to the high number of degrees of freedom in the theory.

It turns out that by using an Effective Fluid Approach\footnote{For an effective fluid description of MG see, for instance, Refs. \cite{Capozziello:2006dj,Nojiri:2006ri}.} things can become easier when implementing $f(R)$ and Horndeski theories in Boltzmann solvers \cite{Arjona:2018jhh,Arjona:2019rfn}. Quasi-static and sub-horizon approximations can be applied to these kinds of theories so that a fluid description is achieved for the effective DE fluid. Fairly general analytical expressions for the equation of state $w_{\rm DE}(z)$, sound speed $c_{s, \rm DE}^2(z,k)$, and anisotropic stress $\pi_{\rm DE}(z,k)$ enable a relatively easy implementation of these kinds of theories in \texttt{CLASS}. Interestingly, this effective fluid approach leads to no significant loss of accuracy when compared to the exact computation of observables in \texttt{Hi-CLASS} and agrees pretty well with other approaches such as the EoS \cite{Pace:2019uow,Pace:2020qpj,Arjona:2019rfn}. 

Vector fields are also present in nature and it is plausible that they might be related to a late-time accelerating universe \cite{Geng:2021jso, Nakamura:2019phn, Nakamura:2018oyy, deFelice:2017paw, PhysRevLett.127.161302, PhysRevD.81.104015, PhysRevD.78.063005, BeltranJimenez:2008enx, BeltranJimenez:2013btb, DeFelice:2016yws, PhysRevD.94.044024, Armendariz-Picon:2004say, Koivisto:2008ig, Koivisto:2008xf, Thorsrud:2012mu, Landim:2016dxh, Gomez:2020sfz, Gomez:2021jbo, Mehrabi:2015lfa, Alvarez:2019ues, Orjuela-Quintana:2020klr, Guarnizo:2020pkj}. After a successful construction of consistent theories for general scalar-tensor interactions, it was shown that a similar procedure can be worked out for vector-tensor interactions, known as generalised Proca theories \cite{Tasinato:2014eka,Heisenberg:2014rta,Allys:2015sht,Allys:2016jaq,BeltranJimenez:2016rff}. It turns out that more general theories having second-order equations of motion and simultaneously including a scalar field and a vector field, namely  Scalar-Vector-Tensor (SVT) theories, were also found \cite{Heisenberg:2018mxx,Heisenberg:2018acv,Kase:2018nwt}. Although SVT theories encompass both Horndeski and generalised Proca theories, and therefore might have an interesting, new, richer phenomenology, they have received little attention thus far in the literature. In fact, as far as we know, there is no public Boltzmann code where SVT theories be fully implemented so that their phenomenology can be investigated in detail \cite{Lagos:2017hdr}. This is a major disadvantage for model testing because we cannot compare theoretical predictions against astrophysical measurements, hence seriously decide about viability of alternative cosmological models. In this work we show that SVT theories can be mapped into an effective fluid so that their phenomenology can be investigated through Boltzmann solvers such as \texttt{CLASS}. We apply quasi-static and sub-horizon approximations to SVT theories and find fairly general analytical expressions for the quantities defining the effective fluid, that is, $w_{\rm DE}(z)$, $c_{s, \rm DE}^2(z,k)$, and $\pi_{\rm DE}(z,k)$.\footnote{While in this work we focus on DE, SVT theories have also been studied in inflation (see, for instance, \cite{Oliveros:2022njz}).} Our approach can be useful to carry out further investigations, for instance, on general EDE and Relativistic Modified Newtonian Dynamics (RMOND) models \cite{Sabla:2022xzj,PhysRevLett.127.161302}.  

The paper is organised as follows. Our notation is set in Section \ref{section:wcdm}. We discuss SVT theories in Section \ref{Section:models}, and in Section \ref{Section:EFA} we explain the effective fluid description. Then we show that our approach allows designing cosmological models having a behaviour in good agreement with observations (Section \ref{Section:Des-SVT}). We give our concluding remarks in Section \ref{Section:conclusions} and provide details of our computations in Appendices \ref{App:1}-\ref{App:4}.  

\section{Perturbations in a general dark energy model}
\label{section:wcdm}

A popular approach to explain the  current accelerated expansion of the universe, as well as other observations, is to assume the existence of yet undetected matter fields generally called as dark energy. In general, these new fields are treated as fluids having equation of state $w$, sound speed $c_s^2$, and anisotropic stress $\sigma$ \cite{Hu:1998kj}. Since we are interested in the late-time dynamics, we consider the existence of a DE fluid aside non-relativistic matter (baryon and dark matter). In addition, we take into consideration the gravitational field through the Einstein-Hilbert action so that 
\be
S = \int \text{d}^{4} x \sqrt{-g} \[ \frac{1}{2\kappa} R + \mc{L}_m + \mathcal{L} \],  \label{eq:action-GR}%
\ee
where $g$ is the determinant of the metric $g_{\mu\nu}$, $R$ is the Ricci scalar, $\kappa\equiv \frac{8\pi G_N}{c^4}$, $\mc{L}_m$ is the matter Lagrangian, and $\mc{L}$ is the Lagrangian of an arbitrary fluid, which we will later identify as dark energy.\footnote{Definitions throughout the paper: speed of light $c=1$; $\kappa=8\pi G_N$ with $G_N$ being the bare Newton's constant; $(-+++)$ for the metric signature; the Riemann and Ricci tensors are denoted respectively as $R^\alpha_{\ \mu\beta\nu}$ and $R_{\mu\nu}=R^\alpha_{\ \mu\alpha\nu}$.} From Eq. \eqref{eq:action-GR}, as it is well-known, we can obtain the Einstein field equations through the principle of least action 
\be
G_{\mu\nu} = \kappa \, \( T^{(m)}_{\mu\nu} + T_{\mu\nu} \),
\label{eq:EE}
\ee
where $G_{\mu\nu}$ is the Einstein tensor, $T^{(m)}_{\mu\nu}$ and $T_{\mu\nu}$ are the  energy-momentum tensors for matter and arbitrary fluid, respectively. Since field equations \eqref{eq:EE} are rather general it is usual to make a few assumptions. Recent analyses \cite{Planck:2018vyg,PhysRevLett.122.171301}  show that the standard cosmological model \lcdm is in very good agreement with observations. The model assumes a flat, linearly perturbed Friedman-Lema\^{i}tre-Robertson-Walker (FLRW) metric that in the Newtonian gauge reads
\be
\text{d} s^2 = a(\eta)^2 \[ - \{1 + 2 \Psi(\boldsymbol{x}, \eta) \} \text{d} \eta^2 + \{ 1 + 2 \Phi(\boldsymbol{x}, \eta) \} \delta_{i j} \text{d} x^i \text{d} x^j \],
\label{eq:FRWpert}
\ee
which takes into account the existence of tiny inhomogeneities in the energy distribution of the universe. In Eq. \eqref{eq:FRWpert} $a$ is the scale factor, $\boldsymbol{x}$ indicates spatial coordinates, $\eta$ is the conformal time, and $\Psi$ and $\Phi$ are the gravitational potentials. Observations and simulations indeed support the assumption that on large enough scales the universe is statistically homogeneous and isotropic \cite{Hogg:2004vw,Ade:2015hxq,Marinoni:2012ba}. Throughout the paper we assume the metric in Eq. \eqref{eq:FRWpert}, unless otherwise specified.

We regard that our DE fluid is an ideal fluid also having tiny perturbations, so that its energy-momentum tensor is given by 
\be
T^\mu_{\ \nu} = P \delta^\mu_{\ \nu} + (\rho + P) U^\mu  U_\nu, 
\label{eq:enten}
\ee
where $P$, $\rho$, and $U^\mu \equiv a^{-1} \left( 1 - \Psi,  \boldsymbol{u} \right)$ are respectively the pressure,  the energy density, the velocity four-vector, and $\boldsymbol{u} = \boldsymbol{x}'$. The linearised energy-momentum tensor then reads 
\bea
T^0_{\ 0} &=& -(\bar{\rho} + \delta \rho), \label{eq:effectTmn1}\\
T^0_{\ i} &=& (\bar{\rho} + \bar{P}) u_i,\label{eq:effectTmnvde}\\
T^i_{\ j} &=& (\bar{P} + \delta P) \delta^i_{\ j} + \Sigma^i_{\  j}, \label{eq:effectTmn}
\eea
where $\bar{\rho} (\eta)$ and $\bar{P} (\eta)$ are the background energy density and pressure of the fluid, while $\delta \rho (\boldsymbol{x}, \eta)$ and $\delta P (\boldsymbol{x}, \eta)$ are their respective perturbations, and $\Sigma^i_{\ j} (\boldsymbol{x}, \eta) \equiv T^i_{\ j} - \delta^i_{\ j} T^k_{\ k} /3$ is its anisotropic stress tensor.\footnote{In our notation,  a prime denotes derivative with respect to the conformal time. In addition, Greek indices run from $0$ to $3$ whereas Latin indices take on values from $1$ to $3$.}

\subsection{Background and linear perturbations}

For the FLRW metric \eqref{eq:FRWpert} the unperturbed Einstein field equations \eqref{eq:EE} read
\be
\mc{H}^2 = \frac{\kappa}{3} a^2 \( \bar{\rho}_m + \bar{\rho} \), \qquad \mc{H}' = - \frac{\kappa}{6} a^2 \left(\bar{\rho}_m + \bar{\rho} + 3 \bar{P} \right),
\label{eq:Friedman-zero-order}
\ee
and describe the background evolution. In Eqs. \eqref{eq:Friedman-zero-order}, 
$\mc{H} \equiv \frac{a'}{a}$ is the conformal Hubble parameter, $\bar{\rho}_m$ is the background density of matter, and we have assumed that matter is pressure-less.\footnote{Note that $\mathcal{H}$ and the Hubble parameter $H$ are related through $\mathcal{H} = a H$.}

Now, we regard linear perturbations to the Einstein field equations \eqref{eq:EE} and work on the Newtonian gauge \eqref{eq:FRWpert}. We find
\be
3 \mc{H} \left( \mc{H}\Psi - \Phi' \right) - k^2 \Phi = \frac{\kappa}{2} a^2 \( \delta T^{0 \, (m)}_{\ 0} + \delta T^0_{\ 0} \), \label{eq:phiprimeeq}
\ee
\be
k^2 \left( \mc{H} \Psi - \Phi' \right) = \frac{\kappa}{2} a^2 \[ \bar{\rho}_m \theta_m + \(\bar{\rho} + \bar{P} \) \theta \],\label{eq:phiprimeeq1}
\ee
\be
- \frac{k^2}{3 a^2} (\Phi + \Psi) + \left( 2  \mc{H}' + \mc{H}^2 \right) \Psi + \mc{H} \left( \Psi' - 2 \Phi' \right) - \Phi''
= \frac{\kappa}{6} a^2 \delta T^{i}_{\ i},
\ee
\be
- k^2 (\Phi + \Psi) = \frac{3\kappa}{2} a^2 (\bar{\rho}+\bar{P})\sigma \label{eq:anisoeq},
\ee
where $\theta\equiv ik^ju_j$ is the velocity divergence, $\theta_m$ is the velocity divergence for matter, $k$ is the wavenumber, and we write the anisotropic stress as  $(\bar{\rho}+\bar{P})\sigma\equiv-(\hat{k}_i\hat{k}_j-\frac13 \delta_{ij})\Sigma^{ij}$. We have assumed that matter has no anisotropic stress, $\sigma_m = 0$, and that its pressure perturbation also vanishes, $\delta T^{i \, (m)}_{\ i} = 0$.
Conservation laws for energy and momentum lead to a couple of differential equations governing the evolution of linear perturbations. If the energy-momentum tensor for our general fluid is conserved, it satisfies  $\nabla_\mu T^\mu_{\ \nu} = 0$, specifically,
\be
\delta' = - (1 + w) \( \theta + 3 \Phi' \) - 3 \mc{H} \left( c_s^2 - w \right) \delta,
\label{eq:cons1}
\ee
\be
\theta' = - \mc{H} (1 - 3 w) \theta - \frac{w'}{1 + w} \theta + \frac{c_s^2}{1 + w} k^2 \delta - k^2 \sigma + k^2 \Psi,
\label{eq:cons2}
\ee
where the equation of state parameter is defined as $w\equiv\frac{\bar{P}}{\bar{\rho}}$, and the sound speed as  \mbox{$c_s^2\equiv\frac{\delta P}{\delta \rho}$}. From Eqs.  \eqref{eq:cons1}-\eqref{eq:cons2}, it becomes clear that when the equation of state crosses $-1$, problems emerge because there is a singularity. The trouble can be solved by a variable transformation:  we use the scalar velocity perturbation $V\equiv i k^jT^0_{\ j}/\rho=(1+w)\theta$ instead of the velocity divergence $\theta$. In terms of this new variable the evolution equations \eqref{eq:cons1}-\eqref{eq:cons2} are
\be
\delta\text{'} = - 3(1 + w) \Phi\text{'} - \frac{V}{a^2 H} - \frac{3}{a}\left(c_s^2 - w \right) \delta,
\label{Eq:evolution-delta}
\ee
\be
V\text{'} = -(1 - 3 w) \frac{V}{a} + \frac{k^2}{a^2 H} c_s^2 \delta + (1 + w)\frac{k^2}{a^2 H} \Psi - \frac{2}{3} \frac{k^2}{a^2 H} \pi,
\label{Eq:evolution-V}
\ee
where we define the anisotropic stress parameter $\pi\equiv\frac32(1+w)\sigma$ and a quote ' denotes a derivative with respect to the scale factor. 

\section{Scalar-Vector-Tensor theories} 
\label{Section:models}

Although plausible, SVT theories have not gotten too much attention over the past years. These theories can accommodate Horndeski as well as generalised Proca theories and might have interesting phenomenology for the late-time universe \cite{Heisenberg:2018mxx}. In this section, we introduce the most general SVT Lagrangian. Let us consider
\begin{equation}
\mc{L} \equiv \sum_{i = 2}^6 \mc{L}_i^\text{SVT} + \sum_{i = 2}^5 \mc{L}_i^\text{ST} + \mc{L}_m, \label{eq:L}
\end{equation}
where terms $\mc{L}_i^\text{ST}$ represent the scalar-tensor interactions, and terms $\mc{L}_i^\text{SVT}$ denote the scalar-vector-tensor interactions. 

A scalar field $\vp$, a vector field $A_\mu$, and the gravitational field $g_{\mu\nu}$ interact with each other through the Lagrangians $\mc{L}_i^\text{SVT}$ taking into account interactions with a broken U(1) gauge symmetry. The Lagrangians for SVT interactions read
\begin{align}
\mc{L}_2^\text{SVT} &= f_2 (\vp, X_1, X_2, X_3, F, Y_1, Y_2, Y_3), \label{eq:svt2}\\
\mc{L}_3^\text{SVT} &= f_3 (\vp, X_3) g^{\mu\nu} S_{\mu\nu} + \tilde{f}_3 (\phi, X_3) A^\mu A^\nu S_{\mu\nu}, \\
\mc{L}_4^\text{SVT} &= f_4 (\vp, X_3) R + f_{4 X_3} (\vp, X_3) \left\{ \( \nabla_\mu A^\mu \)^2 - \nabla_\mu A_\nu \nabla^\nu A^\mu \right\}, \\
\mc{L}_5^\text{SVT} &= f_5 (\vp, X_3) G^{\mu\nu} \nabla_\mu A_\nu + \mc{M}_5^{\mu\nu} \nabla_\mu \nabla_\nu \vp +\mc{N}^{\mu\nu}_5 S_{\mu\nu} \nonumber \\
 &- \frac{1}{6} f_{5 X_3} (\vp, X_3) \left\{ \( \nabla_\mu A^\mu \)^3 - 3 \( \nabla_\mu A^\mu \) \nabla_\rho A_\sigma \nabla^\sigma A^\rho + 2 \nabla_\rho A_\sigma \nabla^\tau A^\rho \nabla^\sigma A_\tau \right\}, \\
\mc{L}_6^\text{SVT} &= f_6 (\vp, X_1) L^{\mu\nu\alpha\beta} F_{\mu\nu} F_{\alpha\beta} + \tilde{f}_6 (\vp, X_3) L^{\mu\nu\alpha\beta} F_{\mu\nu} F_{\alpha\beta}  \nonumber \\
 &+ \mc{M}^{\mu\nu\alpha\beta}_6 \nabla_\mu \nabla_\alpha \vp \nabla_\nu \nabla_\beta \vp + \mc{N}^{\mu\nu\alpha\beta}_6 S_{\mu\alpha} S_{\nu\beta}, \label{eq:svt6}
\end{align}
where we have used the simplified notation $g_{\xi} = \tfrac{\partial g}{\partial \xi}$, for the derivative of any free function $g$ with respect to a scalar $\xi$. Let us define the different terms involved in Eqs. \eqref{eq:svt2}-\eqref{eq:svt6}. Firstly, note that the kinetic term of the scalar field $\vp$, the coupling between $\vp$ and the vector field $A_\mu$, and the quadratic term of $A_\mu$ are defined respectively by
\begin{equation}
X_1 \equiv - \frac{1}{2} \nabla_\mu \vp \nabla^\mu \vp, \quad X_2 \equiv - \frac{1}{2} A_\mu \nabla^\mu \vp, \quad X_3 \equiv - \frac{1}{2} A_\mu A^\mu. \label{eq:x1x2x3}
\end{equation}
Secondly, the antisymmetric strength tensor $F_{\mu\nu}$ and its dual $\tilde{F}^{\mu\nu}$ are constructed from $A_\mu$ as
\begin{equation}
F_{\mu\nu} \equiv \nabla_\mu A_\nu - \nabla_\nu A_\mu, \quad \tilde{F}^{\mu\nu} \equiv \frac{1}{2} \varepsilon^{\mu\nu\alpha\beta} F_{\alpha\beta},
\end{equation}
where $\varepsilon^{\mu\nu\alpha\beta} \equiv \frac{\epsilon^{\mu\nu\alpha\beta}}{\sqrt{-g}}$, $\epsilon^{\mu\nu\alpha\beta}$ is the Levi-Civita symbol. Thirdly, using $F_{\mu\nu}$ we can construct the Lorentz invariant quantities 
\begin{align}
F &\equiv - \frac{1}{4} F_{\mu\nu} F^{\mu\nu}, \nn \\
Y_1 \equiv \nabla_\mu \vp \nabla_\nu \vp F^{\mu\alpha} F^\nu_{\ \alpha}, \qquad Y_2 &\equiv \nabla_\mu \vp A_\nu F^{\mu\alpha} F^\nu_{\ \alpha}, \qquad Y_3 \equiv A_\mu A_\nu F^{\mu\alpha} F^\nu_{\ \alpha}, \label{eq:lorentz-inv} 
\end{align} 
which vanish in the scalar limit when $A_\mu \rightarrow \nabla_\mu \pi$, $\pi$ being a scalar field. Note that quantities in Eqs. \eqref{eq:lorentz-inv} carry the intrinsic vector modes in the Lagrangian $\mc{L}^\text{SVT}_2$. Fourthly, in $\mc{L}^\text{SVT}_3$ we find a symmetric tensor constructed from $A_\mu$ as
\begin{equation}
S_{\mu\nu} \equiv \nabla_\mu A_\nu + \nabla_\nu A_\mu.
\end{equation}
Fifthly, note that the intrinsic vector modes in Lagrangians  $\mc{L}^\text{SVT}_5$ and $\mc{L}^\text{SVT}_6$ are carried by the tensors $\mc{M}$ and $\mc{N}$ given by
\begin{equation}
\mc{M}^{\mu\nu}_5 \equiv \mc{G}_{\rho\sigma}^{h_5} \tilde{F}^{\mu\rho} \tilde{F}^{\nu\sigma}, \quad \mc{N}^{\mu\nu}_5 \equiv \mc{G}_{\rho\sigma}^{\tilde{h}_5} \tilde{F}^{\mu\rho} \tilde{F}^{\nu\sigma},
\end{equation}
\begin{equation}
\mc{M}^{\mu\nu\alpha\beta}_6 \equiv 2 f_{6 X_1} (\vp, X_1) \tilde{F}^{\mu\nu} \tilde{F}^{\alpha\beta}, \quad \mc{N}^{\mu\nu\alpha\beta}_6 \equiv \frac{1}{2} \tilde{f}_{6 X_3} (\vp, X_1) \tilde{F}^{\mu\nu} \tilde{F}^{\alpha\beta},
\end{equation}
where
\begin{align}
\mc{G}^{h_5}_{\rho\sigma} &\equiv h_{51} (\vp, X_i) g_{\rho\sigma} + h_{52} (\vp, X_i) \nabla_\rho \vp \nabla_\sigma \vp + h_{53} (\vp, X_i) A_\rho A_\sigma + h_{54} (\vp, X_i) A_\rho \nabla_\sigma \vp, \label{eq:Gh5}\\
\mc{G}^{\tilde{h}_5}_{\rho\sigma} &\equiv \tilde{h}_{51} (\vp, X_i) g_{\rho\sigma} + \tilde{h}_{52} (\vp, X_i) \nabla_\rho \vp \nabla_\sigma \vp + \tilde{h}_{53} (\vp, X_i) A_\rho A_\sigma + \tilde{h}_{54} (\vp, X_i) A_\rho \nabla_\sigma \vp, \label{eq:Ght5}
\end{align}
are effective metrics containing possible combinations of $g_{\rho\sigma}$, $A_\rho$, and $\nabla_\sigma \vp$, and $X_i$ is a short-hand notation for the set $\{ X_1, X_2, X_3\}$. Finally,  the double dual Riemann tensor is defined as
\begin{equation}
L^{\mu\nu\alpha\beta} \equiv \frac{1}{4} \varepsilon^{\mu\nu\rho\sigma} \varepsilon^{\alpha\beta\gamma\delta} R_{\rho\sigma\gamma\delta}. \label{eq:Lmunualphabeta}
\end{equation}

Scalar-tensor interactions are taken into consideration in Eq. \eqref{eq:L} through the Horndeski theory
\begin{align}
\mc{L}_2^\text{ST} &= G_2 (\vp, X_1), \label{eq:st2}\\
\mc{L}_3^\text{ST} &= - G_3 (\vp, X_1) \square \vp, \\
\mc{L}_4^\text{ST} &= G_4 (\vp, X_1) R + G_{4 X_1} (\vp, X_1) \left\{ \( \square \vp \)^2 - \nabla_\mu \nabla_\nu \vp \nabla^\nu \nabla^\mu \vp \right\}, \\
\mc{L}_5^\text{ST} &= G_5 (\vp, X_1) G^{\mu\nu} \nabla_\mu \nabla_\nu \vp \label{eq:st5}\\
 &- \frac{1}{6} G_{5 X_1} (\vp, X_1) \left\{ \( \square \vp \)^3 - 3 \( \square \vp \) \nabla_\mu \nabla_\nu \vp \nabla^\nu \nabla^\mu \vp + 2 \nabla^\mu \nabla_\sigma \vp \nabla^\sigma \nabla_\rho \vp \nabla^\rho \nabla_\mu \vp \right\} \nonumber. 
\end{align}
In Eqs. \eqref{eq:svt2}-\eqref{eq:svt6}, Eqs. \eqref{eq:Gh5}-\eqref{eq:Ght5}, and Eqs. \eqref{eq:st2}-\eqref{eq:st5}, all the $f_i$, $\tilde{f}_i$, $h_{5i}$, $\tilde{h}_{5i}$, and $G_i$ denote free functions.  

Although in its most general form the theory \eqref{eq:L} has several free functions, it got significantly constrained by the discovery of gravitational waves \cite{Abbott:2017oio,Creminelli:2017sry,Sakstein:2017xjx,Ezquiaga:2017ekz,Baker:2017hug,Amendola:2017orw,Crisostomi:2017pjs,Frusciante:2018jzw,Kase:2018aps,McManus:2016kxu,Lombriser:2015sxa,Copeland:2018yuh,Noller:2018wyv,deRham:2018red}. In the next subsection we explain it with more details.

\subsection{Remaining SVT theories} 
\label{Section:Constraints-FLRW}

An anisotropic expansion of the Universe is strongly disfavoured \cite{PhysRevLett.117.131302}, hence in this paper we restrict ourselves to the FLRW metric in Eq. \eqref{eq:FRWpert}. Our matter fields will also respect constraints on homogeneity and  isotropy, thus we will regard a scalar field and a vector field with the following configurations
\begin{equation}
\vp \equiv \vp (\eta) + \delta \vp(\boldsymbol{x}, \eta), \qquad A_\mu \equiv (A_0 (\eta) + \delta A_0(\boldsymbol{x}, \eta), \delta A_i(\boldsymbol{x}, \eta)).
\label{Field configurations}
\end{equation}
By using definitions in Eqs. \eqref{eq:x1x2x3}-\eqref{eq:Lmunualphabeta}, it is relatively easy to show that up to first order in perturbation theory
\begin{equation}
F_{\mu\nu} = 0, \quad \tilde{F}^{\mu\nu} = 0,
\end{equation}
which implies
\begin{equation}
 F = Y_i = 0, \ \mc{M}^{\mu\nu}_5 = \mc{N}^{\mu\nu}_5 = 0, \ \mc{M}^{\mu\nu\alpha\beta} = \mc{N}^{\mu\nu\alpha\beta} = 0, \ L^{\mu\nu\alpha\beta} F_{\mu\nu} F_{\alpha\beta} = 0.
\end{equation}
As a result, Lagrangians \eqref{eq:svt2}-\eqref{eq:svt6} get reduced. In particular, $\mc{L}_2^\text{SVT}$ becomes $f_2 (\vp, X_1, X_2, X_3)$, in $\mc{L}^\text{SVT}_5$ the terms involving the matrices $\mc{M}$ and $\mc{N}$ vanish, while the Lagrangian $\mc{L}^\text{SVT}_6$ fully disappears.

\subsubsection{Speed of gravitational waves}

Taking into consideration the full Lagrangian of Horndeski theory (see Eqs. \eqref{eq:st2}-\eqref{eq:st5}) and the remaining parts of the Lagrangians \eqref{eq:svt2}-\eqref{eq:svt6}, we can obtain the propagation speed of gravitational waves through the computation of the evolution equation for tensor modes. For this calculation, the perturbed metric reads 
\be
\text{d} s^2 = a^2(\eta) \[ - \text{d} \eta^2 + \{ \delta_{i j} + h_{i j}(\boldsymbol{x}, \eta) \}  \text{d} x^i \text{d} x^j \],
\label{tensor modes}\ee
where $h_{i j}$ are the tensor modes. Using the metric \eqref{tensor modes}, and the configuration for the fields in Eq. \eqref{Field configurations}, we compute the gravitational field equations for the tensor modes up to first order. We get
\be
\( h^i_j \)'' + \( 2 \mc{H} + \gamma_T \) \( h^i_j \)' + c_T^2 k^2 h^i_j = 0,
\ee
where $\gamma_T$ is a drag term due to the expansionary dynamics and 
\begin{equation}
c_T^2 \equiv \frac{f_{4} - \frac{A_{0}^2 A_0' f_{5 X_{3}}}{2 a^4} + G_{4} + \frac{A_{0}^3 f_{5 X_{3}} \mathcal{H}}{2 a^4} - \frac{A_{0} f_{5  \varphi } \vp'}{2 a^2} - \frac{G_{5  \varphi } \vp'^2}{2 a^2} + \frac{G_{5 X_{1}} \mathcal{H} \vp'^3}{2 a^4} - \frac{G_{5 X_{1}} \vp'^2 \varphi''}{2 a^4}}{f_{4} - \frac{A_{0}^2 f_{4 X_{3}}}{a^2} + G_{4} - \frac{A_{0}^3 f_{5 X_{3}} \mathcal{H}}{2 a^4} + \frac{A_{0} f_{5  \varphi } \vp'}{2 a^2} - \frac{G_{4 X_{1}} \vp'^2}{a^2} + \frac{G_{5  \varphi } \vp'^2}{2 a^2} - \frac{G_{5 X_{1}} \mathcal{H} \vp'^3}{2 a^4}}
\end{equation}
is the speed of gravitational waves, which agrees with the result presented in Ref. \cite{Kase:2018nwt}. As previously mentioned, observations indicate that the propagation speed of gravitational waves is practically the speed of light. If we want general SVT theories to satisfy the constraint $c_T^2 = 1$ without fine-tuning,\footnote{In Bayesian statistics models that have to be finely tuned to fit the data are penalised by the Occam factor \cite{Mckay2003}.} the following free functions in the general Lagrangian have to fulfill\footnote{For the choice \eqref{GW constraint}, the drag term $\gamma_T$ vanishes, and we get the usual wave equation for a mass-less field $h_{ij}$ propagating at the speed of light $c_T^2 = 1$.}
\begin{equation}
f_4 = f_4 (\vp), \quad f_5 = \text{constant}, \quad G_4 = G_4 (\vp), \quad G_5 = \text{constant}.
\label{GW constraint}
\end{equation}
Consequently, the remaining SVT theories are given by
\begin{align}
\mc{L}_2^\text{SVT} &= f_2 (\vp, X_1, X_2, X_3), \\
\mc{L}_3^\text{SVT} &= f_3 (\vp, X_3) g^{\mu\nu} S_{\mu\nu} + \tilde{f}_3 (\vp, X_3) A^\mu A^\nu S_{\mu\nu}, \\
\mc{L}_3^\text{ST} &= - G_3 (\vp, X_1) \square \vp, \\
\mc{L}_4^\text{ST} &= G_4 (\vp) R, \label{L4ST}
\end{align} 
and the complete Lagrangian reads
\begin{equation} \label{Eff Lagrangian}
\mc{L} = \mc{L}_2^\text{SVT} + \mc{L}_3^\text{SVT} + \mc{L}_3^\text{ST} + \mc{L}_4^\text{ST}.
\end{equation}
Note that the Lagrangian $\mc{L}_2^\text{ST}$ is contained in $\mc{L}_2^\text{SVT}$, while $\mc{L}_4^\text{SVT}$ is taken into account in $\mc{L}_4^\text{ST}$. Since $G_5$ and $f_5$ are constants, the Lagrangians $\mc{L}_5^\text{SVT}$ and $\mc{L}_5^\text{ST}$ are total derivatives. As a result $\mc{L}_5^\text{SVT}$ and $\mc{L}_5^\text{ST}$ are disregarded in \eqref{Eff Lagrangian}; they do not contribute to the dynamics. In the next sections, we focus on the cosmological implications  of the  Lagrangian \eqref{Eff Lagrangian}. In order to avoid very long expressions within the main text, we provide our results in terms of coefficients defined in the appendices. 

\subsection{Equations of motion} 
\label{Section:Gra-EoM}

Varying the action for the Lagrangian in Eq. \eqref{Eff Lagrangian} with respect to the metric $g^{\mu\nu}$, we obtain the gravitational field equations
\begin{equation} \label{Gra Field Eqs}
\sum_{i = 2}^{3} \mc{G}^{(i)}_{\mu\nu} + \sum_{i = 3}^{4} \mathscr{H}^{(i)}_{\mu\nu} = \frac{1}{2} \, T^{(m)}_{\mu\nu},
\end{equation}
while varying with respect to the scalar field $\vp$, and the vector field $A_\mu$ we get
\begin{equation} 
\sum_{i = 2}^{3} \mc{J}_{i} + \sum_{i = 3}^{4} \mc{K}_{i} = 0, \quad \sum_{i = 2}^{3} \mc{A}^\mu_{\ (i)} = 0,
\label{Gra Scalar Vec Eqs}
\end{equation}
respectively. The terms $\mc{G}^{(i)}_{\mu\nu}$,  $\mc{J}_i$, and $\mc{A}^\mu_{\ (i)}$, are associated with the SVT  Lagrangians $\mc{L}_i^\text{SVT}$ ($i = 2, 3$), while $\mathscr{H}^{(i)}_{\mu\nu}$ and $\mc{K}_i$ are associated to $\mc{L}_i^\text{ST}$ ($i = 3, 4$). The expressions for these terms can be found in the appendices \ref{App: Gral Gravitational Field}, \ref{App: Gral Scalar Field}, and \ref{App: Gral Vector Field}.


The background equations of motion are obtained after replacing the unperturbed FLRW metric \eqref{eq:FRWpert}, and the zeroth-order part of the scalar field and the vector field [see Eqs. \eqref{Field configurations}] in Eqs. \eqref{Gra Field Eqs} and \eqref{Gra Scalar Vec Eqs}. For the gravitational field equations, due to rotational invariance, only the ``time-time'' equation and one of the diagonal ``space-space'' equations are needed, i.e.
\begin{equation}
\sum_{i = 2}^{3} \mc{G}_{0 0}^{(i)} + \sum_{i = 3}^{4} \mathscr{H}_{0 0}^{(i)} = \frac{1}{2} T^{(m)}_{0 0} = \frac{1}{2} a^2 \bar{\rho}_m, \qquad \sum_{i = 2}^{3} \mc{G}_{1 1}^{(i^)} + \sum_{i = 3}^{4} \mathscr{H}_{1 1}^{(i)} = \frac{1}{2} T^{(m)}_{1 1} = 0,
\label{Gral Back Einstein Eqs}
\end{equation}
where $T^{(m)}_{11} = 0$ since we have assumed that matter is a pressure-less fluid. For the scalar field and the vector field we obtain
\begin{equation}
\sum_{i = 2}^{3} \mc{\bar{J}}_i + \sum_{i = 3}^{4} \mc{\bar{K}}_i= 0, \quad \sum_{i = 2}^{3} \mc{\bar{A}}_i = 0,
\label{Back Scalar Vector Eqs}
\end{equation}
where $\mc{A}^0_{\ (i)} \equiv \mc{\bar{A}}_i$, since only the time component of the vector field has dynamics at the background level. The expressions for $\mc{G}_{00}^{(i)}$ and $\mathscr{H}_{00}^{(i)}$ are found in Appendix \ref{App: time-time Back Eq}, $\mc{G}_{11}^{(i)}$ and $\mathscr{H}_{11}^{(i)}$ in Appendix \ref{App: trace space-space Back Eq}, those for $\bar{\mc{J}}_i$, $\bar{\mc{K}}_i$ and $\bar{\mc{A}}_i$ in Appendix \ref{App: Scalar Back Eq}.

Before discussing the linear perturbations of the model, we want to mention that our results differ from those in Ref. \cite{Kase:2018nwt} due to  the choice of the vector field profile. In Ref. \cite{Kase:2018nwt}, the homogeneous vector field is chosen as
\be 
A_\mu \equiv (N(t) A_0(t), 0, 0, 0),
\ee
where $N(t)$ is the lapse function, which is defined in the background  metric as
\be 
\text{d} s^2 = - N^2(t) \text{d} t^2 + a^2(t) \delta_{i j} \text{d} x^i \text{d} x^j.
\label{reduce metric}
\ee 
The choice \eqref{reduce metric} implies that $X_3 = A_0^2/2$, and thus the variation of $\mc{L}^{\text{SVT}}_2$ with respect to $N$ will yield no terms with $f_{2 X_3}$. Furthermore, the Lagrangian $\mc{L}_3^\text{SVT}$ will not contribute to the first Friedman equation. In our case, these terms do appear in the first Friedman equation, as can be seen in Appendix \ref{App: time-time Back Eq}, where $f_{2 X_3}$ can be found and $\mc{G}_{00}^{(3)}$ is not zero. Having clarified this aspect, let us discuss the first order perturbations of the theory.

Since we are only interested in scalar perturbations, we take just the scalar part of the perturbed spatial component of the vector field in Eq. \eqref{Field configurations}, namely, $\delta A_i (\boldsymbol{x}, \eta) \equiv \partial_i \psi (\boldsymbol{x}, \eta)$, where $\psi (\boldsymbol{x}, \eta)$ is a scalar field. Having this in mind, the linear perturbations of the gravitational equations are given by
\begin{align}
0 &= A_1 \frac{\Phi'}{a} + A_2 \frac{ \delta \vp'}{a} + A_3 \frac{k^2}{a^2} \Phi + A_4 \Psi + \( A_5 \frac{k^2}{a^2} - \mu_\vp \) \delta \vp + A_6 \frac{\delta A_0}{a} + A_7 \frac{k^2}{a^2} \psi -  \delta \rho_m, \label{Pert time time Eq}\\
\nn \\
0 &= C_1 \frac{\Phi'}{a} + C_2 \frac{\delta \vp'}{a} + C_3 \Psi + C_4 \delta \vp + C_5 \frac{\delta A_0}{a} + C_6 \psi - \frac{a \bar{\rho}_m V_m}{k^2}, \label{Pert time space Eq}\\
\nn \\
0 &= B_1 \frac{\Phi''}{a^2} + B_2 \frac{\delta \vp''}{a^2} + B_3 \frac{\Phi'}{a} + B_4 \frac{\delta \vp'}{a} + B_5 \frac{\Psi'}{a} + B_6 \frac{k^2}{a^2} \Phi + \( B_7 \frac{k^2}{a^2} + 3 \nu_\vp \) \delta \vp \label{Pert trace Eq} \\ 
 &+ \( B_8 \frac{k^2}{a^2} + B_9 \) \Psi + B_{10} \frac{\delta A'_0}{a^2} + B_{11} \frac{\delta A_0}{a}, \nn \\
\nn \\
0 &= G_4 \( \Psi + \Phi \) + G_{4 \vp} \delta \vp, \label{Long Traceless Eq}
\end{align}
corresponding to the ``time-time'', longitudinal ``time-space'', trace ``space-space'', and longitudinal trace-less ``space-space'' parts of the gravitational field equations \eqref{Gra Field Eqs}, respectively. Here, $\delta \rho_m$ and $V_m$ are the density perturbation and the scalar velocity of matter, respectively. 

The evolution of linear perturbations for the scalar field, for the temporal component, and for the spatial component of the vector field are obtained from \eqref{Gra Scalar Vec Eqs}, respectively giving 
\begin{align}
0 &= D_1 \frac{\Phi''}{a^2} + D_2 \frac{\delta \vp''}{a^2} + D_3 \frac{\Phi'}{a} + D_4 \frac{\delta \vp'}{a} + D_5 \frac{\Psi'}{a} + \( D_7 \frac{k^2}{a^2} + D_8 \) \Phi \label{Pert Scalar Eq}\\
 &+ \( D_9 \frac{k^2}{a^2} - m_\vp^2 \) \delta \vp + \( D_{10} \frac{k^2}{a^2} + D_{11} \) \Psi + D_{12} \frac{\delta A'_0}{a^2} + D_{13} \frac{\delta A_0}{a} + D_{14} \frac{k^2}{a^2} \psi, \nn \\
\nn \\
0 &= F_1 \frac{\Phi'}{a^2} +  F_2 \frac{\delta \vp'}{a^2} + F_3 \frac{\Psi}{a} + F_4 \frac{\delta \vp}{a} + F_5 \frac{\delta A_0}{a^2} + F_6 \frac{k^2}{a^2} \frac{\psi}{a}, \label{Pert Temp Vec Eq}\\
\nn \\
0 &= H_1 \frac{\Psi}{a^2} + H_2 \frac{\delta \vp}{a^2} + H_3 \frac{\delta A_0}{a^3} + H_4 \frac{\psi}{a^2}. \label{Pert Space Vec Eq}
\end{align}
All the coefficients in the first-order equations \eqref{Pert time time Eq}-\eqref{Pert Space Vec Eq} can be found in Appendix \ref{App: Perturbations Coefficients}.

\section{The effective fluid approach} 
\label{Section:EFA}

In this section, we show that it is possible to rearrange the equations previously obtained, in order to define an effective dark energy fluid.\footnote{Effective quantities are also studied in Ref.~\cite{Romano:2018frb}.} First, note that an equation similar to the gravitational field equations
\be
G_{\mu\nu} = \kappa \( T_{\mu\nu}^{(m)} + T_{\mu\nu}^\text{(DE)} \),
\ee
can be obtained in SVT theories if we define the energy-momentum tensor of dark energy as
\be
T_{\mu\nu}^{(\text{DE})} \equiv \frac{1}{\kappa} G_{\mu\nu} - 2 \( \sum_{i = 2}^{3} \mc{G}_{\mu\nu}^{(i)} + \sum_{i = 3}^{4} \mathscr{H}_{\mu\nu}^{(i)} \).
\label{DE Energy Tensor}
\ee
Then, we can extract an effective dark energy density and pressure as
\be
\bar{\rho}_\text{DE} = \frac{T_{00}^\text{(DE)}}{a^2}, \qquad \bar{P}_\text{DE} = \frac{1}{3 a^2} \text{tr} \,  T_{ij}^\text{(DE)},
\ee
\begin{align}
\bar{\rho}_\text{DE} &=- f_{2} + \frac{\vp'^2 f_{2 X_{1}}}{a^2} + \frac{A_{0} \vp'  f_{2 X_{2}}}{a^2} + \frac{A_{0}^2 f_{2 X_{3}}}{a^2} + \frac{2 A_{0} \vp'  f_{3  \varphi }}{a^2} - \frac{2 A_{0}^3 \vp'  \tilde{f}_{3  \varphi }}{a^4} - \frac{\vp'^2 G_{3  \varphi }}{a^2} \nn \\
 &- \frac{6 A_{0}^3 f_{3 X_{3}} \mathcal{H}}{a^4} - \frac{6 A_{0}^3 \tilde{f}_{3} \mathcal{H}}{a^4} + \frac{3 \vp'^3 G_{3 X_{1}} \mathcal{H}}{a^4} - \frac{6 \vp'  G_{4  \varphi } \mathcal{H}}{a^2} - \frac{6 G_{4} \mathcal{H}^2}{a^2} + \frac{3 \mathcal{H}^2}{\kappa a^2},
 \label{eq:den-DE}\\
\bar{P}_\text{DE} &= f_{2} + \frac{2 A_{0}^2 A'_0 f_{3 X_{3}}}{a^4} + \frac{2 A_{0} \vp'  f_{3  \varphi }}{a^2} + \frac{2 A_{0}^2 A'_0 \tilde{f}_{3}}{a^4} - \frac{\vp''  \vp'^2 G_{3 X_{1}}}{a^4} - \frac{\vp'^2 G_{3  \varphi }}{a^2} \nn \\
 &+ \frac{2 \vp''  G_{4  \varphi }}{a^2} + \frac{2 \vp'^2 G_{4  \varphi \varphi }}{a^2} - \frac{2 A_{0}^3 f_{3 X_{3}} \mathcal{H}}{a^4} - \frac{2 A_{0}^3 \tilde{f}_{3} \mathcal{H}}{a^4} + \frac{\vp'^3 G_{3 X_{1}} \mathcal{H}}{a^4} + \frac{2 \vp'  G_{4  \varphi } \mathcal{H}}{a^2} \nn \\
 &+ \frac{2 G_{4} \mathcal{H}^2}{a^2} - \frac{\mathcal{H}^2}{\kappa a^2} + \frac{4 G_{4} \mathcal{H}'}{a^2} - \frac{2 \mathcal{H}'}{\kappa a^2},
 \label{Den and Press DE}
\end{align}
which allows us to characterize the effective dark energy fluid by its equation of state parameter \mbox{$w_\text{DE} \equiv \bar{P}_\text{DE} / \bar{\rho}_\text{DE}$}. The background evolution is governed by the usual Friedman equations in Eq. \eqref{eq:Friedman-zero-order}.

First-order variables may also be extracted from the energy-momentum tensor in Eq. \eqref{DE Energy Tensor}. In general, we obtain expressions with the following structure:
\begin{align}
\delta \rho_\text{DE} &= (\cdots) \delta \vp + (\cdots) \delta \vp' + (\cdots) \Psi + (\cdots) \Phi + (\cdots) \Phi' + (\cdots) \delta A_0 + (\cdots) \psi, \label{Gra Density Pert}\\
\delta P_\text{DE} &= (\cdots) \delta \vp + (\cdots) \delta \vp' + (\cdots) \delta \vp'' + (\cdots) \Psi + (\cdots) \Psi' + (\cdots) \Phi \nn \\
 &+ (\cdots) \Phi' + (\cdots) \Phi'' + (\cdots) \delta A_0 + (\cdots) \delta A_0', \label{Gra Pressure Pert} \\
V_\text{DE} &= (\cdots) \delta \vp + (\cdots) \delta \vp' + (\cdots) \Psi + (\cdots) \Phi' + (\cdots) \delta A_0 + (\cdots) \psi. \label{Gra Velocity Pert}
\end{align}
However, the expressions in ($\cdots$) are cumbersome and it is worthwhile to look for ways to simplify them. Firstly, the quasi-static approximation (QSA) allows us to consider the gravitational potentials $\Phi$ and $\Psi$ as nearly time-independent functions during the matter dominated epoch, in such a way that any time derivative of these potentials can be neglected. Secondly, we assume the so-called sub-horizon approximation (SHA) where only modes deep inside the Hubble horizon are physically interesting, i.e., $k^2 \gg \mc{H}^2$. Under these approximations time derivatives acting on perturbations variables and terms of order $\mc{H} \times$perturbation can also be neglected.\footnote{See the appendix in Ref. \cite{DeFelice:2016yws} for a detailed explanation of QSA and SHA approximations.} For example, applying the SHA in Eq. \eqref{Pert Scalar Eq} we may simplify the coefficient $D_9$ as
\begin{align}
D_9 \delta \vp &= \Big( - f_{2 X_{1}} - \frac{2 \vp''  G_{3 X_{1}}}{a^2} - \frac{\vp''  \vp'^2 G_{3 X_{1} X_{1}}}{a^4} + 2 G_{3  \varphi } - \frac{\vp'^2 G_{3 \varphi X_{1}}}{a^2} - \frac{2 \vp'  G_{3 X_{1}} \mathcal{H} }{a^2} \nn \\
 &+ \frac{\vp'^3 G_{3 X_{1} X_{1}} \mathcal{H} }{a^4} \Big) \delta \vp \nn \\
 &\approx \left( - f_{2 X_{1}} - \frac{2 \vp''  G_{3 X_{1}}}{a^2} - \frac{\vp''  \vp'^2 G_{3 X_{1} X_{1}}}{a^4} + 2 G_{3  \varphi } - \frac{\vp'^2 G_{3 \varphi X_{1}}}{a^2} \right) \delta \vp,
\end{align}
where we neglected terms of order $\mc{H} \delta \vp$. We apply the QSA and the SHA to the linear gravitational field and scalar field equations [Eqs. \eqref{Pert time time Eq}, \eqref{Pert trace Eq}, and \eqref{Pert Scalar Eq}] to obtain
\begin{align}
0 &= A_3 \frac{k^2}{a^2} \Phi + A_5 \frac{k^2}{a^2} \delta \vp + A_7 \frac{k^2}{a^2} \psi -  \delta \rho_m, \label{QSA-SHA 00 Eq} \\
0 &= B_6 \frac{k^2}{a^2} \Phi + B_7 \frac{k^2}{a^2} \delta \vp + B_8 \frac{k^2}{a^2} \Psi, \\
0 &= D_7 \frac{k^2}{a^2} \Phi + \( D_9 \frac{k^2}{a^2} - m_\vp^2 \) \delta \vp + D_{10} \frac{k^2}{a^2} \Psi + D_{14} \frac{k^2}{a^2} \psi, \label{QSA-SHA Scalar Eq}
\end{align}
where we had into account that in some models the mass $m_\vp$ of the scalar field may play a significant role in the past, given that $m_\vp \gg H$ (e.g., quintessence). For the perturbed vector field equations of motion \eqref{Pert Temp Vec Eq}-\eqref{Pert Space Vec Eq}, we only apply  the QSA and neglect derivatives of the fields obtaining
\begin{align}
0 &= F_3 \frac{\Psi}{a} + F_4 \frac{\delta \vp}{a} + F_5 \frac{\delta A_0}{a^2} + F_6 \frac{k^2}{a^2} \frac{\psi}{a},\\
0 &= H_1 \frac{\Psi}{a^2} + H_2 \frac{\delta \vp}{a^2} + H_3 \frac{\delta A_0}{a^3} + H_4 \frac{\psi}{a^2}. \label{Const A0}
\end{align}
If we also applied the SHA to Eq. \eqref{Pert Temp Vec Eq}, we would obtain that the only relevant term would be $\frac{k^2}{a^2} \psi$, yielding the trivial solution $\psi = 0$. In that case, the gravitational field equations and the scalar field equation will not be affected by the presence of the vector field; the system would be reduced to that of Horndeski theory which was already studied in Ref. \cite{Arjona:2019rfn}. Note however that our set of equations \eqref{QSA-SHA 00 Eq}-\eqref{QSA-SHA Scalar Eq} agrees with results in Ref. \cite{Arjona:2019rfn} [see their Eqs. (91)-(93)] when the vector field vanishes. 

Applying the approximations as explained above, we get five algebraic equations \eqref{QSA-SHA 00 Eq}-\eqref{Const A0} that we solve for the five perturbation variables $\Psi$, $\Phi$, $\delta \vp$, $\delta A_0$, and $\psi$.  We obtain 
\begin{equation*}
\delta \vp = \frac{\frac{k^2}{a^2} W_1 + W_2}{\frac{k^4}{a^4} W_3 + \frac{k^2}{a^2} W_4 + W_5} \delta \rho_m,
\end{equation*}
\begin{equation*}
\frac{\delta A_0}{a} = \frac{\frac{k^4}{a^4} W_6 + \frac{k^2}{a^2} W_7 + W_8}{\frac{k^6}{a^6} W_3 + \frac{k^4}{a^4} W_4 + \frac{k^2}{a^2} W_5} \delta \rho_m, \quad \psi = \frac{\frac{k^2}{a^2} W_9 + W_{10}}{\frac{k^6}{a^6} W_3 + \frac{k^4}{a^4} W_4 + \frac{k^2}{a^2} W_5} \delta \rho_m,
\end{equation*}
\begin{equation}
\frac{k^2}{a^2} \Phi = \frac{\frac{k^4}{a^4} W_{11} + \frac{k^2}{a^2} W_{12} + W_{13}}{\frac{k^4}{a^4} W_3 + \frac{k^2}{a^2} W_4 + W_5} \delta \rho_m, \quad \frac{k^2}{a^2} \Psi = -  \frac{\frac{k^4}{a^4} W_{14} + \frac{k^2}{a^2} W_{15} + W_{13}}{\frac{k^4}{a^4} W_3 + \frac{k^2}{a^2} W_4 + W_5} \delta \rho_m,
\label{QSA SHA Variables}
\end{equation}
where we took into account that some of the perturbations coefficients are related (see Appendix \ref{App: Perturbations Coefficients}), 
\be
A_3 = B_6  =  B_8, \quad D_{10} = A_5, \quad D_7 = B_7, \quad D_{14} = H_2, \quad A_7 = H_1, \quad F_6 = H_3.
\ee 
The coefficients $W_i$ ($i = 1, \ldots, 15$) are given in Appendix \ref{App: QSA SHA Coefficients}. Actually, we could have four dynamical degrees of freedom. From Eq. \eqref{Const A0}, we could retrieve $\delta A_0$ and insert it in the approximated linear equations, thus reducing one dynamical degree of freedom. Nonetheless, the procedure we follow might be clearer since the mentioned reduction yields results for $\Phi$, $\Psi$, $\vp$, and $\psi$ much more complicated to handle.

From the expressions for the potentials in Eq. \eqref{QSA SHA Variables}, we can characterize deviations from GR by defining the gravitational slip parameters
\begin{align}
\eta &\equiv \frac{\Psi + \Phi}{\Phi} = \frac{\frac{k^4}{a^4}(W_{11} - W_{14}) + \frac{k^2}{a^2}(W_{12} - W_{15})}{\frac{k^4}{a^4} W_{11} + \frac{k^2}{a^2} W_{12} + W_{13}}, \label{slip eta}\\
\gamma &\equiv - \frac{\Phi}{\Psi} = \frac{\frac{k^4}{a^4} W_{11} + \frac{k^2}{a^2} W_{12} + W_{13}}{\frac{k^4}{a^4} W_{14} + \frac{k^2}{a^2} W_{15} + W_{13}}, \label{slip gamma}
\end{align}
where the GR case corresponds to $\eta = 0$ and $\gamma = 1$. Note that the expressions of the gravitational potentials can be written as Poisson-like equations if we define parameters $G_\text{eff}$ and $Q_\text{eff}$ such that
\be 
\frac{k^2}{a^2} \Psi = - \frac{1}{2} \frac{G_\text{eff}}{G_N} \delta \rho_m, \quad \frac{k^2}{a^2} \Phi = \frac{1}{2} Q_\text{eff} \delta \rho_m.
\label{Poisson Eqs}
\ee
These parameters also characterize modifications to gravity. The GR case corresponds to $G_\text{eff} / G_N = 1$, and $Q_\text{eff} = G_N$, with $\Phi = - \Psi$. In the case of SVT theories, as shown in Eq. \eqref{Long Traceless Eq}, the presence of the scalar field prevents both potential to be opposite. Hence, anisotropic stress in SVT theories is sourced solely by the scalar field whenever the function $G_4 (\vp)$ is not a constant. Note that, under this scheme, generalised Proca theories do not admit anisotropic stress.

As we will show below, the parameter $G_\text{eff}$ largely determines the evolution of the growth of matter perturbations. Using the QSA and the SHA in the Eqs. \eqref{Eq:evolution-delta} and \eqref{Eq:evolution-V} for matter ($w_m = c_{s, m}^2 = \pi_m = 0$), we get 
\be
\delta\text{'}_m \sim - \frac{V_m}{a^2 H}, \qquad V\text{'}_m \sim - \frac{V_m}{a} + \frac{k^2}{a^2 H} \Psi.
\ee
Therefore, differentiating the equation for $\delta\text{'}_m$, inserting $V\text{'}_m$ in that derivative, and using the Poisson equation for $\Psi$ in Eq. \eqref{Poisson Eqs}, the evolution equation for $\delta_m$ will be given by
\be
\delta\text{''}_m(a) + \( \frac{3}{a} + \frac{H\text{'}(a)}{H(a)} \) \delta\text{'}_m(a) - \frac{3}{2} \frac{\Omega_{m 0} G_\text{eff} / G_N}{a^5 H(a)^2 / H_0^2} \delta_m (a)= 0.
\label{Evo Eq CDM}
\ee
Hence, by solving Eq. \eqref{Evo Eq CDM}, we can determine the growth factor. We will work out a fully numerical solution of Eq. \eqref{Evo Eq CDM} in the Sec. \ref{Sec: SVTDES Growth} for a specific $G_\text{eff}$ obtained from our designer SVT model, which we will describe in Sec. \ref{Section:Des-SVT}.

In what follows we will consider perturbations to the effective dark energy under QSA and SHA in two cases: i) non-vanishing anisotropic stress $\pi_{\rm DE}$; ii) $G_4 = \text{constant}$ and hence vanishing anisotropic stress $\pi_{\rm DE}=0$.

\subsection{SVT theories with non-vanishing anisotropic stress}
\label{subsection:without-piDE}

Now, we apply the QSA and the SHA to the quantities in Eqs. \eqref{Gra Density Pert}-\eqref{Gra Velocity Pert}. We proceed as follows. Since the QSA breaks down due to rapid oscillations of the scalar field \cite{Arjona:2019rfn}, we use the trace-less ``space-space'' equation \eqref{Long Traceless Eq} in order to solve for $\delta \vp$ in terms of the gravitational potentials. By differentiating \eqref{Long Traceless Eq}, we can also solve for the derivatives of $\delta \vp$, also applying the QSA and the SHA at the end of the differentiation. For instance, for the first derivative we get
\be
\delta \vp' = - \frac{G_{4 \vp \vp}}{G_{4 \vp}} \vp' \delta \vp - \vp' (\Psi + \Phi).
\label{eq:vp-derivative}
\ee
This is the reason why we discriminate models with and without anisotropic stress: since $G_4 = \text{constant}$ when anisotropic stress vanishes, Eq. \eqref{eq:vp-derivative}  would diverge. When we deal with models having a non-vanishing anisotropic stress we can then replace the potentials using the Poisson equations in Eq. \eqref{Poisson Eqs}, leaving all the expressions in terms of $\delta \rho_m$. We obtain
\begin{equation*}
\delta \rho_\text{DE} = \frac{\frac{k^6}{a^6} Z_1 + \frac{k^4}{a^4} Z_2 + \frac{k^2}{a^2} Z_3 + Z_4}{\frac{k^6}{a^6} Z_5 + \frac{k^4}{a^4} Z_6 + \frac{k^2}{a^2} Z_7} \delta \rho_m, \quad \delta P_\text{DE} = \frac{1}{3 Z_{12}} \frac{\frac{k^6}{a^6} Z_8 + \frac{k^4}{a^4} Z_9 + \frac{k^2}{a^2} Z_{10} + Z_{11}}{\frac{k^6}{a^6} Z_5 + \frac{k^4}{a^4} Z_6 + \frac{k^2}{a^2} Z_7} \delta \rho_m,
\end{equation*}
\begin{equation*}
\frac{a \bar{\rho}_\text{DE}}{k^2} V_\text{DE} = \frac{\frac{k^4}{a^4} Z_{13} + \frac{k^2}{a^2} Z_{14} + Z_{15}}{\frac{k^6}{a^6} Z_5 + \frac{k^4}{a^4} Z_6 + \frac{k^2}{a^2} Z_7} \delta \rho_m, \quad \bar{\rho}_\text{DE} \pi_\text{DE} = \frac{k^2}{a^2} \frac{\tfrac{k^2}{a^2} (W_{14} - W_{11}) + ( W_{15} - W_{12})}{\tfrac{k^4}{a^4} W_3 + \tfrac{k^2}{a^2} W_4 + W_5} \delta \rho_m,
\end{equation*}
\begin{equation}
c_{s, \text{DE}}^2 = \frac{1}{3 Z_{12}} \frac{\tfrac{k^6}{a^6} Z_8 + \tfrac{k^4}{a^4} Z_9 + \tfrac{k^2}{a^2} Z_{10} + Z_{11}}{\tfrac{k^6}{a^6} Z_1 + \tfrac{k^4}{a^4} Z_2 + \tfrac{k^2}{a^2} Z_3 + Z_4}.
\label{Stress Eqs}
\end{equation}
The coefficients $Z_i$ ($i = 1,\dots,15$) are presented in  Appendix \ref{App: Ani Stress Coefficients}. Due to the presence of the anisotropic stress, the sound speed in Eq. \eqref{Stress Eqs} does not fully determine the stability of sub-horizon perturbations. This can be seen by solving Eq. \eqref{eq:cons1} for $\theta$ and substituting the result (and its derivative) into \eqref{eq:cons2}. Doing so, we obtain the following second-order equation for $\delta$ 
\begin{align}
\delta'' &+ (1 - 6 w) \mc{H} \delta' +3 \mc{H} \( \frac{\delta P}{\delta \rho} \)' + 3 \[ (1 - 3 w) \mc{H}^2 + \mc{H}' \] \( \frac{\delta P}{\bar{\rho}} - w \delta \) - 3 \mc{H} w' \delta \nn \\
 &= - 3(1 + w) \[ \Phi'' + \( 1 - 3 w + \frac{w'/\mc{H}}{1 + w} \) \mc{H} \Phi' \] - k^2 \[ (1 + w) \Psi + \frac{\delta P}{\bar{\rho}} - \frac{2}{3} \pi \],
\label{Eq delta DE}
\end{align}
where we have used the relation $\pi = \frac{3}{2} (1 + w) \sigma$. For sub-horizon modes, the last term factorized by $k^2$ in Eq. \eqref{Eq delta DE} is the relevant term which determines the stability of perturbations. Since the potential scales as $\Psi \sim 1 / k^2$ for these modes, therefore, the stability of sub-horizon perturbations is driven mainly by an effective sound speed defined as \cite{Cardona:2014iba}
\be 
c_{s, \text{eff}}^2 \equiv c_{s, \text{DE}}^2 - \frac{2}{3} \frac{\bar{\rho}_\text{DE} \pi_\text{DE}}{\delta \rho_\text{DE}}.
\ee 

Thus far the discussion has been quite general, providing analytical expressions for the field perturbations $\delta \vp$, $\delta A_0$, $\psi$, and the potentials $\Phi$ and $\Psi$ in Eqs. \eqref{QSA SHA Variables}. Now we want to test our equations against known results in literature, namely: $f(R)$ theories, quintessence, and generalised Proca. We also indicate some possible, minimal modifications in the context of SVT theories. 

\begin{itemize}
\item $\boldsymbol{f(R)}$ \textbf{Theories}
\end{itemize}
Through a conformal transformation, $f(R)$ theories can be seen as a theory for a scalar field non-minimally coupled to $R$. This theory will be contained in SVT theories if we do the following identification
\be
f_2 = - \frac{R F - f}{2}, \quad f_{2 \vp} = - \frac{R}{2}, \quad f_{2 \vp \vp} = - \frac{1}{2 F_R}, \quad G_4 = \frac{F}{2}, \quad G_{4 \vp} = \frac{1}{2}, 
\label{functions f(R)}
\ee
where $F \equiv f_R = \sqrt{\kappa} \vp$, $F_R \equiv f_{R R}$; unspecified  derivatives and free-functions are set to zero, and we have assumed $\kappa = 1$. Replacing \eqref{functions f(R)} in Eqs. \eqref{eq:den-DE}-\eqref{Den and Press DE}, we get the well-known expressions for  the density and pressure of DE in $f(R)$ theories:
\be
a^2 \bar{\rho}_\text{DE} = - \frac{a^2 f}{2} + 3 \mc{H}^2 + 3 F \mc{H}' - 3 \mc{H} F',
\label{eq:f(R)-density}
\ee 
\be
a^2 \bar{P}_\text{DE} = \frac{a^2 f}{2} - (1 + 2 F) \mc{H}^2 - (2 + F) \mc{H}' + \mc{H} F' + F''.
\label{f(R) pressure}
\ee 
Under the QSA and the SHA, replacing the above functions [Eq. \eqref{functions f(R)}] in Eqs. \eqref{QSA SHA Variables} we get for the perturbation variables 
\begin{equation*} 
\delta \vp = \frac{F_R}{F + 3 \tfrac{k^2}{a^2} F_R} \delta \rho_m, \quad \delta A_0 = 0, \quad \psi = 0,
\end{equation*}
\be 
\Psi = - \frac{F + 4 \tfrac{k^2}{a^2} F_R}{2 \tfrac{k^2}{a^2} F^2 + 6 \tfrac{k^4}{a^4} F F_R} \delta \rho_m, \quad \Phi = \frac{F + 2 \tfrac{k^2}{a^2} F_R}{2 \tfrac{k^2}{a^2} F^2 + 6 \tfrac{k^4}{a^4} F F_R} \delta \rho_m.
\ee 
Since $\Phi \neq - \Psi$, the slip parameters are not constants and Eqs. \eqref{slip eta} and \eqref{slip gamma} become
\be 
\eta = - \frac{2 \tfrac{k^2}{a^2} F_R}{F + 2 \tfrac{k^2}{a^2} F_R}, \quad \gamma = \tfrac{F + 2 \tfrac{k^2}{a^2} F_R}{F + 4 \tfrac{k^2}{a^2} F_R}.
\ee 
The effective DE perturbed quantities in Eq. \eqref{Stress Eqs} take on
\begin{align}
\delta \rho_\text{DE} = \frac{(1 - F) F + (2 - 3 F) \tfrac{k^2}{a^2} F_R}{F (F + 3 \tfrac{k^2}{a^2} F_R)} \delta \rho_m&, \quad \delta P_\text{DE} = \frac{1}{3 F} \frac{2 \tfrac{k^4}{a^4} F_R + 15 \tfrac{k^2}{a^4} F_R F'' + \tfrac{3 F F''}{a^2}}{ 3 \tfrac{k^4}{a^4} F_R + \tfrac{k^2}{a^2} F} \delta \rho_m, \nn \\
\bar{\rho}_\text{DE} V_\text{DE} = \frac{1}{2 F} \frac{(F + 6 \tfrac{k^2}{a^2} F_R) F'}{F + 3 \tfrac{k^2}{a^2} F_R} \delta \rho_m&, \quad \bar{\rho}_\text{DE} \pi_\text{DE} = \frac{\frac{k^2}{a^2} F_R}{F^2 + 3 \tfrac{k^2}{a^2} F F_R} \delta \rho_m.
\end{align}
These results are in perfect agreement with those reported in Refs. \cite{Arjona:2018jhh, Arjona:2019rfn, Tsujikawa:2007gd}.

\begin{itemize}
\item $\boldsymbol{f(R) \ + }$ \textbf{Cubic Vector Interactions}
\end{itemize}

We can have minimal modifications to $f(R)$ theories by adding cubic interactions coming from the vector sector in SVT theories. If we assume $f_3 = \text{constant}$, we see that in order to get non trivial solutions, $\tilde{f}_3$ must depend at least on $\vp$. Assuming Eqs. \eqref{functions f(R)}, $f_3 = 0$, $\tilde{f}_3 = \tilde{f}_3 (\vp)$, and $\kappa = 1$, from the background equation of motion for the scalar and vector fields  \eqref{Back Scalar Vector Eqs}, we obtain
\be 
2 \frac{A_0^2}{a^4} \tilde{f}_{3 \vp} \( A'_0 - \mc{H} A_0 \) = 0, \quad - 2 \frac{A_0^2}{a^4} \( \tilde{f}_{3 \vp} F' + 3 \mc{H} \tilde{f}_3 \) = 0,
\ee 
and therefore,
\be 
3 \mc{H} \tilde{f}_3 = - \tilde{f}_{3 \vp} F', \quad A'_0 = \mc{H} A_0.
\label{fRA solutions}
\ee 
Note that the right-hand side expression in the last equation can be recast as $(\frac{A_0}{a})'=0$. 
In generalised Proca theories, a vector field fulfilling this condition characterizes de Sitter solutions \cite{DeFelice:2016yws}, which is not the case in this vector $f(R)$ theory. Using Eq. \eqref{functions f(R)} and Eq. \eqref{fRA solutions} in Eqs. \eqref{eq:den-DE}-\eqref{Den and Press DE}, we get the same density and pressure given in Eqs. \eqref{eq:f(R)-density} and \eqref{f(R) pressure}, meaning that the background evolution of this new model is not modified by the inclusion of the vector field. However, perturbations do get a contribution from the vector field. Replacing the new configuration in Eqs. \eqref{functions f(R)} and \eqref{fRA solutions} in the expressions in Eq. \eqref{Stress Eqs}, we get
\begin{align}
\delta \rho_\text{DE} &= \frac{(1 - F) F + (2 - 3 F) \tfrac{k^2}{a^2} F_R  + 2 \frac{F_R F' A_0^3 \tilde{f}_{3 \vp} }{a^4} \( 1 + \tfrac{a^2 F}{2 k^2 F_R} \)}{F \{ F + 3 \tfrac{k^2}{a^2} F_R - 2 F_R \[ 4 (1 + \tfrac{a^2 F }{4 k^2 F_R} )\tilde{f}_{3 \vp} + F \tilde{f}_{3 \vp \vp}\] \frac{F' A_0^3}{a^4 F}\} } \delta \rho_m, \nn \\
\delta P_\text{DE} &= \frac{1}{3 F} \frac{2 \tfrac{k^4}{a^4} F_R + 15 \tfrac{k^2}{a^4} F_R F'' + \tfrac{3 F F''}{a^2} - 2 \tilde{f}_{3 \vp} \frac{F F' A_0^3}{a^4} (1 + 4 \frac{F_R k^2}{F a^2})}{3 \tfrac{k^4}{a^4} F_R + \tfrac{k^2}{a^2} F - 2 \frac{F' A_0^3 F_R k^2}{a^6 F} \[ 4 \( 1 + \tfrac{F a^2}{4 F_R k^2} \) \tilde{f}_{3 \vp} + F \tilde{f}_{3 \vp \vp} \] } \delta \rho_m, \nn \\
\bar{\rho}_\text{DE} V_\text{DE} &= \frac{1}{2 F} \frac{(F + 6 \tfrac{k^2}{a^2} F_R) F'}{F + 3 \tfrac{k^2}{a^2} F_R - 2 \frac{F_R F' A_0^3}{F a^4} \[ 4 \( 1 + \tfrac{F a^2}{4 F_R k^2} \) \tilde{f}_{3 \vp} + F \tilde{f}_{3 \vp \vp} \]} \delta \rho_m, \nonumber \\
\bar{\rho}_\text{DE} \pi_\text{DE} &= \frac{\frac{k^2}{a^2} F_R}{F^2 + 3 \tfrac{k^2}{a^2} F F_R -  2 \frac{F' A_0^3 F_R }{a^4} \[ 4 \( 1 + \tfrac{F a^2}{4 F_R k^2} \) \tilde{f}_{3 \vp} + F \tilde{f}_{3 \vp \vp} \] } \delta \rho_m.
\end{align}

The phenomenology of these kinds of models could therefore add new features or help addressing current discrepancies in cosmological parameters as discussed, for instance, in Ref. \cite{Sabla:2022xzj}.

\subsection{SVT theories with vanishing anisotropic stress}

If the DE anisotropic stress vanishes, then $\Phi = - \Psi$ and from Eq. \eqref{Long Traceless Eq} $G_4$ does not depend on the scalar field. For the sake of simplicity, we assume $G_4 = 1/2\kappa$ so that  $\mc{L}_4^\text{ST}$ in Eq. \eqref{L4ST} equals the Einstein-Hilbert Lagrangian. Applying these assumptions in Eqs. \eqref{Gra Density Pert}-\eqref{Gra Velocity Pert} as well as the QSA and the SHA, we get 
\begin{equation*}
\delta \rho_\text{DE} = \frac{\frac{k^6}{a^6} Y_1 + \frac{k^4}{a^4} Y_2 + \frac{k^2}{a^2} Y_3 + Y_4}{\frac{k^6}{a^6} Y_5 + \frac{k^4}{a^4} Y_6 + \frac{k^2}{a^2} Y_7} \delta \rho_m, \quad \delta P_\text{DE} = \frac{1}{3}  \frac{\frac{k^4}{a^4} Y_8 + \frac{k^2}{a^2} Y_9 + Y_{10}}{\frac{k^6}{a^6} Y_5 + \frac{k^4}{a^4} Y_6 + \frac{k^2}{a^2} Y_7} \delta \rho_m,
\end{equation*}
\begin{equation}
\frac{a \bar{\rho}_\text{DE}}{k^2} V_\text{DE} = \frac{\frac{k^4}{a^4} Y_{11} + \frac{k^2}{a^2} Y_{12} + Y_{13}}{\frac{k^6}{a^6} Y_5 + \frac{k^4}{a^4} Y_6 + \frac{k^2}{a^2} Y_7} \delta \rho_m, \quad c_{s, \text{DE}}^2 = \frac{1}{3} \frac{\frac{k^4}{a^4} Y_8 + \frac{k^2}{a^2} Y_{9} + Y_{10}}{\frac{k^6}{a^6} Y_1 + \frac{k^4}{a^4} Y_2 + \frac{k^2}{a^2} Y_3 + Y_4}.
\label{No Stress Eqs}
\end{equation}
The coefficients $Y_i$, ($i = 1,\dots,14$) are presented in  Appendix \ref{App: No Ani Stress Coefficients}. Next we present a few examples of SVT theories with vanishing DE anisotropic stress. 

\begin{itemize}
\item \textbf{Quintessence}
\end{itemize}

The typical Lagrangian of a quintessence scalar field can be recovered by defining
\be
f_2 = X_1 - V(\vp), \quad f_{2 \vp} = - V_\vp, \quad f_{2 \vp \vp} = - V_{\vp \vp}, \quad G_4 = \frac{1}{2},
\label{QS functions}
\ee 
while any other function vanishes, and $\kappa = 1$. Using the definitions  \eqref{QS functions} in Eqs. \eqref{eq:den-DE} and \eqref{Den and Press DE} we get the usual density and pressure
\be 
\bar{\rho}_\text{DE} = X_1 + V, \quad \bar{P}_\text{DE} = X_1 - V.
\label{Den QS}
\ee 
The full perturbations in Eqs. \eqref{Gra Density Pert}-\eqref{Gra Velocity Pert} are simply given by
\be 
\delta \rho_\text{DE} = \frac{\vp' \delta \vp'}{a^2} + V_\vp \delta \vp - \frac{\vp'^2}{a^2} \Psi, \quad \delta P_\text{DE} = \frac{\vp' \delta \vp'}{a^2} - V_\vp \delta \vp - \frac{\vp'^2}{a^2} \Psi,
\label{Perts QS}
\ee 
\be 
\bar{\rho}_\text{DE} V_\text{DE} = k^2 a^{-1} \frac{\vp' \delta \vp}{a}, \quad c_{s, \text{DE}}^2 = \frac{\delta P_\text{DE}}{\delta \rho_\text{DE}}.
\label{V QS}
\ee 
Note however that, under the SHA and the QSA, Eqs. \eqref{No Stress Eqs} provide simplified expressions
\be 
\delta \rho_\text{DE} = \delta P_\text{DE} = \frac{\vp'^2}{2 k^2} \delta \rho_m, \quad \bar{\rho}_\text{DE} V_\text{DE} = 0, \quad c_{s, \text{DE}}^2 = 1.
\label{QSA SHA Quintessence}
\ee 
These results agree with those reported in Refs. \cite{Arjona:2019rfn}.

\begin{itemize}
\item \textbf{Quintessence + Cubic Vector Interactions}
\end{itemize}

The phenomenology in the previous example can become more interesting by introducing a vector field, as we did for $f(R)$ theories in Subsection \ref{subsection:without-piDE}. Replacing the quintessence functions \eqref{QS functions} in the background equation of motion for the scalar and vector fields in Eq. \eqref{Back Scalar Vector Eqs}, and assuming that $f_3 = 0$ and $\tilde{f}_3 = \tilde{f}_3 (\vp)$, we get
\be 
\vp'' + 2 \mc{H} \vp' + a^2 V_\vp - 2 \tilde{f}_{3 \vp} \frac{A_0^2}{a^2} \( A_0' - \mc{H} A_0 \)= 0, \quad - 2 \frac{A_0^2}{a^4} \( \tilde{f}_{3 \vp} \vp' + 3 \mc{H} \tilde{f}_3 \) = 0. 
\label{eq:qcvi}
\ee 
We see that the usual Klein-Gordon equation for a scalar field is recovered when the vector field fulfills $A'_0 = \mc{H} A_0$. Using the expression in the right-hand side of Eq. \eqref{eq:qcvi} and the functions \eqref{QS functions}  in Eqs. \eqref{eq:den-DE}-\eqref{Den and Press DE}, we get that the background density and pressure are
\be 
\bar{\rho}_\text{DE} = X_1 + V, \quad \bar{P}_\text{DE} = X_1 - V - \frac{2}{3} \frac{A_0^2 \vp'}{a^4 \mc{H}} \tilde{f}_{3 \vp} (A'_0 - \mc{H} A_0).
\ee 
Under the QSA and the SHA, the sound speed is now more involved due to non trivial contributions to the pressure perturbation in Eq. \eqref{No Stress Eqs}
\be
\delta \rho_\text{DE} = \frac{\vp'^2 + 2\tfrac{A_0^3 \vp' \tilde{f}_{3 \vp}}{a^2}}{2 k^2 - 4 \tfrac{A_0^3 \vp' \tilde{f}_{3 \vp}}{a^2} } \delta \rho_m, \quad 
\delta P_\text{DE} = \frac{\vp'^2 - \tfrac{4 A_0^2 A_0' \vp' \tilde{f}_{3 \vp}}{3 a^2 \mc{H}}}{2 k^2 - 4 \tfrac{A_0^3 \vp' \tilde{f}_{3 \vp}}{a^2} } \delta \rho_m,
\ee 
\be 
\bar{\rho}_\text{DE} V_\text{DE} = - \frac{1}{a k^2} \frac{2 A_0^2 \tilde{f}_{3 \vp} \vp' A_0'}{k^2 - 2 \tfrac{A_0^3 \vp' \tilde{f}_{3\vp} }{a^2} } \delta \rho_m, \quad 
c_{s, \text{DE}}^2 = \frac{\vp'^2 - \tfrac{4 A_0^2 A_0' \vp' \tilde{f}_{3 \vp}}{3 a^2 \mc{H}}}{\vp'^2 + 2\tfrac{A_0^3 \vp' \tilde{f}_{3 \vp}}{a^2}}.
\ee 
Note that in this case the velocity perturbation does not vanish and  the sound speed in general $c_{s, \text{DE}}^2 \neq 1$.

\begin{itemize}
\item \textbf{Generalised Proca}
\end{itemize}

Generalised Proca theories are obtained from the SVT Lagrangian \eqref{Eff Lagrangian} by assuming that
\be 
f_2 = f_2 (X_3), \quad f_3 \rightarrow \frac{1}{2} f_3 (X_3), \quad G_4 = \frac{1}{2},
\label{GP functions}
\ee 
while all the other unspecified functions vanish, due to the constraint coming from  the speed of gravitational waves, namely $c_T^2 = 1$ \cite{DeFelice:2016yws}, and we have assumed $\kappa = 1$. The second Friedman equation and the vector field equation of motion obtained from Eqs. \eqref{Gral Back Einstein Eqs} and \eqref{Back Scalar Vector Eqs}, using \eqref{GP functions}, are the following constraints 
\be
f_2 - \frac{A_0^2 f_{2 X_3}}{3 a^2} + \frac{\mc{H}^2}{a^2} + 2 \frac{\mc{H}'}{a^2} + \frac{A_0 A'_0 f_{2 X_3}}{2 a^2 \mc{H}} = 0, \quad \frac{A_0}{a^2} f_{2 X_3} - \frac{3 \mc{H} A_0^2}{a^4} f_{3 X_3} = 0.
\label{eq:vec-proca}
\ee
Using the equation in the right-hand side of Eqs. \eqref{eq:vec-proca} to eliminate $f_{3 X_3}$, the corresponding background density and pressure for generalised Proca model  [see Eqs. \eqref{eq:den-DE} and \eqref{Den and Press DE}] are given by
\be
\bar{\rho}_\text{DE} = - f_2, \quad \bar{P}_\text{DE} = f_2 - \frac{A_0 (A_0' - \mc{H} A_0)}{3 a^2 \mc{H}} f_{2 X_3}.
\ee 
The full perturbations in Eqs. \eqref{Gra Density Pert}-\eqref{Gra Velocity Pert} for the effective DE fluid read
\begin{align}
\delta \rho_\text{DE} &= \( \frac{A_{0}^3 f_{2 X_{3} X_{3}}}{a^4} - \frac{2 A_{0} f_{2 X_{3}}}{a^2} - \frac{3 A_{0}^4 f_{3 X_{3} X_{3}} \mathcal{H}}{a^6} \) \delta A_{0}  - \frac{A_{0} f_{2 X_{3}}}{3 \mathcal{H}} \frac{k^2}{a^2} \psi  \nn \\
 &+ \frac{A_{0}^2 f_{2 X_{3}}}{a^2 \mathcal{H}} \Psi' + \( \frac{3 A_{0}^2 f_{2 X_{3}}}{a^2} - \frac{A_{0}^4 f_{2 X_{3} X_{3}}}{a^4} + \frac{3 A_{0}^5 f_{3 X_{3} X_{3}} \mathcal{H}}{a^6} \) \Psi, \\
\bar{\rho}_\text{DE} V_\text{DE} &= \frac{k^2}{a^2} \frac{f_{2 X_3}}{3 \mc{H}}(\delta A_0 - A_0 \Psi), \\
\delta P_\text{DE} &= \frac{A_{0} f_{2 X_{3}}}{3 a^2 \mathcal{H}} \delta A_0' + \(\frac{A_{0}^3 A_0' f_{3 X_{3} X_{3}}}{a^6} + \frac{2 A_0' f_{2 X_{3}}}{3 a^2 \mathcal{H}} - \frac{A_{0}^4 f_{3 X_{3} X_{3}} \mathcal{H}}{a^6} \) \delta A_{0} \nn \\
 &+ \(\frac{A_{0}^2 f_{2 X_{3}}}{3 a^2} - \frac{A_{0}^4 A_0' f_{3 X_{3} X_{3}}}{a^6} - \frac{4 A_{0} A_0' f_{2 X_{3}}}{3 a^2 \mathcal{H}} + \frac{A_{0}^5 f_{3 X_{3} X_{3}} \mathcal{H}}{a^6} \) \Psi - \frac{A_{0}^2 f_{2 X_{3}}}{3 a^2 \mathcal{H}} \Psi'.
\label{Perts Proca}
\end{align}
These results are very similar to those reported in Ref. \cite{Heisenberg:2020xak}, but they are not equal since differences arise due to a different choice for the vector field profile. Under the QSA and the SHA, these perturbed quantities take the following simple form
\begin{equation*}
\delta \rho_\text{DE} = - \frac{A_0^3 f_{2 X_3}}{A_0^3 f_{2 X_3} + 2 k^2 A_0} \delta \rho_m, \quad \bar{\rho}_\text{DE} V_\text{DE} = \frac{A_0 f_{2 X_3} A'_0 }{A_0^2 f_{2 X_3} + 2 k^2} \mc{H} \delta \rho_m \approx 0,
\end{equation*}
\be
\delta P_\text{DE} = \frac{2}{3} \frac{A_0^2 f_{2 X_3} - 3 a^2 f_2}{A_0^2 f_{2 X_3} + 2 k^2} \delta \rho_m, \quad c_{s, \text{DE}}^2 = - \frac{2}{3} + \frac{2 a^2 f_2}{A_0^2 f_{2 X_3}},
\label{eq: Proca QSA}
\ee
where we have used the equation in the left-hand side of Eqs. \eqref{eq:vec-proca} to eliminate $A'_0$. Then, under these approximations, DE in generalised Proca theories is on its rest-frame, and the sound speed $c_{s, \text{DE}}^2$ is different from 1.

In Refs. \cite{DeFelice:2016yws, Geng:2021jso}, authors investigated a particular model where the free functions are given by
\be 
f_2 = b X_3^m, \quad f_3 = \frac{1}{2} c X_3^n, \quad G_4 = \frac{1}{2},
\label{eq:gpexample}
\ee
where $b$, $c$, $m$, $n$, are constants. This power law Proca model has a phantom equation of state of dark energy when $A_0^p \propto H^{-1}$, with $p \equiv 2(n - m) + 1$. Because in the next example we want to show how the introduction of a scalar field can change the dynamics of the generalised Proca model \eqref{eq:gpexample}, we will assume $b = c = -1$, $m=1$   $n = 5/2$.

From the second Friedman equation and the equation of motion for the vector field in Eqs. \eqref{eq:vec-proca}, and using \eqref{eq:gpexample}, we get
\be
\frac{A'_0}{a^2} = - \frac{2 \sqrt{2} a^3}{5 A_0^3} + \frac{\mc{H} A_0}{a^2} + \frac{4 \sqrt{2} a^3}{5 A_0^5}\( \mc{H}^2 + 2 \mc{H}' \), \quad \frac{A_0}{a} = \frac{2^{5/8} a^{1/4}}{15^{1/4} \mc{H}^{1/4}}.
\ee
Under these assumptions, the approximated perturbations \eqref{eq: Proca QSA} read 
\begin{equation}
\delta \rho_\text{DE} = - \frac{2^{1/4} a^{5/2}}{2^{1/4} a^{5/2} - \sqrt{15 \mc{H}} k^2} \delta \rho_m, \quad \delta P_\text{DE} = - \frac{1}{3} \frac{2^{1/4} a^{5/2}}{2^{1/4} a^{5/2} - \sqrt{15 \mc{H}} k^2} \delta \rho_m, \quad c_{s, \text{DE}}^2 = \frac{1}{3}.
\label{eq: QSA Proca Perts}
\end{equation}

\begin{itemize}
\item \textbf{Generalised Proca + Scalar Interactions}
\end{itemize}

We modify the generalised Proca model \eqref{eq:gpexample} in the following way
\be 
f_2 = b X_3^m + X_1.
\ee 
Taking the same powers, namely, $b = c = -1 $, $ m = 1$, $n = 5/2$, from Eqs. \eqref{No Stress Eqs} the approximated perturbations in Eqs. \eqref{eq: Proca QSA} are modified as
\begin{equation*}
\delta \rho_\text{DE} = - \frac{2^{5/4} a^{5/2} + \sqrt{15 \mc{H}} \vp'^2}{2^{5/4} a^{5/2} - 2 \sqrt{15 \mc{H}} k^2} \delta \rho_m, \quad c_{s, \text{DE}}^2 = \frac{1}{3} \frac{2^{5/4} a^{5/2}}{2^{5/4} a^{5/2} + \sqrt{15 \mc{H}} \vp'^2},
\end{equation*}
while $\delta P_\text{DE}$ is the same given in Eq. \eqref{eq: QSA Proca Perts}. Therefore, in this case the sound speed of scalar perturbations $c_{s, \text{DE}}^2$ is in general time-dependent.

\section{Designer SVT} \label{Section:Des-SVT}

For the SVT theories regarded in  this work, Eqs. \eqref{Stress Eqs} and \eqref{No Stress Eqs} represent analytical expressions for the effective DE perturbations under the QSA and the SHA. General SVT theories have several free functions (i.e., $f_2 (\vp, X_1, X_2, X_3)$, $f_3 (\vp, X_3)$, $\tilde{f}_3 (\vp, X_3)$, $G_3 (\vp, X_1)$, and $G_4 (\vp)$) which could be useful for unravelling conundrums in the standard cosmological model. In this section, we will show an example of how our effective fluid approach to SVT theories makes it possible to design a cosmological model matching the background evolution in the \lcdm model while having non-vanishing DE perturbations. We will designate this model as SVTDES.

\subsection{Designer procedure}

To begin with, note that using the Leibniz rule, the general equation of motion for the scalar field in the left-hand side of Eq. \eqref{Gra Scalar Vec Eqs} can be recast as 
\begin{equation}
\nabla^\mu J_\mu = \mc{K}_\vp,
\end{equation}
where 
\begin{align}
J_\mu &= \( - f_{2 X_1} + G_{3 X_1} \square \vp + 2 G_{3 \vp} \) \nabla_\mu \vp + G_{3 X_1} \nabla_\mu X_1 \nonumber \\
 &+ \( - \frac{1}{2} f_{2 X_2} - 2 f_{3 \vp} + 4 X_3 \tilde{f}_{3 \vp} \) A_\mu, \label{eq:J_mu} \\
\mc{K}_\vp &= f_{2 \vp} - 2 A_\mu \nabla^\mu f_{3 \vp} - 2 \tilde{f}_{3 \vp} \( A_\nu \nabla^\mu A^\nu + A^\nu \nabla^\mu A_\nu \) A_\mu + 4 X_3 A_\mu \nabla^\mu \tilde{f}_{3 \vp} \nonumber \\
 & + \nabla^\mu G_{3 \vp} \nabla_\mu \vp + G_{4 \vp} R, \label{eq:Kphi}
\end{align}
where $\square \equiv \nabla_\rho \nabla^\rho$ is the usual Laplacian operator. Substituting the background configuration for the fields [see Eq. \eqref{Field configurations}] in the Eq. \eqref{eq:J_mu}, we find that only the temporal component of the current $J_\mu$ does not vanish 
\begin{equation}
J_0 \equiv J (\eta) = \( - f_{2 X_1} - 3 \frac{\mc{H} \vp'}{a^2} G_{3 X_1} + 2 G_{3 \vp} \) \vp' + \( - \frac{1}{2} f_{2 X_2} - 2 f_{3 \vp} + 2 \frac{A_0^2}{a^2} \tilde{f}_{3 \vp} \) A_0,
\label{eq:Jmu}
\end{equation}
and satisfies the differential equation
\begin{equation}
J' + 2 \mc{H} J + a^2 \mc{K}_\vp = 0.
\label{eq:dfJ}
\end{equation}
When $\mc{K}_\vp = 0$, the solution of Eq. \eqref{eq:dfJ} is simply
\begin{equation} 
J(\eta) = - \frac{J_c}{a^2},
\label{Eq:Sol J}
\end{equation}
where $J_c$ is a constant. If $J_c = 0$, then the system is on the attractor solution. If $J_c \neq 0$, then the system is out of the attractor and interesting phenomenology might emerge. We will assume that $J_c$ is small, and as we will see later, it will serve as a parameter tracking deviations from $\Lambda$CDM.  

Avoiding fine-tuning of the functions, the term $\mc{K}_\vp$ can be zero if we demand
\begin{equation}
f_{2 \vp} = 0, \quad f_{3 \vp} = 0, \quad \tilde{f}_{3 \vp} = 0, \quad G_{3 \vp} = 0, \quad G_{4 \vp} = 0,
\label{zero functions}
\end{equation}
which implies that the remaining SVT free-functions must be of the form
\begin{equation}
f_2 = f_2 (X_1, X_2, X_3), \quad f_3 = f_3 (X_3), \quad \tilde{f}_3 = \tilde{f}_3 (X_3), \quad G_3 = G_3 (X_1), \quad G_4 = \text{constant}.
\label{Simp Functions}
\end{equation}
From now on we will assume $G_4 = 1 / 2$ and $\kappa = 1$. Note that conditions \eqref{zero functions}-\eqref{Simp Functions} on the effective DE density \eqref{eq:den-DE} and pressure \eqref{Den and Press DE} yield an effective DE equation of state
\be 
w_{\rm DE} = \frac{f_{2} + \frac{(f_{3 X_{3}}+\tilde{f}_{3})}{a^4} (2 A_{0}^2 A'_0 - 2 A_{0}^3 \mathcal{H})  + (\vp'^3 \mathcal{H} - \vp''  \vp'^2 )\frac{G_{3 X_{1}}}{a^4}}{- f_{2} + \frac{\vp'^2 f_{2 X_{1}}}{a^2} + \frac{A_{0} \vp'  f_{2 X_{2}}}{a^2} + \frac{A_{0}^2 f_{2 X_{3}}}{a^2} - \frac{6 A_{0}^3 \mathcal{H}}{a^4}(f_{3 X_{3}}+\tilde{f}_{3}) + \frac{3 \vp'^3 G_{3 X_{1}} \mathcal{H}}{a^4}}.
\label{eq:EoS-designer}
\ee
Since $G_4$ is a constant, Eq. \eqref{Long Traceless Eq} implies $\Phi = - \Psi$,  or in other words, we are designing a model with no anisotropic stress. Note that using the conditions \eqref{Simp Functions}, we can rewrite the Friedman equation [left-hand side of \eqref{Gral Back Einstein Eqs}], the equation of motion for the scalar field [Eqs. \eqref{eq:Jmu} and \eqref{Eq:Sol J}], and the equation of motion for the vector field [right-hand side of \eqref{Back Scalar Vector Eqs}], respectively as
\begin{align}
0 &= - \frac{\mc{H}^2}{a^2} + H_0^2 \frac{\Omega_{m 0}}{a^3} - \frac{1}{3} f_2 + \frac{2}{3} X_1 f_{2 X_1} + \frac{2}{3} X_2 f_{2 X_2} + \frac{2}{3} X_3 f_{2 X_3}  \nonumber \\
 &+ 2 \sqrt{2} X_1^{3 / 2} \frac{\mc{H}}{a} G_{3 X_1} - 4 \sqrt{2} X_3^{3 / 2} \frac{\mc{H}}{a} \( f_{3 X_3} + \tilde{f}_3 \), \label{DES Friedman Eq}\\
0 &= \frac{J_c}{a^3} - 6 X_1 \frac{\mc{H}}{a} G_{3 X_1} - \sqrt{2} X_1^{1 / 2} f_{2 X_1} - \frac{\sqrt{2}}{2} X_3^{1 / 2} f_{2 X_2}, \label{DES Scalar Eq} \\
0 &= - \sqrt{2} X_3^{1 / 2} f_{2 X_3} + 12 X_3 \frac{\mc{H}}{a} \(f_{3 X_3} +  \tilde{f}_3 \) - \frac{\sqrt{2}}{2} X_1^{1 / 2} f_{2 X_2}, \label{DES Vector Eq}
\end{align}
where we have defined the density parameter of matter $\Omega_{m 0} \equiv \rho_{m 0} / 3 H_0^2 \approx 0.3$; $\rho_{m 0}$ is the density of matter today and $H_0$ is the Hubble constant. In Eqs. \eqref{DES Friedman Eq}-\eqref{DES Vector Eq} we replaced the fields $\vp'$ and $A_0$ by the variables
\be
X_1 = \frac{\vp'^2}{2 a^2}, \quad X_2 = \frac{\vp' A_0}{2 a^2}, \quad X_3 = \frac{A_0^2}{2 a^2}. 
\ee
We are interested in a particular model in SVT theories whose background evolution matches identically that of $\Lambda$CDM, where the Hubble parameter is given by
\be
\mc{H}^2 = a^2 H_0^2 (\Omega_{m 0} a^{-3} + \Omega_{\Lambda 0}),
\label{lcdm H}
\ee
where $\Omega_{\Lambda 0} \approx 0.7$ is the density parameter of dark energy today. From Eq. \eqref{eq:EoS-designer} we can see that $f_{3 X_{3}}+\tilde{f}_{3}=0$, $G_{3 X_1}=0$, and $f_{2}=\mathrm{constant}$ imply $w_{\rm DE} = -1$ as in the standard cosmological model. In this case, the vector field equation of motion \eqref{DES Vector Eq} is trivially satisfied while from Eq. \eqref{DES Scalar Eq} we see that the scalar field is on the attractor solution $J_c=0$. Furthermore, the Friedman equation \eqref{DES Friedman Eq} and its solution for $\Lambda$CDM \eqref{lcdm H} allow us to determine $f_2 = - 3 H_0^2 \Omega_{\Lambda 0}$. Such a model has vanishing DE perturbations. Next we will show that there exists a SVT model matching $\Lambda$CDM background while having non-vanishing DE perturbations.

For SVT theories assuming Eq. \eqref{Simp Functions}, in general $\bar{\rho}_\text{DE}$ depends on $X_1$, $X_2$, and $X_3$, and thus from the Friedman equation \eqref{DES Friedman Eq} we see that $\mc{H}$ might also be also a function of these terms, i.e., $\mc{H} = \mc{H} (X_1, X_2, X_3)$. We can consider a further simplification, by noting that $X_2^2 = X_1 X_3$ up to first order in perturbations, therefore we can assume $f_2 = f_2 (X_1, X_3)$ and $\mc{H} = \mc{H}(X_1, X_3)$. These assumptions imply $f_{2 X_2} = 0$ and using Eqs. \eqref{DES Friedman Eq}-\eqref{DES Vector Eq}  we find 
\be
f_2 (X_1, X_3) = - 3 H_0^2 \Omega_{\Lambda 0} + \frac{J_c \sqrt{2} X_1^{1/2} }{\Omega_{m 0}} \[ \frac{H^2}{H_0^2} - \Omega_{\Lambda 0} \], \label{f2 Eq}
\ee
\be
f_{2 X_1} = \frac{J_c H^2}{\sqrt{2} H_0^2 X_1^{1/2} \Omega_{m 0} } - \frac{J_c \Omega_{\Lambda 0}}{\sqrt{2} X_1^{1/2} \Omega_{m 0} } + \frac{2 \sqrt{2} H J_c X_1^{1/2} H_{X_1} }{H_0^2 \Omega_{m 0}} 
\ee 
\be 
f_{2 X_3} = \frac{2 \sqrt{2} H J_c X_1^{1/2} H_{X_3}}{H_0^2 \Omega_{m 0}}
\ee 
\be 
G_{3 X_1} = - \frac{2}{3} \frac{J_c H_{X_1}}{H_0^2 \Omega_{m0}}, \label{G3X1 Eq}
\ee
\be 
 f_{3 X_3} + \tilde{f}_3  = \frac{1}{3} \frac{J_c X_1^{1/2} H_{X_3}}{X_3^{1/2} H_0^2 \Omega_{m 0}},\label{SVTDES H 1}
\ee
where we have used the expression for the Hubble parameter in the standard model $H^2 = H_0^2 (\Omega_{m 0} a^{-3} + \Omega_{\Lambda 0})$. We can now assume that the Hubble parameter can be written in terms of $X_1$ and $X_3$ as 
\be
H(X_1,X_3) = H_0 \(\frac{X_1}{H_0^2}\)^{-n} \(\frac{X_3}{H_0^2}\)^{-m},
\label{H X1 X3}
\ee
where $n$ and $m$ are constants. Note that the units in the previous expression are correct, given that $[X_1] = [X_3] = H_0^2$. For the sake of simplicity, we further assume $f_{3} = 0$ which in turn defines $\tilde{f}_3$ from Eq. \eqref{SVTDES H 1} and keeps alive the vector interactions in the model.\footnote{Under the conditions \eqref{zero functions}, the cubic interactions of SVT theories are present at the first-order level only through the combination $f_{3 X_3} + \tilde{f}_3$. See Appendix \ref{App:1} where the perturbations coefficients in Eqs. \eqref{Pert time time Eq}-\eqref{Pert Space Vec Eq} are shown.} From Eqs. \eqref{f2 Eq}-\eqref{H X1 X3} it is possible to obtain expressions for $f_2, f_{2 X_1}, f_{2 X_3}, G_{3 X_1}, \tilde{f}_3$ in terms of $X_1$ and $X_3$. In order to close the system, we assume that $X_1$ and $X_3$ depend on $H$ as
\be
X_1 = \frac{X_{10}}{H^p}, \ X_3 = \frac{X_{30}}{H^q}, 
\label{ansatz 3}
\ee
where $[X_{10}] = H_0^{p+2}$, $[X_{30}] = H_0^{2+q}$, $p$ and $q$ are constants. Thus, the problem of finding a model in SVT theories with the same background as \lcdm is reduced to find an appropriate set of parameters $\{ n, m, p, q \}$. From the effective DE density \eqref{eq:den-DE} and Eqs. \eqref{lcdm H}-\eqref{ansatz 3}   we obtain 
\begin{align}
\rho_{\rm DE} &= 3 H_0^2 \Omega_{\Lambda 0} \nn \\
 &+ \frac{4\sqrt{2} H_0^2 (m+n) \tilde{J}}{\Omega_{m 0}} \[ \frac{\Omega_{m 0}}{a^{-3}} + \Omega_{\Lambda 0} \]^{\frac{(2n - 1)p + 2 m q + 2}{4}} \left( 1 - \[ \frac{\Omega_{m 0}}{a^{-3}} + \Omega_{\Lambda 0}\]^{\frac{np+mq-1}{2}}\right),
\end{align}
so that for
\be
n p + m q = 1,
\ee
the model matches the $\Lambda$CDM background evolution, while having non-vanishing perturbations: $\bar{\rho}_\text{DE} = 3 H_0^2 \Omega_{\Lambda 0}, \quad \bar{P}_\text{DE} = - \bar{\rho}_\text{DE}, \quad w_\text{DE} = -1$. We choose  
\be 
n = 1, \quad m = -2, \quad p = 2, \quad q = \frac{1}{2},
\label{SVTDES Parameters}
\ee
as a suitable set of parameters yielding manageable expressions for the perturbations in Eq. \eqref{No Stress Eqs}. Then, the choice \eqref{SVTDES Parameters} defines our SVTDES model 
\begin{equation*}
f_2 = -3 H_0^2 \Omega_{\Lambda 0} + \frac{\sqrt{2} H_0 \tilde{J} X_1^{1/2} }{\Omega_{m 0}} \[ \( \frac{X_1}{H_0^2} \)^{-2} \( \frac{X_3}{H_0^2}  \)^{4} - \Omega_{\Lambda 0} \],
\end{equation*}
\begin{equation*}
f_{3} = 0, \quad \tilde{f}_3 = \frac{2 X_1^{1/2} \tilde{J} \( \frac{X_1}{H_0^2} \)^{-1} \( \frac{X_3}{H_0^2} \)^2   }{3 X_3^{3/2} \Omega_{m 0}}, 
\end{equation*}
\be 
G_{3 X_1} = \frac{2 \tilde{J} \( \frac{X_1}{H_0^2} \)^{-2} \( \frac{X_3}{H_0^2} \)^{2} }{3 H_0^2 \Omega_{m 0}}, \quad G_4 = \frac{1}{2}.
\label{SVTDES I}
\ee
where $\tilde{J} \equiv J_c / H_0$ is dimensionless. The equations of motion for the scalar field \eqref{DES Scalar Eq} and the vector field \eqref{DES Vector Eq} are trivially satisfied for the SVTDES model. Having defined the background evolution for the SVTDES model, we will focus on the evolution of perturbations which are defined by the coefficients $Y_i$ in the Appendix \ref{App: No Ani Stress Coefficients}. The non-vanishing $Y_i$ for the SVTDES model \eqref{SVTDES I} are the coefficients $Y_1, Y_2, Y_3, Y_5, Y_6, Y_8, Y_9, Y_{11}, Y_{12}$, which yield
\begin{align}
\delta \rho_\text{DE} &\approx \frac{2 \sqrt{2} \tilde{J}}{21 \Omega_{m 0}} \sqrt{a \( \Omega_{m0} + a^3 \Omega_{\Lambda 0}\)} \nn \\
 &\times \[ \frac{28 a}{29 \Omega_{m 0} + 20 a^3 \Omega_{\Lambda 0}} + 3 H_0^2 \( \frac{5}{k^2} - \frac{96 a}{8 a k^2 + 273 H_0^2 \Omega_{m 0} + 336 a^3 H_0^2 \Omega_{\Lambda 0}} \) \] \delta \rho_m,
\label{eq:svtdesdrho}
\end{align}
\begin{align}
\delta P_\text{DE} &\approx \frac{a H_0^2 \tilde{J} \delta \rho_m}{3 \sqrt{2} k^2 \Omega_{m 0} \sqrt{a \( \Omega_{m0} + a^3 \Omega_{\Lambda 0}\)} ( 8 a k^2 + 273 H_0^2 \Omega_{m 0} + 336 a^3 H_0^2 \Omega_{\Lambda 0})} \\
 &\times \[ 13 \Omega_{m0} \( 8 a k^2 + 615 H_0^2 \Omega_{m 0} \) + 4 a^3 (8 a k^2 + 3669 H_0^2 \Omega_{m 0}) \Omega_{\Lambda 0} + 5952 a^6 H_0^2 \Omega_{\Lambda 0}^2 \], \nn
\end{align}
\begin{align}
\label{SVTDES velocity}
\bar{\rho}_\text{DE} V_\text{DE} &\approx - \frac{16 \sqrt{2} a H_0 \tilde{J} \delta \rho_m}{3 \Omega_{m 0} (29 \Omega_{m 0} + 20 a^3 \Omega_{\Lambda 0}) ( 8 a k^2 + 273 H_0^2 \Omega_{m 0} + 336 a^3 H_0^2 \Omega_{\Lambda 0})} \\
 &\times \( \Omega_{m 0} \(20 a k^2 + 1161 H_0^2 \Omega_{m 0} \) + a^3 \( 8 a k^2 + 1791 H_0^2 \Omega_{m 0} \) \Omega_{\Lambda 0}  + 576 a^6 H_0^2 \Omega_{\Lambda 0}^2 \), \nn
\end{align}
\begin{align}
c_{s, \text{DE}}^2 &= \frac{H_0^2 (29 \Omega_{m 0} + 20 a^3 \Omega_{\Lambda 0})}{4 \( \Omega_{m 0} + a^3 \Omega_{\Lambda 0} \)} \nn \\
 &\times \( 13 \Omega_{m 0} \(8 a k^2 + 615 H_0^2 \Omega_{m 0} \) + 4 a^3 \( 8 a k^2 + 3669 H_0^2 \Omega_{m 0} \) \Omega_{\Lambda 0}  + 5952 a^6 H_0^2 \Omega_{\Lambda 0}^2 \) \nn \\
 &\times[ 32 a^2 k^4 + 396 a H_0^2 k^2 \Omega_{m 0} + 16965 H_0^4 \Omega_{m 0}^2 \nn \\
 &+ 36 a^3 H_0^2 (24 a k^2 + 905 H_0^2 \Omega_{m 0}) \Omega_{\Lambda 0} + 14400 a^6 H_0^4 \Omega_{\Lambda 0}^2 ]^{-1},
\label{SVTDES cs2}
\end{align}
where we have assumed $\tilde{J} \ll 1$. We have replaced $X_1$ and $X_3$ in terms of $H$ using Eqs. \eqref{ansatz 3} and \eqref{SVTDES Parameters},  then $H$ can be written in terms of $a$ by using Eq. \eqref{lcdm H}. It becomes clear that $\delta \rho_\text{DE}$, $\delta P_\text{DE}$, and $V_\text{DE}$,  vanish for $\tilde{J} =0$, therefore we recover $\Lambda$CDM, i.e., there are no dark energy perturbations. In the following subsections, we will explore the cosmological implications of the SVTDES model \eqref{SVTDES I}, always tracking deviations from \lcdm through the parameter $\tilde{J}$.  

\subsection{Evolution of matter and dark energy perturbations}

In this subsection, we numerically solve the differential equations in Eqs. \eqref{Eq:evolution-delta} and \eqref{Eq:evolution-V} for matter perturbations, i.e., for $\delta_m$ and $V_m$. 

\begin{figure}[t!]
\includegraphics[width = 0.48\textwidth]{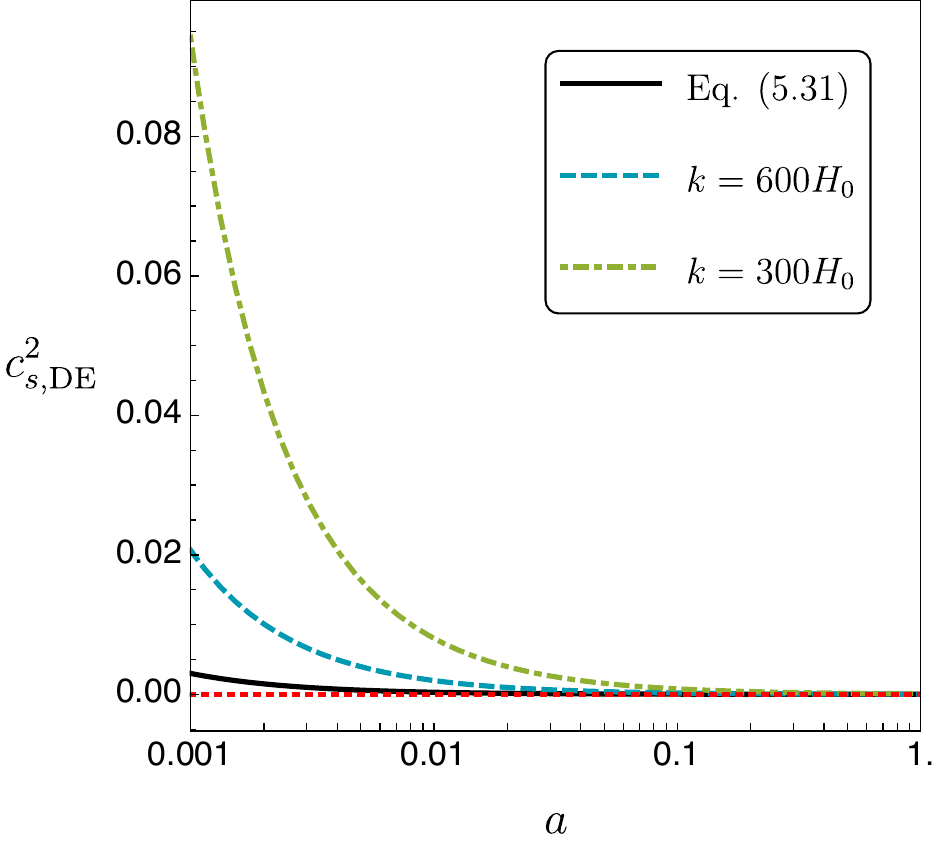}
\hfill
\includegraphics[width = 0.48\textwidth]{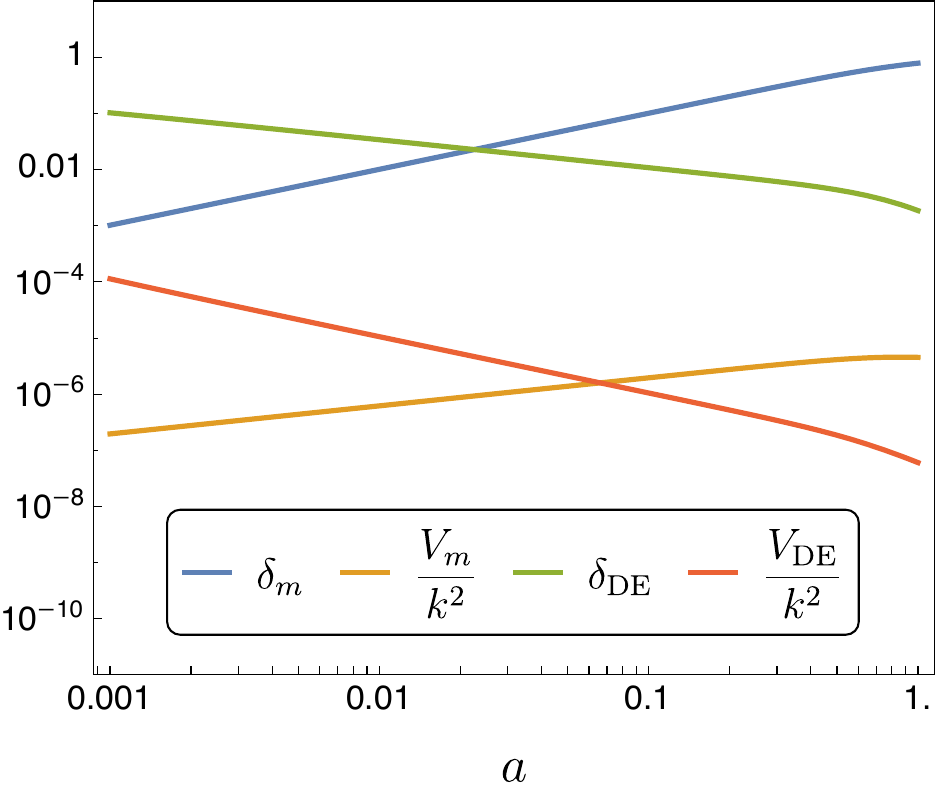}
\caption{Left: Evolution of the DE sound speed $c_{s, \text{DE}}^2$ for the modes $k = 600 H_0$ and $k = 300 H_0$. For the two values considered, we can see $c_{s, \text{DE}}^2$ is positive during the whole evolution, and it gets greater values than those in the black solid line, which marks a rough bound above which the SHA can be safely applied. Right: Evolution of $\delta_m$, $\delta_\text{DE}$, $V_m / k^2$, and $V_\text{DE} / k^2$ (their absolute values) for $\tilde{J} = 0.01$. The other parameters used in this figure are $\Omega_{m0} = 0.3$, $\Omega_{\Lambda 0} = 0.7$, and $k = 300 H_0$. The initial conditions are obtained from Eq. \eqref{eq:madom}.}
\label{Evo of Perts}
\end{figure}

To begin with, we check the stability of DE perturbations. Since $G_4$ is a constant for the SVTDES model, Eq. \eqref{Long Traceless Eq} implies that there is no anisotropic stress, and thus we do not have to consider an effective sound speed. The sound speed of DE perturbations in Eq. \eqref{No Stress Eqs} is the key quantity driving the stability of  perturbations, and for the SVTDES model in the Newtonian gauge it is given by the Eq. \eqref{SVTDES cs2} which interestingly does not depend on $\tilde{J}$. We show the evolution of $c_{s, \text{DE}}^2$ as a function of $a$, for a few values of $k$, in the left panel of Fig.~\ref{Evo of Perts}. We would like to make some comments about the behaviour of $c_{s, \text{DE}}^2$. First, we can see that the squared sound speed is positive during the whole evolution, assuring that our SVTDES model avoids Laplacian instabilities. Second, the SHA indeed applies for modes well within the sound horizon, i.e., for modes such that \cite{PhysRevD.94.044024}
\be
c_{s, \text{DE}}^2 k^2 \gg \mc{H}^2.
\label{eq:de-sound-horizon}
\ee 
Therefore, the SHA breaks down for $c_{s, \text{DE}}^2 \approx 0$. We can make a rough estimate of how small $c_{s, \text{DE}}^2$ can be so that the SHA be justifiable. Since co-moving wavenumbers relevant to the observations of large-scale structures lay in the range $30 H_0 \lesssim k \lesssim 600 H_0$ \cite{SDSS:2003tbn}, and it is reasonable to assume that during matter domination $H^2 \approx H_0^2 \Omega_{m 0} a^{-3}$, from \eqref{eq:de-sound-horizon} and using $k \sim 300 H_0$ we get  
\be
c_{s, \text{DE}}^2 \gtrsim 3 \times 10^{-6} a^{-1},
\label{cs2 bound}
\ee
which provides a rough bound for modes inside the sound horizon. In Fig. \ref{Evo of Perts}, the solid black line  shows the relation $c_{s, \text{DE}}^2 = 3 \times 10^{-6} a^{-1}$. We see that the values taken by $c_{s, \text{DE}}^2$ for the two values considered, namely, $k = 600 H_0$ and $k = 300 H_0$ (blue dashed line and green dot-dashed line, respectively), are higher than those taken in the black line, therefore, the SHA can be safely applied. Third, note that earlier than the regime of validity of our treatment (i.e., matter dominance) as well as for modes $k \approx H_0$, DE perturbations propagate with speed greater than the speed of light.  

Now, we focus on the differential equations \eqref{Eq:evolution-delta} and \eqref{Eq:evolution-V} for matter perturbations. For pressure-less matter we have $w_m = 0$, $\pi_m = 0$, and $c_{s, m}^2 = 0$. Hence, matter perturbations equations read 
\be
\delta\text{'}_m = - \frac{V_m (a)}{a^2 H (a)} - 3 \Phi\text{'}(a), \quad V\text{'}_m = - \frac{V_m (a)}{a} - \frac{k^2}{a^2} \frac{\Phi (a)}{H(a)},
\label{matter perturbations}
\ee
The evolution equations for matter perturbations in Eq. \eqref{matter perturbations} couple to the DE perturbations through the gravitational potential $\Phi$. From Eqs. \eqref{eq:phiprimeeq}-\eqref{eq:phiprimeeq1}, we can eliminate $\Phi'$ and obtain
\be 
\Phi(a) = \frac{a^2}{2 k^2} \left\{ 3 H_0^2 \Omega_{m 0} a^{-3} \( \delta_m + \frac{3 a H}{k^2} V_m \) + \bar{\rho}_\text{DE} \( \delta_\text{DE} + \frac{3 a H}{k^2} V_\text{DE} \) \right\}.
\label{eq:phi-of-a}
\ee
Since the background of the SVTDES model is equivalent to that of $\Lambda$CDM, the Hubble parameter is given by Eq. \eqref{lcdm H}, and the density of dark energy is $\bar{\rho}_\text{DE} = 3 H_0^2 \Omega_{\Lambda 0}$, hence \eqref{eq:phi-of-a} is simplified to
\be
\Phi (a) = \frac{3}{2} \frac{H_0^2}{a k^2} \left\{ \Omega_{m 0} \( \delta_m + \frac{3 a H}{k^2} V_m \) + \Omega_{\Lambda 0} a^3 \( \delta_\text{DE} + \frac{3 a H}{k^2} V_\text{DE} \) \right\}.
\ee
In order to solve Eqs. \eqref{matter perturbations}, we need to determine $\delta_\text{DE}$ and $V_\text{DE}$. Our effective fluid approach allowed us to find analytical expressions for the perturbations $\delta \rho_\text{DE}$ and $V_\text{DE}$ which for our SVTDES model \eqref{SVTDES I} are given by Eqs. \eqref{eq:svtdesdrho}-\eqref{SVTDES velocity}, respectively. 

The initial conditions required to solve Eqs. \eqref{matter perturbations} are set by the following expressions
\be
\delta_{m, i} = \delta_i\, a_i \(1 + 3 \frac{a_i^2 H^2(a_i)}{k^2} \), \quad V_{m, i} = - \delta_i\, H_0\, \Omega_{m 0}\, a_i^{1/2} ,
\label{eq:madom}
\ee
corresponding to the standard solutions of Eqs. \eqref{matter perturbations} for $\delta_m$ and $V_m$  in matter dominance, i.e., assuming that $H^2 = H_0^2 \Omega_{m 0} a^{-3}$. The overall factor $\delta_i$ is set to unity, and we choose $a_i = 10^{-3}$, ensuring initial conditions well within the matter epoch, right after decoupling. 

The evolution of $\delta_m$, $V_m/k^2$, $\delta_\text{DE}$, and $V_\text{DE}/k^2$ (their absolute values) are depicted on the right panel of Fig.~\ref{Evo of Perts}. Note that the velocity perturbation is $u$, which is defined through $u_i \equiv - \partial_i u$ for scalar perturbations [see Eq. \eqref{eq:effectTmnvde}]. The relation of the velocity perturbation to the scalar velocity and the velocity divergence is $V \propto \theta = i k^j u_j = k^2 u$, and then $u \propto V / k^2$.

\subsection{Solution for the growth factor}
\label{Sec: SVTDES Growth}

As explained in Sec. \ref{Section:EFA}, under the SHA and the QSA, the parameter $G_\text{eff}$ plays an important role in the growth of structure, as can be seen in Eq. \eqref{Evo Eq CDM}. In this subsection, we explore possible changes in the parameter $f\sigma_8$ within the SVTDES model due to variations in the strength of gravity which are encoded in the parameter $G_\text{eff}$.

\begin{figure}[t!]
\includegraphics[width = 0.45\textwidth]{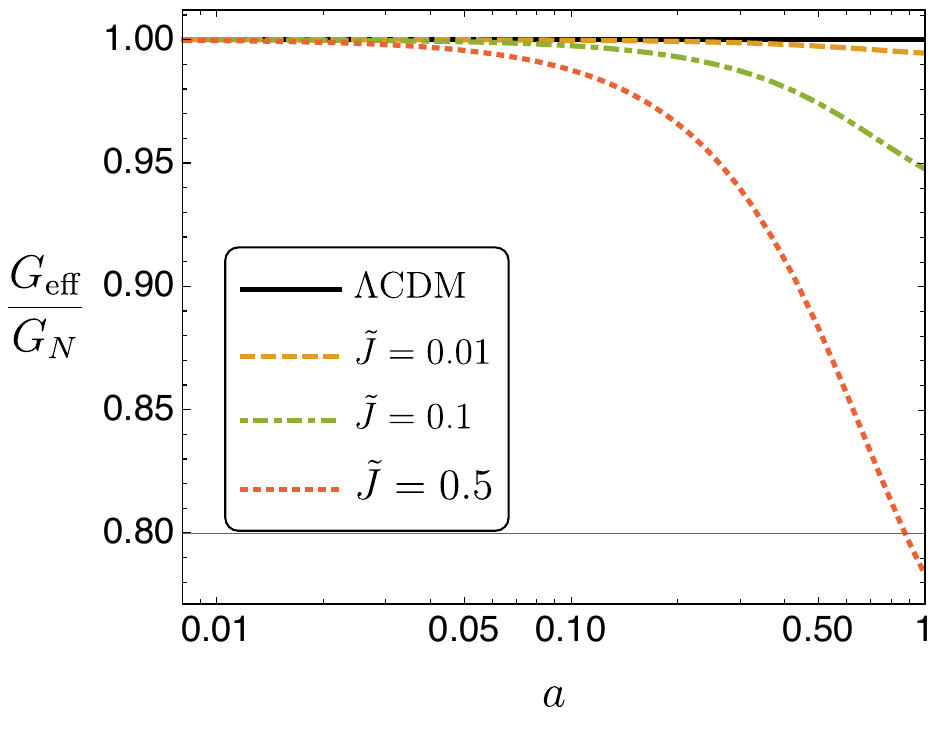}
\hfill
\includegraphics[width = 0.5\textwidth]{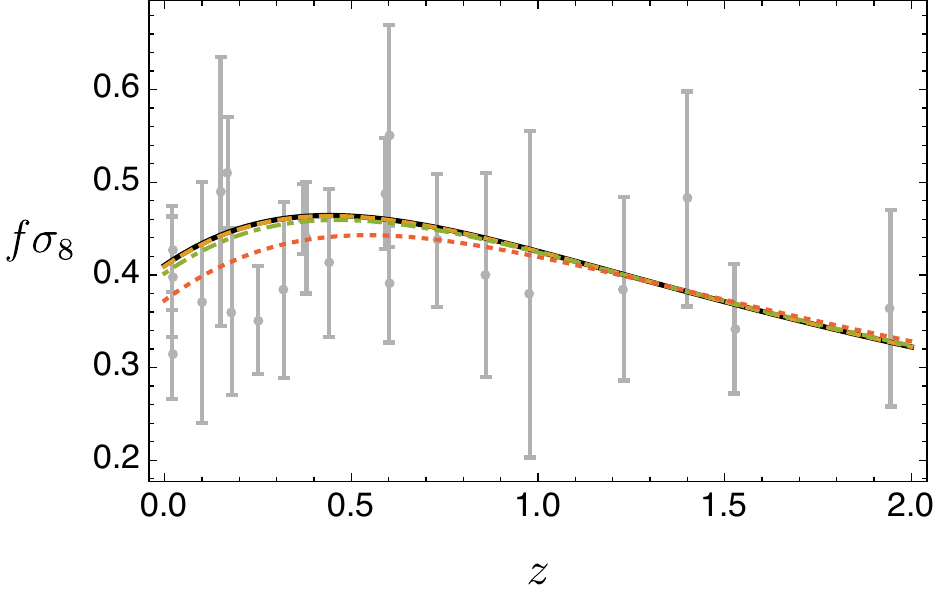}
\caption{Left: Deviations of $G_\text{eff}$ from $G_N$ from decoupling to today for different values of the parameter $\tilde{J}$. We can see that deviations from GR are only at late-times (occurring around the dark energy transition at $z \approx 0.3$), and smaller for $\tilde{J} \approx 0$ as expected. Observe that for $\tilde{J} = 0.5$, the difference between $G_\text{eff}$ and $G_N$ is around $20\%$ at the present time. Right: Evolution of $f\sigma_8 (z)$ for the same values of $\tilde{J}$ shown in the left panel. In the  case $\tilde{J} = 0.5$, deviations of SVTDES from \lcdm are fairly noticeable. For the SVTDES model gravity is weaker than in the standard cosmological model leading to a less efficient late-time matter clustering. Other parameters used in the figures are $\Omega_{m0} = 0.3$, $\Omega_{\Lambda 0} = 0.7$, $\sigma_8 = 0.8$, and $k = 300 H_0$.}
\label{Growth}
\end{figure}

From Eqs. \eqref{QSA SHA Variables} and \eqref{Poisson Eqs}, we obtain the following analytical expression for $G_\text{eff}$ under the QSA and the SHA 
\be
\frac{G_\text{eff}}{G_N} = 2 \frac{\frac{k^4}{a^4} W_{14} + \frac{k^2}{a^2} W_{15} + W_{13}}{\frac{k^4}{a^4} W_3 + \frac{k^2}{a^2} W_4 + W_5},
\label{Geff}
\ee 
where the coefficients $W_i$ are given in the Appendix \ref{App: QSA SHA Coefficients}. Replacing the SVTDES model [Eqs. \eqref{SVTDES I}] in the parameter $G_\text{eff}$ in Eq. \eqref{Geff}, using the Hubble parameter of \lcdm given in \eqref{lcdm H}, and assuming some values for the parameter $\tilde{J}$, namely, $\tilde{J} = 0.01$, $0.1$, $0.5$, we can numerically solve the differential equation for $\delta_m$ in Eq. \eqref{Evo Eq CDM}, where the initial conditions are set as $\delta_m (a_i) = a_i$ and $\delta\text{'}_m (a_i) = 1$ for a value of the scale factor $a_i$ deep in the matter era ($a_i \sim 10^{-3}$). Other parameters used in the numerical solutions are $\Omega_{m0} = 0.3$, $\Omega_{\Lambda 0} = 0.7$, and $k = 300 H_0$. In order to compare with observations, from this numerical solution we compute the $f\sigma_8$ function, which is defined as
\be 
f\sigma_8 (a) \equiv \sigma_8 \frac{a \, \delta\text{'}_m (a)}{\delta_m (a = 1)},
\label{eq:fs8}
\ee  
where $\sigma_8 \sim 0.8$ is the expected RMS over-density in a sphere of co-moving radius equal to $8 h^{-1}$ Mpc, $h$ being the normalized Hubble parameter. The results for the different values of $\tilde{J}$, aside the \lcdm case $\tilde{J}=0$, are shown in Fig. \ref{Growth}. In the left panel of Fig. \ref{Growth} we can see that for $\tilde{J} = 0.5$, the modifications to GR, i.e., deviations of $G_\text{eff}$ from $G_N$, are fairly noticeable at late-times. This difference translates to a weaker gravity when DE becomes relevant in the cosmic budget, which leads to a different evolution of the growth factor. In the right panel of Fig. \ref{Growth}, we plot $f\sigma_8 (z)$ versus the data compilation from Ref. \cite{Sagredo:2018ahx}. For $\tilde{J} = 0.5$, we can see that $f\sigma_8$ for SVTDES has a strong departure from $\Lambda$CDM (black solid curve) indicating a less efficient matter clustering in comparison with the standard model. Weaker gravity can be helpful in understanding the discrepancy in $S_8$ between low- and high-redshift probes. Gravity strength also decreases for smaller values of the parameter $\tilde{J}=0.1,\,0.01$, but differences with respect to \lcdm are hardly significant. 

\subsection{CMB angular power spectrum and matter power spectrum}
\label{Section: CMB - Pk}

Having studied the evolution of matter perturbations in the previous subsections, here we present our results for the CMB power spectrum and the linear matter power spectrum. The advantage of the effective fluid approach is that it allows a relatively easy implementation of the SVTDES model in Boltzmann solvers. In its default version, Boltzmann codes usually have already a DE fluid implemented and parameterised by an equation of state $w$, sound speed in the fluid rest-frame $\hat{c}_s^2$, and vanishing anisotropic stress $\pi=0$. In this work, we have computed the effective fluid quantities describing fairly general SVT theories.  

We chose to carry out the implementation of the SVTDES model in the Boltzmann solver \texttt{CLASS}.\footnote{Version v3.2.0} Since our model matches the \lcdm \, background  evolution ($w_\text{DE} = - 1$) and has vanishing anisotropic stress ($\pi_{\rm DE}=0$), we decided to perform the smallest number of modifications in the code. It turns out that only one modification in the module \texttt{perturbations.c} is required: i) the scalar velocity $V_\text{DE}$ \eqref{SVTDES velocity} modifies the equation for $\Phi'$ in the function \texttt{perturbations\_einstein}. 

\begin{figure}[t!]
\includegraphics[width = 0.49\textwidth]{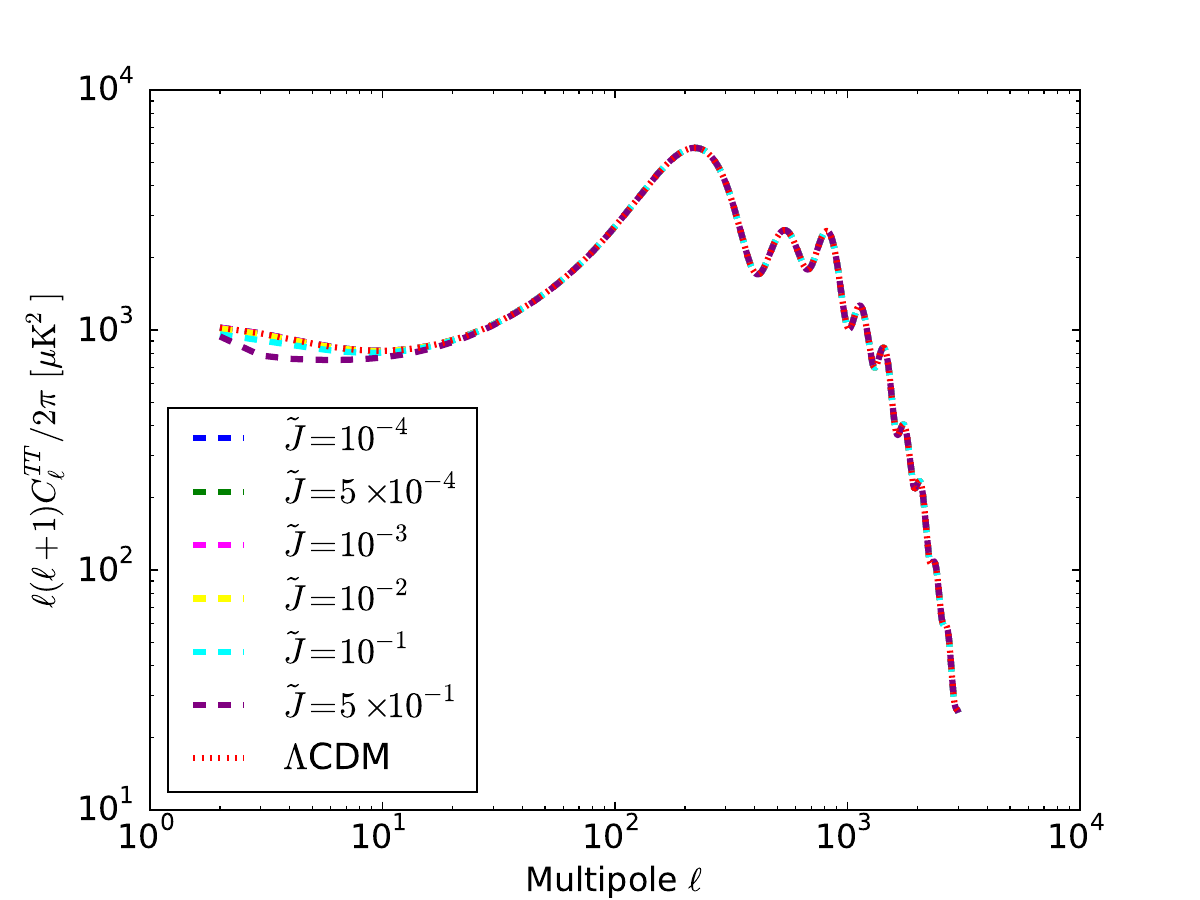}
\hfill
\includegraphics[width = 0.49\textwidth]{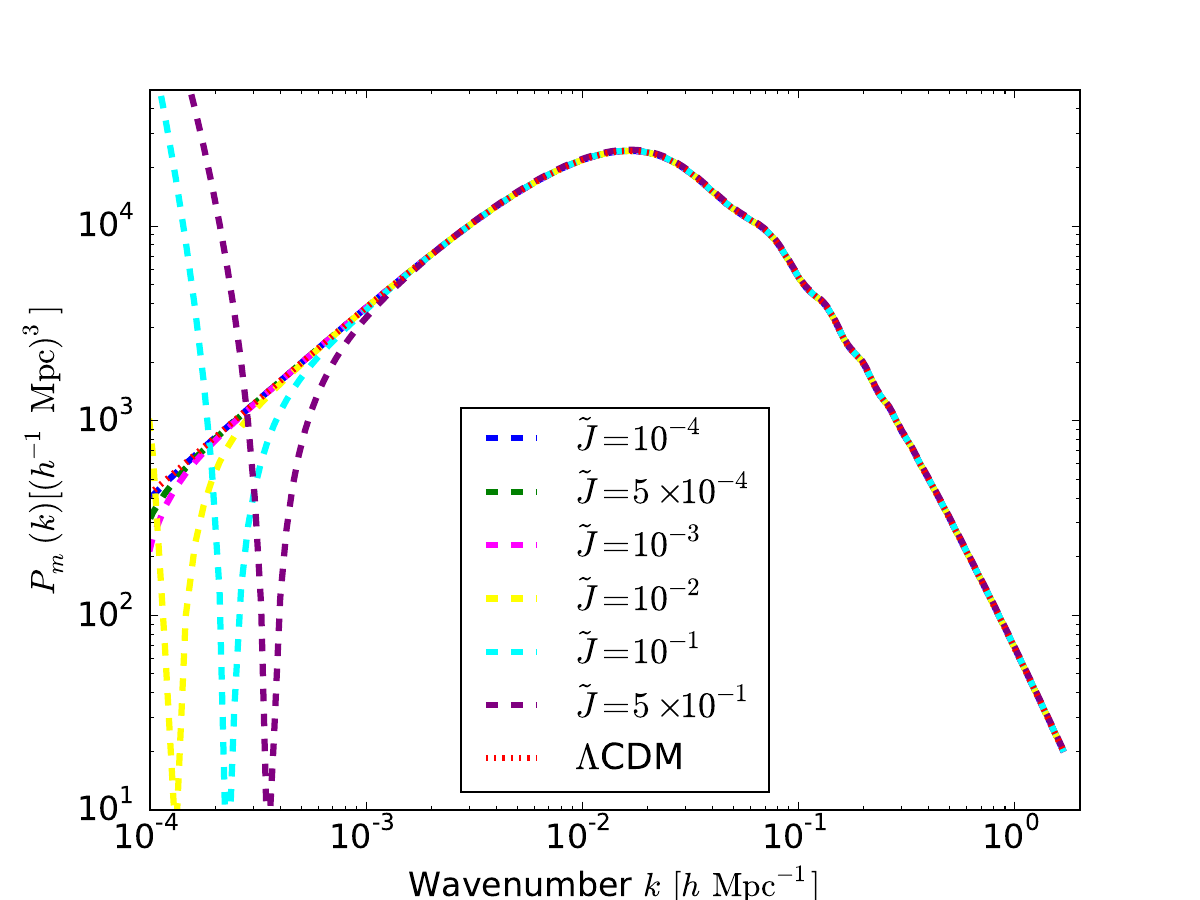}
\caption{Left: TT CMB angular power spectrum for the SVTDES model (dashed lines), and $\Lambda$CDM (dotted, red line). Right: Linear matter power spectrum at $z=0$ for the SVTDES model (dashed lines), and $\Lambda$CDM (dotted, red line). While the CMB angular power spectrum is almost identical in SVTDES and $\Lambda$CDM for small $\tilde{J}$, the matter power spectrum differs on very large scales where  SHA cannot be applied: for late-times and modes as large as $k \sim 10^{-4} \, h$ Mpc$^{-1}$ QSA and SHA cease to be valid. These plots were generated by slightly modifying \texttt{CLASS}, assuming $n_s = 0.9649$, $\ln 10^{10}A_s = 3.044$, $h = 0.6736$, $\omega_{\rm b} = 0.02237$, $\omega_{\rm cdm}=0.1200$, $\tau = 0.0544$, $\sum m_\nu = 0.06 \mathrm{eV}$  and $\tilde{J}$ as indicated in the legend.}
\label{TT_Pk}
\end{figure}

The CMB temperature power spectrum and the linear matter power spectrum are shown in Fig.~\ref{TT_Pk}. Perturbation equations were solved by using the cosmological parameters from the 2018 Planck baseline result \cite{Planck:2018vyg}: scalar spectrum power-law index $n_s = 0.9649$, Log power of the primordial curvature  perturbations $\ln 10^{10} A_s = 3.044$, reduced Hubble parameter $h = 0.6736$, baryon density today $\omega_{\rm b} = 0.02237$, cold dark matter density today $\omega_{\rm cdm} = 0.1200$, Thomson scattering optical depth due to reionization $\tau=0.0544$, sum of neutrino masses in eV $\sum m_\nu = 0.06$, and some values of the SVTDES parameter $\tilde{J}$. We also plot the standard $\Lambda$CDM results for reference.  As it can be seen in the left panel of Fig.~\ref{TT_Pk}, the match in the TT CMB angular power spectrum between SVTDES and $\Lambda$CDM is almost perfect. In the right panel of Fig.~\ref{TT_Pk} we can see that the agreement in the linear matter power spectrum is also quite good for SVTDES and \lcdm models. Nonetheless, for modes $k \sim 10^{-4}-10^{-3} \, h$ Mpc$^{-1}$ there is a departure from $\Lambda$CDM. Since we are working under the SHA and QSA, a big deviation is expected on large scales and late-times where the approximations are not valid.

\subsubsection{Sound speed in the rest-frame}
\label{App: rest-frame}

When implementing DE fluids in \texttt{CLASS}, it is important to bear in mind that the code uses the \textit{co-moving sound speed} $\hat{c}_s^2$, i.e., the sound speed in the rest-frame of the fluid, which, in general, is given by
\be 
\hat{c}_s^2 = \frac{\delta P^C}{\delta \rho^C},
\ee
where $\delta P^C$ and $\delta \rho^C$ are the pressure and density perturbations computed in the co-moving gauge. These quantities are related to quantities in the Newtonian gauge (the gauge where our main results were derived) through the following transformation rules \cite{Bean:2003fb}
\be
\delta \rho^C = \delta \rho^N + \bar{\rho}' \frac{\theta^N}{k^2}, \quad \delta P^C = \delta P^N + \bar{P}' \frac{\theta^N}{k^2},
\ee 
where the superscript $N$ denotes a quantity computed in the Newtonian gauge. Therefore, the sound speed in the rest frame will be given by
\be
\hat{c}_s^2 = \frac{\delta P^N + \bar{P}' \theta^N / k^2}{\delta \rho^N + \bar{\rho}' \theta^N / k^2}.
\label{rest frame cs2}
\ee 
Since the background pressure and density of our SVTDES model are constants, we see from the last equations that the sound speed in Eq. \eqref{SVTDES cs2} is actually equivalent to the sound speed in the rest frame. However, this is not the case in general. Let us take the quintessence field as an example. Replacing the full perturbations \eqref{Perts QS}-\eqref{V QS} in Eq. \eqref{rest frame cs2}, computing the derivatives of $\bar{\rho}_\text{DE}$ and $\bar{P}_\text{DE}$ in Eqs. \eqref{Den QS}, and using the relation $\vp'' = - 2 \mc{H} \vp' - a^2 V_\vp$, we find that
\be
\hat{c}_s^2 = 1.
\ee 
In this case, the result is equal to $c_{s, \text{DE}}^2$ computed under the QSA and SHA [see Eq. \eqref{QSA SHA Quintessence}] since $V_\text{DE} = 0$ for this particular model, but it is substantially different to $c_{s, \text{DE}}^2$ given in Eq. \eqref{V QS} in the Newtonian gauge. Another interesting example concerns $f(R)$ theories where, in general, DE sound speed might depend on both time and scale. For instance, in Ref. \cite{Arjona:2018jhh} we can find expressions [see their Eqs. $(61)-(68)$] under QSA and SHA for the Hu \& Sawicki model that allow us to obtain $\hat{c}_s^2 \neq c_{s, \text{DE}}^2 $.

In summary, \texttt{CLASS} uses in their computations the sound speed in the co-moving gauge which is related to the Newtonian gauge by Eq. \eqref{rest frame cs2}. Therefore, for a specific SVT model, we have to replace $\delta \rho_\text{DE}$, $\delta P_\text{DE}$, $V_\text{DE} = (1 + w_\text{DE}) \theta_\text{DE}$ from Eq. \eqref{Stress Eqs} or Eq. \eqref{No Stress Eqs} depending whether or not the model has a vanishing DE anisotropic stress, and compute the derivatives of the background density and pressure in Eqs. \eqref{eq:den-DE} and \eqref{Den and Press DE}.

\section{Conclusions}
\label{Section:conclusions}

Both scalar and vector fields are present in nature and it is reasonable that they might provide explanations for shortcomings in the standard cosmological model \lcdm. In this work we investigated fairly general scalar-vector-tensor theories having second order equations of motion: SVT theories encompass both Horndeski and generalised Proca Lagrangians. Although these kinds of theories might provide new, interesting phenomenology for cosmology, they have been overlooked in the literature.      
SVT theories have various free functions taking in all relevant  interactions, hence possibly richer phenomenology than in the standard model. Nevertheless, more degrees of freedom come along with more complicate equations of motion. Even though complexity in SVT theories is reduced thanks to the constraint in the propagation speed of gravitational waves $c_T^2=1$, equations of motion remain intricate enough to find general analytical or numerical solutions.  

Here, we applied an effective fluid approach to SVT theories satisfying $c_T^2=1$. In order to decrease the complexity in the equations of motion, we carefully performed both sub-horizon and quasi-static approximations. As a result, we obtained analytical expressions  describing the effective dark energy fluid, namely, equation of state $w(a)$, squared sound speed $c_s^2(a,k)$, and anisotropic stress $\pi(a,k)$. Equations \eqref{Stress Eqs} and \eqref{No Stress Eqs} summarise our main results for the behaviour of perturbations, while from Eqs. \eqref{eq:den-DE}-\eqref{Den and Press DE} the equation of state is obtained.  

Our analytical expressions allowed us to retrieve well known results (e.g., quintessence and $f(R)$). Moreover, we also proposed extensions to these popular theories which exemplify possible, new phenomenology, for instance, changes in quantities driving the perturbations such as the sound speed and anisotropic stress. 

An interesting aspect of our investigation is that it makes it possible to design cosmological models satisfying certain conditions. As an example, we found a SVT model (dubbed SVTDES in the main text) exactly matching the background behaviour in the standard cosmological model $\Lambda$CDM, while having non-vanishing dark energy perturbations. Our effective fluid approach and the analytical solutions for the effective dark energy perturbations made it possible a relatively easy implementation of SVTDES in the Boltzmann solver CLASS. Having a code computing numerical solutions for perturbation equations in SVT cosmological models is relevant because it allows testing against measurements, e.g., CMB angular power spectra, matter power spectrum. There is however no public Boltzmann solver including a fully numerical implementation of SVT models, that is, using neither QSA nor SHA. Therefore, our results might be helpful as a reference for future exact computations testing the limitations of QSA and SHA. Since our SVTDES model has one additional parameter with respect to $\Lambda$CDM, in a model comparison it would be penalised by the Bayesian evidence. However, our example also shows that exploring the construction of cosmological models satisfying additional conditions might be well worth an investigation. Given the current discrepancies in cosmological parameters such as $H_0$ and $\sigma_8$, theories providing non trivial behaviour for $\pi$ and $c_s^2$ could alleviate the tensions while not being affected by the Occam's razor \cite{Sabla:2022xzj}.   

\section*{Acknowledgements}

We are grateful to Jose Palacios for careful reading the manuscript and providing comments. WC acknowledges financial support from the S\~{a}o Paulo Research Foundation (FAPESP) through grant \#2021/10290-2. This research was supported by resources supplied by the Center for Scientific Computing (NCC/GridUNESP) of the S\~{a}o Paulo State University (UNESP). BOQ and CAVT are supported by Patrimonio Autónomo - Fondo Nacional de Financiamiento  para  la  Ciencia,  la  Tecnología  y  la  Innovación  Francisco  José  de  Caldas  (MINCIENCIAS - COLOMBIA)  Grant  No.   110685269447  RC-80740-465-2020,  projects  69723 and 69553. 

\section*{Numerical codes}

Modified \texttt{CLASS} code reproducing results in this work can be found in the GitHub branch \texttt{svt} of the repository  \href{https://github.com/wilmarcardonac/EFCLASS.git}{\texttt{EFCLASS}}. A large part of the calculations in this paper were carried out using several  \texttt{Mathematica} packages, like \texttt{xPand}. The notebooks showing these computations can be found in the GitHub repository \href{https://github.com/BayronO/SVT}{\texttt{SVT}}.


\appendix

\section{General equations of motion}
\label{App:1}

\subsection{Gravitational field equations}
\label{App: Gral Gravitational Field}

Coefficients in Eq.\eqref{Gra Field Eqs}:
\begin{align}
\mc{G}^{(2)}_{\mu\nu} &= - \frac{1}{2} f_{2 X_{3}} A_{\mu} A_{\nu} - \frac{1}{2} f_{2} g_{\mu \nu} - \frac{1}{4} f_{2 X_{2}} A_{\nu} \nabla_{\mu}\varphi - \frac{1}{4} f_{2 X_{2}} A_{\mu} \nabla_{\nu}\varphi - \frac{1}{2} f_{2 X_{1}} \nabla_{\mu}\varphi \nabla_{\nu}\varphi, \\
\mc{G}^{(3)}_{\mu\nu} &= - \tilde{f}_{3 \varphi } A^{\alpha} A_{\mu} A_{\nu} \nabla_{\alpha}\varphi + f_{3 \varphi } A^{\alpha} g_{\mu \nu} \nabla_{\alpha}\varphi - f_{3 X_{3}} A_{\mu} A_{\nu} \nabla_{\alpha}A^{\alpha} - \tilde{f}_{3} A_{\mu} A_{\nu} \nabla_{\alpha}A^{\alpha} \nn \\
 &- f_{3 X_{3}} A^{\alpha} A^{\beta} g_{\mu \nu} \nabla_{\beta}A_{\alpha} - \tilde{f}_{3} A^{\alpha} A^{\beta} g_{\mu \nu} \nabla_{\beta}A_{\alpha} - f_{3 \varphi } A_{\nu} \nabla_{\mu}\varphi + f_{3 X_{3}} A^{\alpha} A_{\nu} \nabla_{\mu}A_{\alpha} \nn \\
 &+ \tilde{f}_{3} A^{\alpha} A_{\nu} \nabla_{\mu}A_{\alpha} - f_{3 \varphi } A_{\mu} \nabla_{\nu}\varphi + f_{3 X_{3}} A^{\alpha} A_{\mu} \nabla_{\nu}A_{\alpha} + \tilde{f}_{3} A^{\alpha} A_{\mu} \nabla_{\nu}A_{\alpha},  \\
\mathscr{H}^{(3)}_{\mu\nu} &= - \frac{1}{2} G_{3 \varphi} g_{\mu \nu} \nabla_{\alpha}\varphi \nabla^{\alpha}\varphi + \frac{1}{2} G_{3 X_{1}} g_{\mu \nu} \nabla^{\alpha}\varphi \nabla_{\beta}\nabla_{\alpha}\varphi \nabla^{\beta}\varphi - \frac{1}{2} G_{3 X_{1}} \nabla_{\alpha}\nabla_{\nu}\varphi \nabla^{\alpha}\varphi \nabla_{\mu}\varphi \nn \\
 &- \frac{1}{2} G_{3 X_{1}} \nabla_{\alpha}\nabla_{\mu}\varphi \nabla^{\alpha}\varphi \nabla_{\nu}\varphi + G_{3 \varphi} \nabla_{\mu}\varphi \nabla_{\nu}\varphi + \frac{1}{2} G_{3 X_{1}} \nabla_{\alpha}\nabla^{\alpha}\varphi \nabla_{\mu}\varphi \nabla_{\nu}\varphi,   \\
\mathscr{H}^{(4)}_{\mu\nu} &= G_{4} G_{\mu \nu} + G_{4 \varphi} g_{\mu \nu} \nabla _{\alpha}\nabla^{\alpha}\varphi + G_{4 \varphi \varphi} g_{\mu \nu} \nabla _{\alpha}\varphi \nabla^{\alpha}\varphi - G_{4 \varphi \varphi} \nabla _{\mu}\varphi \nabla _{\nu}\varphi - G_{4 \varphi} \nabla _{\nu}\nabla _{\mu}\varphi.
\end{align}

\subsection{Scalar field equation of motion}
\label{App: Gral Scalar Field}

Coefficients on left-hand side of Eqs.\eqref{Gra Scalar Vec Eqs}:
\begin{align}
\mc{J}_2 &= f_{2 \varphi } + \frac{1}{2} f_{2 \varphi X_{2}} A^{\alpha} \nabla_{\alpha}\varphi + \frac{1}{2} f_{2 X_{2}} \nabla_{\alpha}A^{\alpha} + f_{2 X_{1}} \nabla_{\alpha}\nabla^{\alpha}\varphi + f_{2 \varphi X_{1}} \nabla_{\alpha}\varphi \nabla^{\alpha}\varphi \nn \\
 &- \frac{1}{4} f_{2 X_{2} X_{2}} A^{\alpha} \nabla_{\alpha}A^{\beta} \nabla_{\beta}\varphi - \frac{1}{2} f_{2 X_{2} X_{3}} A^{\alpha} A^{\beta} \nabla_{\beta}A_{\alpha} - \frac{1}{4} f_{2 X_{2} X_{2}} A^{\alpha} A^{\beta} \nabla_{\beta}\nabla_{\alpha}\varphi \nn \\
 &- f_{2 X_{1} X_{2}} A^{\alpha} \nabla_{\beta}\nabla_{\alpha}\varphi \nabla^{\beta}\varphi - f_{2 X_{1} X_{1}} \nabla^{\alpha}\varphi \nabla_{\beta}\nabla_{\alpha}\varphi \nabla^{\beta}\varphi - f_{2 X_{1} X_{3}} A^{\alpha} \nabla_{\beta}\varphi \nabla^{\beta}A_{\alpha} \nn \\
 &- \frac{1}{2} f_{2 X_{1} X_{2}} \nabla_{\alpha}\varphi \nabla_{\beta}\varphi \nabla^{\beta}A^{\alpha},  \\
\mc{J}_3 &= 2 f_{3 \varphi} \nabla_{\alpha}A^{\alpha} + 2 \tilde{f}_{3 \varphi} A^{\alpha} A^{\beta} \nabla_{\beta}A_{\alpha},
\end{align}
\begin{align}
\mc{K}_3 &= -2 G_{3 \varphi } \nabla_{\alpha}\nabla^{\alpha}\varphi - G_{3 \varphi \varphi } \nabla_{\alpha}\varphi \nabla^{\alpha}\varphi - G_{3 X_{1}} \nabla_{\alpha}\nabla_{\beta}\nabla^{\beta}\varphi \nabla^{\alpha}\varphi - G_{3 X_{1}} \nabla_{\alpha}\nabla^{\alpha}\varphi \nabla_{\beta}\nabla^{\beta}\varphi \nn \\
&- G_{3 \varphi X_{1}} \nabla_{\alpha}\varphi \nabla^{\alpha}\varphi \nabla_{\beta}\nabla^{\beta}\varphi + G_{3 X_{1}} \nabla^{\alpha}\varphi \nabla_{\beta}\nabla^{\beta}\nabla_{\alpha}\varphi + 2 G_{3 \varphi X_{1}} \nabla^{\alpha}\varphi \nabla_{\beta}\nabla_{\alpha}\varphi \nabla^{\beta}\varphi \nn \\
 &+ G_{3 X_{1}} \nabla_{\beta}\nabla_{\alpha}\varphi \nabla^{\beta}\nabla^{\alpha}\varphi + G_{3 X_{1} X_{1}} \nabla^{\alpha}\varphi \nabla_{\beta}\nabla_{\alpha}\varphi \nabla^{\beta}\varphi \nabla_{\gamma}\nabla^{\gamma}\varphi \nn \\
 &- G_{3 X_{1} X_{1}} \nabla^{\alpha}\varphi \nabla^{\beta}\varphi \nabla_{\gamma}\nabla_{\beta}\varphi \nabla^{\gamma}\nabla_{\alpha}\varphi,  \\
\mc{K}_4 &= G_{4 \varphi} R.
\end{align}

\subsection{Vector field equation of motion}
\label{App: Gral Vector Field}

Coefficients on right-hand side of Eqs.\eqref{Gra Scalar Vec Eqs}:
\begin{align}
\mc{A}^\mu_{\ (2)} &= - f_{2 X_{3}} A^{\mu} - \frac{1}{2} f_{2 X_{2}} \nabla^{\mu}\varphi, \\
\mc{A}^\mu_{\ (3)} &= -2 \tilde{f}_{3 \varphi } A^{\alpha} A^{\mu} \nabla_{\alpha}\varphi -2 f_{3 X_{3}} A^{\mu} \nabla_{\alpha}A^{\alpha} -2 \tilde{f}_{3} A^{\mu} \nabla_{\alpha}A^{\alpha} -2 f_{3 \varphi } \nabla^{\mu}\varphi \nn \\
 &+ 2 f_{3 X_{3}} A^{\alpha} \nabla^{\mu}A_{\alpha} + 2 \tilde{f}_{3} A^{\alpha} \nabla^{\mu}A_{\alpha}.
\end{align}

\section{Background equations of motion}

\subsection{``Time-Time'' equation}
\label{App: time-time Back Eq}
Coefficients on the left-hand side of Eq. \eqref{Gral Back Einstein Eqs}:
\begin{align}
\mc{G}_{0 0}^{(2)} &= \frac{1}{2} a^2 f_{2} - \frac{1}{2} \vp'^2 f_{2 X_{1}} - \frac{1}{2} A_{0} \vp'  f_{2 X_{2}} - \frac{1}{2} A_{0}^2 f_{2 X_{3}},  \\
\mc{G}_{0 0}^{(3)} &= - A_{0} \vp' f_{3 \varphi} + \frac{A_{0}^3 \vp' \tilde{f}_{3 \varphi}}{a^2} + \frac{3 A_{0}^3 f_{3 X_{3}} \mathcal{H}}{a^2} + \frac{3 A_{0}^3 \tilde{f}_{3} \mathcal{H}}{a^2},  \\
\mathscr{H}_{00}^{(3)} &= \frac{1}{2}  \vp'^2 G_{3  \varphi } - \frac{3  \vp'^3 G_{3 X_{1}} \mathcal{H}}{2 a^2},  \\
\mathscr{H}_{00}^{(4)} &= 3 \mathcal{H} (\vp'  G_{4  \varphi } + G_{4} \mathcal{H}).
\end{align}

\subsection{``Space-Space'' equation}
\label{App: trace space-space Back Eq}

Coefficients on the right-hand side of Eq. \eqref{Gral Back Einstein Eqs}:
\begin{align}
\mc{G}_{11}^{(2)} &= - \frac{1}{2} a^2 f_{2}, \\
\mc{G}_{11}^{(3)} &= - \frac{A_{0}^2 A_0' f_{3 X_{3}}}{a^2} - A_{0} \vp' f_{3 \varphi} - \frac{A_{0}^2 A_0' \tilde{f}_{3}}{a^2} + \frac{A_{0}^3 f_{3 X_{3}} \mathcal{H}}{a^2} + \frac{A_{0}^3 \tilde{f}_{3} \mathcal{H}}{a^2}, \\
\mathscr{H}_{11}^{(3)} &= \frac{\vp'' \vp'^2 G_{3 X_{1}}}{2 a^2} + \frac{1}{2} \vp'^2 G_{3 \varphi } - \frac{\vp'^3 G_{3 X_{1}} \mathcal{H}}{2 a^2}, \\
\mathscr{H}_{11}^{(4)} &= - \vp'^2 G_{4 \varphi \varphi} - G_{4 \varphi} (\vp'' + \vp' \mathcal{H}) - G_{4} (\mathcal{H}^2 + 2 \mathcal{H}'). 
\end{align}

\subsection{Background equation of motion for the scalar and vector fields}
\label{App: Scalar Back Eq}

Coefficients on the left-hand side of Eq. \eqref{Back Scalar Vector Eqs}:
\begin{align}
\mc{\bar{J}}_2 &= - \frac{\vp''  f_{2 X_{1}}}{a^2} - \frac{\vp''  \vp'^2 f_{2 X_{1} X_{1}}}{a^4} - \frac{A_{0} \vp''  \vp'  f_{2 X_{1} X_{2}}}{a^4} - \frac{A_0' \vp'^2 f_{2 X_{1} X_{2}}}{2 a^4} - \frac{A_{0} A_0' \vp'  f_{2 X_{1} X_{3}}}{a^4} \nn \\
 &- \frac{A_0' f_{2 X_{2}}}{2 a^2} - \frac{A_{0}^2 \vp''  f_{2 X_{2} X_{2}}}{4 a^4} - \frac{A_{0} A_0' \vp'  f_{2 X_{2} X_{2}}}{4 a^4} - \frac{A_{0}^2 A_0' f_{2 X_{2} X_{3}}}{2 a^4} + f_{2  \varphi } - \frac{\vp'^2 f_{2 \varphi X_{1}}}{a^2} \nn \\
 &- \frac{A_{0} \vp'  f_{2 \varphi X_{2}}}{2 a^2} - \frac{2 \vp'  f_{2 X_{1}} \mathcal{H}}{a^2} + \frac{\vp'^3 f_{2 X_{1} X_{1}} \mathcal{H}}{a^4} + \frac{3 A_{0} \vp'^2 f_{2 X_{1} X_{2}} \mathcal{H}}{2 a^4} + \frac{A_{0}^2 \vp'  f_{2 X_{1} X_{3}} \mathcal{H}}{a^4} \nn \\
 &- \frac{A_{0} f_{2 X_{2}} \mathcal{H}}{a^2} + \frac{A_{0}^2 \vp'  f_{2 X_{2} X_{2}} \mathcal{H}}{2 a^4} + \frac{A_{0}^3 f_{2 X_{2} X_{3}} \mathcal{H}}{2 a^4},  \\
\mc{\bar{J}}_3 &= - \frac{2 A_0' f_{3 \varphi}}{a^2} + \frac{2 A_{0}^2 A_0' \tilde{f}_{3 \varphi}}{a^4} - \frac{4 A_{0} f_{3 \varphi} \mathcal{H}}{a^2} - \frac{2 A_{0}^3 \tilde{f}_{3 \varphi} \mathcal{H}}{a^4},  \\
\bar{\mc{K}}_3 &= \frac{2 \vp'' G_{3 \varphi}}{a^2} + \frac{\vp'' \vp'^2 G_{3 \varphi X_{1}}}{a^4} + \frac{\vp'^2 G_{3 \varphi \varphi}}{a^2} - \frac{6 \vp'' \vp' G_{3 X_{1}} \mathcal{H}}{a^4} - \frac{3 \vp'' \vp'^3 G_{3 X_{1} X_{1}} \mathcal{H}}{a^6} \nn \\
 &+ \frac{4 \vp' G_{3 \varphi} \mathcal{H}}{a^2} - \frac{4 \vp'^3 G_{3 \varphi X_{1}} \mathcal{H}}{a^4} + \frac{3 \vp'^4 G_{3 X_{1} X_{1}} \mathcal{H}^2}{a^6} - \frac{3 \vp'^2 G_{3 X_{1}} \mathcal{H}'}{a^4}, \\
\bar{\mc{K}}_4 &= \frac{6 G_{4  \varphi} (\mathcal{H}^2 + \mathcal{H}')}{a^2}.
\end{align}

Coefficients on the right-hand side of Eq. \eqref{Back Scalar Vector Eqs}:
\begin{align}
\mc{\bar{A}}_2 &= \frac{\vp'  f_{2 X_{2}}}{2 a^2} + \frac{A_{0} f_{2 X_{3}}}{a^2},  \\
\mc{\bar{A}}_3 &= \frac{2 \vp'  f_{3  \varphi }}{a^2} - \frac{2 A_{0}^2 \vp'  \tilde{f}_{3  \varphi }}{a^4} - \frac{6 A_{0}^2 f_{3 X_{3}} \mathcal{H}}{a^4} - \frac{6 A_{0}^2 \tilde{f}_{3} \mathcal{H}}{a^4}.
\end{align}

\section{Linear perturbations: coefficients}
\label{App: Perturbations Coefficients}

\subsection{``Time-Time'' equation}

Coefficients in Eq. \eqref{Pert time time Eq}:
\begin{align}
A_1 &= \frac{6 A_{0}^3 f_{3 X_{3}}}{a^3} + \frac{6 A_{0}^3 \tilde{f}_{3}}{a^3} - \frac{3 \vp'^3 G_{3 X_{1}}}{a^3} + \frac{6 \vp'  G_{4  \varphi }}{a} + \frac{12 G_{4} \mathcal{H}}{a}, \\
A_2 &= - \frac{\vp'  f_{2 X_{1}}}{a} - \frac{\vp'^3 f_{2 X_{1} X_{1}}}{a^3} - \frac{3 A_{0} \vp'^2 f_{2 X_{1} X_{2}}}{2 a^3} - \frac{A_{0}^2 \vp'  f_{2 X_{1} X_{3}}}{a^3} - \frac{A_{0} f_{2 X_{2}}}{2 a} \nn \\
 &- \frac{A_{0}^2 \vp'  f_{2 X_{2} X_{2}}}{2 a^3} - \frac{A_{0}^3 f_{2 X_{2} X_{3}}}{2 a^3} - \frac{2 A_{0} f_{3  \varphi}}{a} + \frac{2 A_{0}^3 \tilde{f}_{3  \varphi}}{a^3} + \frac{2 \vp'  G_{3  \varphi}}{a} + \frac{\vp'^3 G_{3 \varphi X_{1}}}{a^3} \nn \\
 &- \frac{9 \vp'^2 G_{3 X_{1}} \mathcal{H}}{a^3} - \frac{3 \vp'^4 G_{3 X_{1} X_{1}} \mathcal{H}}{a^5} + \frac{6 G_{4  \varphi} \mathcal{H}}{a}, \\
A_3 &= 4 G_{4}, 
\end{align}
\begin{align}
A_4 &= \frac{\vp'^2 f_{2 X_{1}}}{a^2} + \frac{\vp'^4 f_{2 X_{1} X_{1}}}{a^4} + \frac{2 A_{0} \vp'^3 f_{2 X_{1} X_{2}}}{a^4} + \frac{2 A_{0}^2 \vp'^2 f_{2 X_{1} X_{3}}}{a^4} + \frac{A_{0} \vp'  f_{2 X_{2}}}{a^2} \nn \\
 &+ \frac{A_{0}^2 \vp'^2 f_{2 X_{2} X_{2}}}{a^4} + \frac{2 A_{0}^3 \vp'  f_{2 X_{2} X_{3}}}{a^4} + \frac{A_{0}^2 f_{2 X_{3}}}{a^2} + \frac{A_{0}^4 f_{2 X_{3} X_{3}}}{a^4} + \frac{4 A_{0} \vp'  f_{3  \varphi }}{a^2} \nn \\
 &+ \frac{2 A_{0}^3 \vp'  f_{3 \varphi X_{3}}}{a^4} - \frac{8 A_{0}^3 \vp'  \tilde{f}_{3  \varphi }}{a^4} - \frac{2 \vp'^2 G_{3  \varphi }}{a^2} - \frac{\vp'^4 G_{3 \varphi X_{1}}}{a^4} - \frac{2 A_{0}^5 \vp'  \tilde{f}_{3 \varphi X_{3}}}{a^6} \nn \\
 &- \frac{24 A_{0}^3 f_{3 X_{3}} \mathcal{H} }{a^4} - \frac{6 A_{0}^5 f_{3 X_{3} X_{3}} \mathcal{H} }{a^6} - \frac{24 A_{0}^3 \tilde{f}_{3} \mathcal{H} }{a^4} + \frac{12 \vp'^3 G_{3 X_{1}} \mathcal{H} }{a^4} + \frac{3 \vp'^5 G_{3 X_{1} X_{1}} \mathcal{H} }{a^6} \nn \\
 &- \frac{6 A_{0}^5 \tilde{f}_{3 X_{3}} \mathcal{H} }{a^6} - \frac{12 \vp'  G_{4  \varphi } \mathcal{H} }{a^2} - \frac{12 G_{4} \mathcal{H}^2}{a^2}, \\
A_5 &= - \frac{\vp'^2 G_{3 X_{1}}}{a^2} + 2 G_{4  \varphi}, \\
A_6 &= - \frac{\vp'^3 f_{2 X_{1} X_{2}}}{2 a^3} - \frac{A_{0} \vp'^2 f_{2 X_{1} X_{3}}}{a^3} - \frac{\vp'  f_{2 X_{2}}}{2 a} - \frac{A_{0} \vp'^2 f_{2 X_{2} X_{2}}}{2 a^3} - \frac{3 A_{0}^2 \vp'  f_{2 X_{2} X_{3}}}{2 a^3} \nn \\
 &- \frac{A_{0} f_{2 X_{3}}}{a} - \frac{A_{0}^3 f_{2 X_{3} X_{3}}}{a^3} - \frac{2 \vp'  f_{3  \varphi}}{a} - \frac{2 A_{0}^2 \vp'  f_{3 \varphi X_{3}}}{a^3} + \frac{6 A_{0}^2 \vp'  \tilde{f}_{3  \varphi}}{a^3} \nn \\
 &+ \frac{2 A_{0}^4 \vp'  \tilde{f}_{3 \varphi X_{3}}}{a^5} + \frac{18 A_{0}^2 f_{3 X_{3}} \mathcal{H}}{a^3} + \frac{6 A_{0}^4 f_{3 X_{3} X_{3}} \mathcal{H}}{a^5} + \frac{18 A_{0}^2 \tilde{f}_{3} \mathcal{H}}{a^3} + \frac{6 A_{0}^4 \tilde{f}_{3 X_{3}}\mathcal{H}}{a^5}, \\
A_7 &= \frac{2 A_{0}^2 f_{3 X_{3}}}{a^2} + \frac{2 A_{0}^2 \tilde{f}_{3}}{a^2}, \\
\mu_\vp &= - f_{2  \varphi} + \frac{\vp'^2 f_{2 \varphi X_{1}}}{a^2} + \frac{A_{0} \vp'  f_{2 \varphi X_{2}}}{a^2} + \frac{A_{0}^2 f_{2 \varphi X_{3}}}{a^2} + \frac{2 A_{0} \vp'  f_{3  \varphi \varphi}}{a^2} \nn \\
 &- \frac{2 A_{0}^3 \vp'  \tilde{f}_{3  \varphi \varphi}}{a^4} - \frac{\vp'^2 G_{3  \varphi \varphi}}{a^2} - \frac{6 A_{0}^3 f_{3 \varphi X_{3}} \mathcal{H}}{a^4} - \frac{6 A_{0}^3 \tilde{f}_{3  \varphi} \mathcal{H}}{a^4} + \frac{3 \vp'^3 G_{3 \varphi X_{1}} \mathcal{H}}{a^4} \nn \\
 &- \frac{6 \vp'  G_{4  \varphi \varphi} \mathcal{H}}{a^2} - \frac{6 G_{4  \varphi} \mathcal{H}^2}{a^2}.
\end{align}

\subsection{Longitudinal ``Time-Space'' equation}

Coefficients in Eq. \eqref{Pert time space Eq}:
\begin{align}
C_1 &= 4 G_4, \\
C_2 &= - \frac{\vp'^2 G_{3 X_{1}}}{a^2} + 2 G_{4  \varphi}, \\
C_3 &= - \frac{2 A_{0}^3 f_{3 X_{3}}}{a^3} - \frac{2 A_{0}^3 \tilde{f}_{3}}{a^3} + \frac{\vp'^3 G_{3 X_{1}}}{a^3} - \frac{2 \vp'  G_{4  \varphi }}{a} - \frac{4 G_{4} \mathcal{H}}{a}, \\
C_4 &= \frac{\vp'  f_{2 X_{1}}}{a} + \frac{A_{0} f_{2 X_{2}}}{2 a} + \frac{2 A_{0} f_{3  \varphi}}{a} - \frac{2 \vp'  G_{3  \varphi}}{a} + \frac{2 \vp'  G_{4  \varphi \varphi}}{a} + \frac{3 \vp'^2 G_{3 X_{1}} \mathcal{H}}{a^3} - \frac{2 G_{4  \varphi} \mathcal{H}}{a}, \\
C_5 &= \frac{2 A_{0}^2 f_{3 X_{3}}}{a^2} + \frac{2 A_{0}^2 \tilde{f}_{3}}{a^2}, \\
C_6 &= \frac{\vp'  f_{2 X_{2}}}{2 a} + \frac{A_{0} f_{2 X_{3}}}{a} + \frac{2 \vp'  f_{3  \varphi}}{a} - \frac{2 A_{0}^2 \vp'  \tilde{f}_{3  \varphi}}{a^3} - \frac{6 A_{0}^2 f_{3 X_{3}} \mathcal{H}}{a^3} - \frac{6 A_{0}^2 \tilde{f}_{3} \mathcal{H}}{a^3}.
\end{align}

\subsection{Trace ``Space-Space'' equation}
Coefficients in Eq. \eqref{Pert trace Eq}:
\begin{align}
B_1 &= 12 G_4, \\
B_2 &= - \frac{3 \vp'^2 G_{3 X_{1}}}{a^2} + 6 G_{4  \varphi}, \\
B_3 &= \frac{12 \vp'  G_{4  \varphi}}{a} + \frac{24 G_{4} \mathcal{H}}{a}, \\
B_4 &= \frac{3 \vp'  f_{2 X_{1}}}{a} + \frac{3 A_{0} f_{2 X_{2}}}{2 a} + \frac{6 A_{0} f_{3  \varphi }}{a} - \frac{6 \vp''  \vp'  G_{3 X_{1}}}{a^3} - \frac{3 \vp''  \vp'^3 G_{3 X_{1} X_{1}}}{a^5} - \frac{6 \vp'  G_{3  \varphi }}{a} \nn \\
 &- \frac{3 \vp'^3 G_{3 \varphi X_{1}}}{a^3} + \frac{12 \vp'  G_{4  \varphi \varphi }}{a} + \frac{9 \vp'^2 G_{3 X_{1}} \mathcal{H}}{a^3} + \frac{3 \vp'^4 G_{3 X_{1} X_{1}} \mathcal{H}}{a^5} + \frac{6 G_{4  \varphi } \mathcal{H}}{a}, \\
B_5 &= - \frac{6 A_{0}^3 f_{3 X_{3}}}{a^3} - \frac{6 A_{0}^3 \tilde{f}_{3}}{a^3} + \frac{3 \vp'^3 G_{3 X_{1}}}{a^3} - \frac{6 \vp'  G_{4  \varphi }}{a} - \frac{12 G_{4} \mathcal{H}}{a}, \\
B_6 &= 4 G_4, \\
B_7 &= 4 G_{4 \vp}, \\
B_8 &= 4 G_4, \\
B_9 &= - \frac{3 \vp'^2 f_{2 X_{1}}}{a^2} - \frac{3 A_{0} \vp'  f_{2 X_{2}}}{a^2} - \frac{3 A_{0}^2 f_{2 X_{3}}}{a^2} - \frac{24 A_{0}^2 A'_0 f_{3 X_{3}}}{a^4} - \frac{6 A_{0}^4 A'_0 f_{3 X_{3} X_{3}}}{a^6} \nn \\
 &- \frac{12 A_{0} \vp'  f_{3  \varphi }}{a^2} - \frac{6 A_{0}^3 \vp'  f_{3 \varphi X_{3}}}{a^4} - \frac{24 A_{0}^2 A'_0 \tilde{f}_{3}}{a^4} + \frac{12 \vp''  \vp'^2 G_{3 X_{1}}}{a^4} + \frac{3 \vp''  \vp'^4 G_{3 X_{1} X_{1}}}{a^6} \nn \\
 &- \frac{6 A_{0}^4 A'_0 \tilde{f}_{3 X_{3}}}{a^6} + \frac{6 \vp'^2 G_{3  \varphi }}{a^2} + \frac{3 \vp'^4 G_{3 \varphi X_{1}}}{a^4} - \frac{12 \vp''  G_{4  \varphi }}{a^2} - \frac{12 \vp'^2 G_{4  \varphi \varphi }}{a^2} \nn \\
 &+ \frac{24 A_{0}^3 f_{3 X_{3}} \mathcal{H}}{a^4} + \frac{6 A_{0}^5 f_{3 X_{3} X_{3}} \mathcal{H}}{a^6} + \frac{24 A_{0}^3 \tilde{f}_{3} \mathcal{H}}{a^4} - \frac{12 \vp'^3 G_{3 X_{1}} \mathcal{H}}{a^4} - \frac{3 \vp'^5 G_{3 X_{1} X_{1}} \mathcal{H}}{a^6} \nn \\
 &+ \frac{6 A_{0}^5 \tilde{f}_{3 X_{3}} \mathcal{H}}{a^6} - \frac{12 \vp'  G_{4  \varphi } \mathcal{H}}{a^2} - \frac{12 G_{4} \mathcal{H}^2}{a^2} - \frac{24 G_{4} \mathcal{H}'}{a^2}, \\
B_{10} &= \frac{6 A_{0}^2 f_{3 X_{3}}}{a^2} + \frac{6 A_{0}^2 \tilde{f}_{3}}{a^2},
\end{align}
 
 \begin{align}
B_{11} &= \frac{3 \vp'  f_{2 X_{2}}}{2 a} + \frac{3 A_{0} f_{2 X_{3}}}{a} + \frac{12 A_{0} A'_0 f_{3 X_{3}}}{a^3} + \frac{6 A_{0}^3 A'_0 f_{3 X_{3} X_{3}}}{a^5} + \frac{6 \vp'  f_{3  \varphi }}{a} \nn \\
 &+ \frac{6 A_{0}^2 \vp'  f_{3 \varphi X_{3}}}{a^3} + \frac{12 A_{0} A'_0 \tilde{f}_{3}}{a^3} + \frac{6 A_{0}^3 A'_0 \tilde{f}_{3 X_{3}}}{a^5} - \frac{18 A_{0}^2 f_{3 X_{3}} \mathcal{H}}{a^3} - \frac{6 A_{0}^4 f_{3 X_{3} X_{3}} \mathcal{H}}{a^5} \nn \\
 &- \frac{18 A_{0}^2 \tilde{f}_{3} \mathcal{H}}{a^3} - \frac{6 A_{0}^4 \tilde{f}_{3 X_{3}} \mathcal{H}}{a^5}, \\
\nn \\
\nu_\vp &= f_{2  \varphi } + \frac{2 A_{0}^2 A'_0 f_{3 \varphi X_{3}}}{a^4} + \frac{2 A_{0} \vp'  f_{3  \varphi \varphi }}{a^2} + \frac{2 A_{0}^2 A'_0 \tilde{f}_{3  \varphi }}{a^4} - \frac{\vp''  \vp'^2 G_{3 \varphi X_{1}}}{a^4} - \frac{\vp'^2 G_{3  \varphi \varphi }}{a^2} \nn \\
 &+ \frac{2 \vp''  G_{4  \varphi \varphi }}{a^2} + \frac{2 \vp'^2 G_{4  \varphi \varphi \varphi }}{a^2} - \frac{2 A_{0}^3 f_{3 \varphi X_{3}} \mathcal{H}}{a^4} - \frac{2 A_{0}^3 \tilde{f}_{3  \varphi } \mathcal{H}}{a^4} + \frac{\vp'^3 G_{3 \varphi X_{1}} \mathcal{H}}{a^4} \nn \\
 &+ \frac{2 \vp'  G_{4  \varphi \varphi } \mathcal{H}}{a^2} + \frac{2 G_{4  \varphi } \mathcal{H}^2}{a^2} + \frac{4 G_{4  \varphi } \mathcal{H}'}{a^2}.
\end{align}
We also found the  coefficient
\begin{align}
B_0 &= 6 f_{2} + \frac{12 A_{0}^2 A'_0 f_{3 X_{3}}}{a^4} + \frac{12 A_{0} \vp'  f_{3  \varphi }}{a^2} + \frac{12 A_{0}^2 A'_0 \tilde{f}_{3}}{a^4} - \frac{6 \vp''  \vp'^2 G_{3 X_{1}}}{a^4} - \frac{6 \vp'^2 G_{3  \varphi }}{a^2} \nn \\
 &+ \frac{12 \vp''  G_{4  \varphi }}{a^2} + \frac{12 \vp'^2 G_{4  \varphi \varphi }}{a^2} - \frac{12 A_{0}^3 f_{3 X_{3}} \mathcal{H}}{a^4} - \frac{12 A_{0}^3 \tilde{f}_{3} \mathcal{H}}{a^4} + \frac{6 \vp'^3 G_{3 X_{1}} \mathcal{H}}{a^4} \nn \\
 &+ \frac{12 \vp'  G_{4  \varphi } \mathcal{H}}{a^2} + \frac{12 G_{4} \mathcal{H}^2}{a^2} + \frac{24 G_{4} \mathcal{H}'}{a^2},\end{align}
accompanying $\Phi$ in Eq. \eqref{Pert trace Eq}. However, it vanishes when using the equation in the right-hand side of Eq. \eqref{Gral Back Einstein Eqs} to eliminate $G_4$.

\subsection{Scalar field equation of motion}
Coefficients in Eq. \eqref{Pert Scalar Eq}
\begin{align}
D_1 &= - \frac{3 \vp'^2 G_{3 X_{1}}}{a^2} + 6 G_{4  \varphi }, \\
D_2 &= - f_{2 X_{1}} - \frac{\vp'^2 f_{2 X_{1} X_{1}}}{a^2} - \frac{A_{0} \vp'  f_{2 X_{1} X_{2}}}{a^2} - \frac{A_{0}^2 f_{2 X_{2} X_{2}}}{4 a^2} + 2 G_{3  \varphi } + \frac{\vp'^2 G_{3 \varphi X_{1}}}{a^2} \nn \\
 &- \frac{6 \vp'  G_{3 X_{1}} \mathcal{H}}{a^2} - \frac{3 \vp'^3 G_{3 X_{1} X_{1}} \mathcal{H}}{a^4}, \\
D_3 &= - \frac{3 \vp'  f_{2 X_{1}}}{a} - \frac{3 A_{0} f_{2 X_{2}}}{2 a} - \frac{6 A_{0} f_{3  \varphi }}{a} - \frac{6 \vp''  \vp'  G_{3 X_{1}}}{a^3} - \frac{3 \vp''  \vp'^3 G_{3 X_{1} X_{1}}}{a^5} + \frac{6 \vp'  G_{3  \varphi }}{a} \nn \\
 &- \frac{3 \vp'^3 G_{3 \varphi X_{1}}}{a^3} - \frac{9 \vp'^2 G_{3 X_{1}} \mathcal{H} }{a^3} + \frac{3 \vp'^4 G_{3 X_{1} X_{1}} \mathcal{H} }{a^5} + \frac{18 G_{4  \varphi } \mathcal{H} }{a},\\
D_4 &= - \frac{3 \vp''  \vp'  f_{2 X_{1} X_{1}}}{a^3} - \frac{\vp''  \vp'^3 f_{2 X_{1} X_{1} X_{1}}}{a^5} - \frac{3 A_{0} \vp''  \vp'^2 f_{2 X_{1} X_{1} X_{2}}}{2 a^5} - \frac{A'_0 \vp'^3 f_{2 X_{1} X_{1} X_{2}}}{2 a^5}  \nn \\
 &- \frac{A_{0}A'_0 \vp'^2 f_{2 X_{1} X_{1} X_{3}}}{a^5} - \frac{3 A_{0} \vp''  f_{2 X_{1} X_{2}}}{2 a^3} - \frac{3A'_0 \vp'  f_{2 X_{1} X_{2}}}{2 a^3} - \frac{3 A_{0}^2 \vp''  \vp'  f_{2 X_{1} X_{2} X_{2}}}{4 a^5} \nn \\
 &- \frac{A_{0}A'_0 \vp'^2 f_{2 X_{1} X_{2} X_{2}}}{2 a^5} - \frac{A_{0}^2A'_0 \vp'  f_{2 X_{1} X_{2} X_{3}}}{a^5} - \frac{A_{0}A'_0 f_{2 X_{1} X_{3}}}{a^3} - \frac{A_{0}A'_0 f_{2 X_{2} X_{2}}}{2 a^3} \nn \\
 &- \frac{A_{0}^3 \vp''  f_{2 X_{2} X_{2} X_{2}}}{8 a^5} - \frac{A_{0}^2A'_0 \vp'  f_{2 X_{2} X_{2} X_{2}}}{8 a^5} - \frac{A_{0}^3A'_0 f_{2 X_{2} X_{2} X_{3}}}{4 a^5} - \frac{\vp'  f_{2 \varphi X_{1}}}{a} - \frac{\vp'^3 f_{2 \varphi X_{1} X_{1}}}{a^3} \nn \\
 &- \frac{A_{0} \vp'^2 f_{2 \varphi X_{1} X_{2}}}{a^3} - \frac{A_{0}^2 \vp'  f_{2 \varphi X_{2} X_{2}}}{4 a^3} + \frac{4 \vp''  \vp'  G_{3 \varphi X_{1}}}{a^3} + \frac{\vp''  \vp'^3 G_{3 \varphi X_{1} X_{1}}}{a^5} + \frac{2 \vp'  G_{3  \varphi \varphi }}{a} \nn \\
 &+ \frac{\vp'^3 G_{3 \varphi \varphi X_{1}}}{a^3} - \frac{2 f_{2 X_{1}} \mathcal{H}}{a} + \frac{\vp'^2 f_{2 X_{1} X_{1}} \mathcal{H}}{a^3} + \frac{\vp'^4 f_{2 X_{1} X_{1} X_{1}} \mathcal{H}}{a^5} + \frac{2 A_{0} \vp'^3 f_{2 X_{1} X_{1} X_{2}} \mathcal{H}}{a^5} \nn \\
 &+ \frac{A_{0}^2 \vp'^2 f_{2 X_{1} X_{1} X_{3}} \mathcal{H}}{a^5} + \frac{A_{0} \vp'  f_{2 X_{1} X_{2}} \mathcal{H}}{a^3} + \frac{5 A_{0}^2 \vp'^2 f_{2 X_{1} X_{2} X_{2}} \mathcal{H}}{4 a^5} + \frac{A_{0}^3 \vp'  f_{2 X_{1} X_{2} X_{3}} \mathcal{H}}{a^5} \nn \\
 &+ \frac{A_{0}^3 \vp'  f_{2 X_{2} X_{2} X_{2}} \mathcal{H}}{4 a^5} + \frac{A_{0}^4 f_{2 X_{2} X_{2} X_{3}} \mathcal{H}}{4 a^5} - \frac{6 \vp''  G_{3 X_{1}} \mathcal{H}}{a^3} - \frac{15 \vp''  \vp'^2 G_{3 X_{1} X_{1}} \mathcal{H}}{a^5} \nn \\
 &- \frac{3 \vp''  \vp'^4 G_{3 X_{1} X_{1} X_{1}} \mathcal{H}}{a^7} + \frac{4 G_{3  \varphi } \mathcal{H}}{a} - \frac{8 \vp'^2 G_{3 \varphi X_{1}} \mathcal{H}}{a^3} - \frac{4 \vp'^4 G_{3 \varphi X_{1} X_{1}} \mathcal{H}}{a^5} + \frac{12 \vp'^3 G_{3 X_{1} X_{1}} \mathcal{H}^2}{a^5} \nn \\
 &+ \frac{3 \vp'^5 G_{3 X_{1} X_{1} X_{1}} \mathcal{H}^2}{a^7} - \frac{6 \vp'  G_{3 X_{1}} \mathcal{H}'}{a^3} - \frac{3 \vp'^3 G_{3 X_{1} X_{1}} \mathcal{H}'}{a^5} + \frac{A_{0}^2 f_{2 X_{1} X_{3}} \mathcal{H}}{a^3}, 
\end{align}

\begin{align}
D_5 &= \frac{\vp'  f_{2 X_{1}}}{a} + \frac{\vp'^3 f_{2 X_{1} X_{1}}}{a^3} + \frac{3 A_{0} \vp'^2 f_{2 X_{1} X_{2}}}{2 a^3} + \frac{A_{0}^2 \vp'  f_{2 X_{1} X_{3}}}{a^3} + \frac{A_{0} f_{2 X_{2}}}{2 a} + \frac{A_{0}^2 \vp'  f_{2 X_{2} X_{2}}}{2 a^3} \nn \\
 &+ \frac{A_{0}^3 f_{2 X_{2} X_{3}}}{2 a^3} + \frac{2 A_{0} f_{3  \varphi }}{a} - \frac{2 A_{0}^3 \tilde{f}_{3  \varphi }}{a^3} - \frac{2 \vp'  G_{3  \varphi }}{a} - \frac{\vp'^3 G_{3 \varphi X_{1}}}{a^3} + \frac{9 \vp'^2 G_{3 X_{1}} \mathcal{H}}{a^3} \nn \\
 &+ \frac{3 \vp'^4 G_{3 X_{1} X_{1}} \mathcal{H}}{a^5} - \frac{6 G_{4  \varphi } \mathcal{H}}{a}, \\
D_7 &= 4 G_{4  \varphi }, \\
D_9 &= - f_{2 X_{1}} - \frac{2 \vp''  G_{3 X_{1}}}{a^2} - \frac{\vp''  \vp'^2 G_{3 X_{1} X_{1}}}{a^4} + 2 G_{3  \varphi } - \frac{\vp'^2 G_{3 \varphi X_{1}}}{a^2} - \frac{2 \vp'  G_{3 X_{1}} \mathcal{H} }{a^2} \nn \\
 &+ \frac{\vp'^3 G_{3 X_{1} X_{1}} \mathcal{H} }{a^4}, \\
D_{10} &= - \frac{\vp'^2 G_{3 X_{1}}}{a^2} + 2 G_{4  \varphi}, \\
D_{12} &= - \frac{\vp'^2 f_{2 X_{1} X_{2}}}{2 a^2} - \frac{A_{0} \vp'  f_{2 X_{1} X_{3}}}{a^2} - \frac{1}{2} f_{2 X_{2}} - \frac{A_{0} \vp'  f_{2 X_{2} X_{2}}}{4 a^2} - \frac{A_{0}^2 f_{2 X_{2} X_{3}}}{2 a^2} \nn \\
 &- 2 f_{3  \varphi } + \frac{2 A_{0}^2 \tilde{f}_{3  \varphi }}{a^2},\\
D_{13} &= - \frac{\vp''  \vp'^3 f_{2 X_{1} X_{1} X_{2}}}{2 a^5} - \frac{A_{0} \vp''  \vp'^2 f_{2 X_{1} X_{1} X_{3}}}{a^5} - \frac{3 \vp''  \vp'  f_{2 X_{1} X_{2}}}{2 a^3} - \frac{A_{0} \vp''  \vp'^2 f_{2 X_{1} X_{2} X_{2}}}{2 a^5} \nn \\
 &- \frac{A'_0 \vp'^3 f_{2 X_{1} X_{2} X_{2}}}{4 a^5} - \frac{A_{0}^2 \vp''  \vp'  f_{2 X_{1} X_{2} X_{3}}}{a^5} - \frac{A_{0} A'_0 \vp'^2 f_{2 X_{1} X_{2} X_{3}}}{a^5} - \frac{A_{0} \vp''  f_{2 X_{1} X_{3}}}{a^3} \nn \\
 &- \frac{A'_0 \vp'  f_{2 X_{1} X_{3}}}{a^3} - \frac{A_{0}^2 A'_0 \vp'  f_{2 X_{1} X_{3} X_{3}}}{a^5} - \frac{A_{0} \vp''  f_{2 X_{2} X_{2}}}{2 a^3} - \frac{A'_0 \vp'  f_{2 X_{2} X_{2}}}{2 a^3} - \frac{A_{0}^2 \vp''  \vp'  f_{2 X_{2} X_{2} X_{2}}}{8 a^5} \nn \\
 &- \frac{A_{0} A'_0 \vp'^2 f_{2 X_{2} X_{2} X_{2}}}{8 a^5} - \frac{A_{0}^3 \vp''  f_{2 X_{2} X_{2} X_{3}}}{4 a^5} - \frac{A_{0}^2 A'_0 \vp'  f_{2 X_{2} X_{2} X_{3}}}{2 a^5} - \frac{3 A_{0} A'_0 f_{2 X_{2} X_{3}}}{2 a^3} \nn \\
 &- \frac{A_{0}^3 A'_0 f_{2 X_{2} X_{3} X_{3}}}{2 a^5} - \frac{\vp'^3 f_{2 \varphi X_{1} X_{2}}}{2 a^3} - \frac{A_{0} \vp'^2 f_{2 \varphi X_{1} X_{3}}}{a^3} - \frac{A_{0} \vp'^2 f_{2 \varphi X_{2} X_{2}}}{4 a^3} - \frac{A_{0}^2 \vp'  f_{2 \varphi X_{2} X_{3}}}{2 a^3} \nn \\
 &+ \frac{A_{0} f_{2 \varphi X_{3}}}{a} - \frac{2 A_{0} A'_0 f_{3 \varphi X_{3}}}{a^3} + \frac{4 A_{0} A'_0 \tilde{f}_{3  \varphi }}{a^3} + \frac{2 A_{0}^3 A'_0 \tilde{f}_{3 \varphi X_{3}}}{a^5} + \frac{\vp'^4 f_{2 X_{1} X_{1} X_{2}} \mathcal{H} }{2 a^5} \nn \\
 &+ \frac{A_{0} \vp'^3 f_{2 X_{1} X_{1} X_{3}} \mathcal{H} }{a^5} + \frac{\vp'^2 f_{2 X_{1} X_{2}} \mathcal{H} }{2 a^3} + \frac{3 A_{0} \vp'^3 f_{2 X_{1} X_{2} X_{2}} \mathcal{H} }{4 a^5} + \frac{2 A_{0}^2 \vp'^2 f_{2 X_{1} X_{2} X_{3}} \mathcal{H} }{a^5} \nn \\
 &+ \frac{A_{0}^3 \vp'  f_{2 X_{1} X_{3} X_{3}} \mathcal{H} }{a^5} - \frac{f_{2 X_{2}} \mathcal{H} }{a} + \frac{A_{0} \vp'  f_{2 X_{2} X_{2}} \mathcal{H} }{2 a^3} + \frac{A_{0}^2 \vp'^2 f_{2 X_{2} X_{2} X_{2}} \mathcal{H} }{4 a^5} \nn \\
 &+ \frac{3 A_{0}^3 \vp'  f_{2 X_{2} X_{2} X_{3}} \mathcal{H} }{4 a^5} + \frac{A_{0}^2 f_{2 X_{2} X_{3}} \mathcal{H} }{2 a^3} + \frac{A_{0}^4 f_{2 X_{2} X_{3} X_{3}} \mathcal{H} }{2 a^5} - \frac{4 f_{3  \varphi } \mathcal{H} }{a} - \frac{4 A_{0}^2 f_{3 \varphi X_{3}} \mathcal{H} }{a^3} \nn \\
 &- \frac{6 A_{0}^2 \tilde{f}_{3  \varphi } \mathcal{H} }{a^3} - \frac{2 A_{0}^4 \tilde{f}_{3 \varphi X_{3}} \mathcal{H} }{a^5},  \\
D_{14} &= - \frac{1}{2} f_{2 X_{2}} -2 f_{3  \varphi },  
\end{align}
\begin{align}
D_{11} &= \frac{2 \vp''  f_{2 X_{1}}}{a^2} + \frac{5 \vp''  \vp'^2 f_{2 X_{1} X_{1}}}{a^4} + \frac{\vp''  \vp'^4 f_{2 X_{1} X_{1} X_{1}}}{a^6} + \frac{2 A_{0} \vp''  \vp'^3 f_{2 X_{1} X_{1} X_{2}}}{a^6} \nn \\
 &+ \frac{A'_0 \vp'^4 f_{2 X_{1} X_{1} X_{2}}}{2 a^6} + \frac{A_{0}^2 \vp''  \vp'^2 f_{2 X_{1} X_{1} X_{3}}}{a^6} + \frac{A_{0} A'_0 \vp'^3 f_{2 X_{1} X_{1} X_{3}}}{a^6} + \frac{5 A_{0} \vp''  \vp'  f_{2 X_{1} X_{2}}}{a^4} \nn \\
 &+ \frac{5 A'_0 \vp'^2 f_{2 X_{1} X_{2}}}{2 a^4} + \frac{5 A_{0}^2 \vp''  \vp'^2 f_{2 X_{1} X_{2} X_{2}}}{4 a^6} + \frac{3 A_{0} A'_0 \vp'^3 f_{2 X_{1} X_{2} X_{2}}}{4 a^6} + \frac{A_{0}^3 \vp''  \vp'  f_{2 X_{1} X_{2} X_{3}}}{a^6} \nn \\
 &+ \frac{2 A_{0}^2 A'_0 \vp'^2 f_{2 X_{1} X_{2} X_{3}}}{a^6} + \frac{A_{0}^2 \vp''  f_{2 X_{1} X_{3}}}{a^4} + \frac{4 A_{0} A'_0 \vp'  f_{2 X_{1} X_{3}}}{a^4} + \frac{A_{0}^3 A'_0 \vp'  f_{2 X_{1} X_{3} X_{3}}}{a^6} \nn \\
 &+ \frac{A'_0 f_{2 X_{2}}}{a^2} + \frac{A_{0}^2 \vp''  f_{2 X_{2} X_{2}}}{a^4} + \frac{3 A_{0} A'_0 \vp'  f_{2 X_{2} X_{2}}}{2 a^4} + \frac{A_{0}^3 \vp''  \vp'  f_{2 X_{2} X_{2} X_{2}}}{4 a^6} \nn \\
 &+ \frac{A_{0}^2 A'_0 \vp'^2 f_{2 X_{2} X_{2} X_{2}}}{4 a^6} + \frac{A_{0}^4 \vp''  f_{2 X_{2} X_{2} X_{3}}}{4 a^6} + \frac{3 A_{0}^3 A'_0 \vp'  f_{2 X_{2} X_{2} X_{3}}}{4 a^6} + \frac{5 A_{0}^2 A'_0 f_{2 X_{2} X_{3}}}{2 a^4} \nn \\
 &+ \frac{A_{0}^4 A'_0 f_{2 X_{2} X_{3} X_{3}}}{2 a^6} + \frac{\vp'^2 f_{2 \varphi X_{1}}}{a^2} + \frac{\vp'^4 f_{2 \varphi X_{1} X_{1}}}{a^4} + \frac{3 A_{0} \vp'^3 f_{2 \varphi X_{1} X_{2}}}{2 a^4} \nn \\
 &+ \frac{A_{0}^2 \vp'^2 f_{2 \varphi X_{1} X_{3}}}{a^4} + \frac{A_{0}^2 \vp'^2 f_{2 \varphi X_{2} X_{2}}}{2 a^4} + \frac{A_{0}^3 \vp'  f_{2 \varphi X_{2} X_{3}}}{2 a^4} - \frac{A_{0}^2 f_{2 \varphi X_{3}}}{a^2} \nn \\
 &+ \frac{4 A'_0 f_{3  \varphi }}{a^2} + \frac{2 A_{0}^2 A'_0 f_{3 \varphi X_{3}}}{a^4} - \frac{8 A_{0}^2 A'_0 \tilde{f}_{3  \varphi }}{a^4} - \frac{4 \vp''  G_{3  \varphi }}{a^2} \nn \\
 &- \frac{6 \vp''  \vp'^2 G_{3 \varphi X_{1}}}{a^4} - \frac{\vp''  \vp'^4 G_{3 \varphi X_{1} X_{1}}}{a^6} - \frac{2 A_{0}^4 A'_0 \tilde{f}_{3 \varphi X_{3}}}{a^6} - \frac{2 \vp'^2 G_{3  \varphi \varphi }}{a^2} \nn \\
 &- \frac{\vp'^4 G_{3 \varphi \varphi X_{1}}}{a^4} + \frac{4 \vp'  f_{2 X_{1}} \mathcal{H}}{a^2} - \frac{2 \vp'^3 f_{2 X_{1} X_{1}} \mathcal{H}}{a^4} - \frac{\vp'^5 f_{2 X_{1} X_{1} X_{1}} \mathcal{H}}{a^6} \nn \\
 &- \frac{5 A_{0} \vp'^4 f_{2 X_{1} X_{1} X_{2}} \mathcal{H}}{2 a^6} - \frac{2 A_{0}^2 \vp'^3 f_{2 X_{1} X_{1} X_{3}} \mathcal{H}}{a^6} - \frac{3 A_{0} \vp'^2 f_{2 X_{1} X_{2}} \mathcal{H}}{a^4} - \frac{2 A_{0}^2 \vp'^3 f_{2 X_{1} X_{2} X_{2}} \mathcal{H}}{a^6} \nn \\
 &- \frac{3 A_{0}^3 \vp'^2 f_{2 X_{1} X_{2} X_{3}} \mathcal{H}}{a^6} - \frac{2 A_{0}^2 \vp'  f_{2 X_{1} X_{3}} \mathcal{H}}{a^4} - \frac{A_{0}^4 \vp'  f_{2 X_{1} X_{3} X_{3}} \mathcal{H}}{a^6} + \frac{2 A_{0} f_{2 X_{2}} \mathcal{H}}{a^2} \nn \\
 &- \frac{A_{0}^2 \vp'  f_{2 X_{2} X_{2}} \mathcal{H}}{a^4} - \frac{A_{0}^3 \vp'^2 f_{2 X_{2} X_{2} X_{2}} \mathcal{H}}{2 a^6} - \frac{A_{0}^4 \vp'  f_{2 X_{2} X_{2} X_{3}} \mathcal{H}}{a^6} - \frac{A_{0}^3 f_{2 X_{2} X_{3}} \mathcal{H}}{a^4} \nn \\
 &- \frac{A_{0}^5 f_{2 X_{2} X_{3} X_{3}} \mathcal{H}}{2 a^6} + \frac{8 A_{0} f_{3  \varphi } \mathcal{H}}{a^2} + \frac{4 A_{0}^3 f_{3 \varphi X_{3}} \mathcal{H}}{a^4} + \frac{24 \vp''  \vp'  G_{3 X_{1}} \mathcal{H}}{a^4} \nn \\
 &+ \frac{24 \vp''  \vp'^3 G_{3 X_{1} X_{1}} \mathcal{H}}{a^6} + \frac{3 \vp''  \vp'^5 G_{3 X_{1} X_{1} X_{1}} \mathcal{H}}{a^8} + \frac{8 A_{0}^3 \tilde{f}_{3  \varphi } \mathcal{H}}{a^4} - \frac{8 \vp'  G_{3  \varphi } \mathcal{H}}{a^2} \nn \\
 &+ \frac{12 \vp'^3 G_{3 \varphi X_{1}} \mathcal{H}}{a^4} + \frac{4 \vp'^5 G_{3 \varphi X_{1} X_{1}} \mathcal{H}}{a^6} + \frac{2 A_{0}^5 \tilde{f}_{3 \varphi X_{3}} \mathcal{H}}{a^6} - \frac{18 \vp'^4 G_{3 X_{1} X_{1}} \mathcal{H}^2}{a^6} \nn \\
 &- \frac{3 \vp'^6 G_{3 X_{1} X_{1} X_{1}} \mathcal{H}^2}{a^8} - \frac{12 G_{4  \varphi } \mathcal{H}^2}{a^2} + \frac{12 \vp'^2 G_{3 X_{1}} \mathcal{H}'}{a^4} + \frac{3 \vp'^4 G_{3 X_{1} X_{1}} \mathcal{H}'}{a^6} - \frac{12 G_{4  \varphi } \mathcal{H}'}{a^2}, 
\end{align}

\begin{align}
m_\vp^2 &= \frac{\vp''  f_{2 \varphi X_{1}}}{a^2} + \frac{\vp''  \vp'^2 f_{2 \varphi X_{1} X_{1}}}{a^4} + \frac{A_{0} \vp''  \vp'  f_{2 \varphi X_{1} X_{2}}}{a^4} + \frac{A'_0 \vp'^2 f_{2 \varphi X_{1} X_{2}}}{2 a^4} + \frac{A_{0} A'_0 \vp'  f_{2 \varphi X_{1} X_{3}}}{a^4} \nn \\
 &+ \frac{A'_0 f_{2 \varphi X_{2}}}{2 a^2} + \frac{A_{0}^2 \vp''  f_{2 \varphi X_{2} X_{2}}}{4 a^4} + \frac{A_{0} A'_0 \vp'  f_{2 \varphi X_{2} X_{2}}}{4 a^4} + \frac{A_{0}^2 A'_0 f_{2 \varphi X_{2} X_{3}}}{2 a^4} - f_{2  \varphi \varphi } + \frac{\vp'^2 f_{2 \varphi \varphi X_{1}}}{a^2} \nn \\
 &+ \frac{A_{0} \vp'  f_{2 \varphi \varphi X_{2}}}{2 a^2} + \frac{2 A'_0 f_{3  \varphi \varphi }}{a^2} - \frac{2 A_{0}^2 A'_0 \tilde{f}_{3  \varphi \varphi }}{a^4} - \frac{2 \vp''  G_{3  \varphi \varphi }}{a^2} - \frac{\vp''  \vp'^2 G_{3 \varphi \varphi X_{1}}}{a^4} - \frac{\vp'^2 G_{3  \varphi \varphi \varphi }}{a^2} \nn \\
 &+ \frac{2 \vp'  f_{2 \varphi X_{1}} \mathcal{H}}{a^2} - \frac{\vp'^3 f_{2 \varphi X_{1} X_{1}} \mathcal{H}}{a^4} - \frac{3 A_{0} \vp'^2 f_{2 \varphi X_{1} X_{2}} \mathcal{H}}{2 a^4} - \frac{A_{0}^2 \vp'  f_{2 \varphi X_{1} X_{3}} \mathcal{H}}{a^4} + \frac{A_{0} f_{2 \varphi X_{2}} \mathcal{H}}{a^2} \nn \\
 &- \frac{A_{0}^2 \vp'  f_{2 \varphi X_{2} X_{2}} \mathcal{H}}{2 a^4} - \frac{A_{0}^3 f_{2 \varphi X_{2} X_{3}} \mathcal{H}}{2 a^4} + \frac{4 A_{0} f_{3  \varphi \varphi } \mathcal{H}}{a^2} + \frac{6 \vp''  \vp'  G_{3 \varphi X_{1}} \mathcal{H}}{a^4} + \frac{3 \vp''  \vp'^3 G_{3 \varphi X_{1} X_{1}} \mathcal{H}}{a^6} \nn \\
 &+ \frac{2 A_{0}^3 \tilde{f}_{3  \varphi \varphi } \mathcal{H}}{a^4} - \frac{4 \vp'  G_{3  \varphi \varphi } \mathcal{H}}{a^2} + \frac{4 \vp'^3 G_{3 \varphi \varphi X_{1}} \mathcal{H}}{a^4} - \frac{3 \vp'^4 G_{3 \varphi X_{1} X_{1}} \mathcal{H}^2}{a^6} - \frac{6 G_{4  \varphi \varphi } \mathcal{H}^2}{a^2} \nn \\
 &+ \frac{3 \vp'^2 G_{3 \varphi X_{1}} \mathcal{H}'}{a^4} - \frac{6 G_{4  \varphi \varphi } \mathcal{H}'}{a^2}.
\end{align}

\subsection{``Time'' vector field equation of motion}
Coefficients in Eq. \eqref{Pert Temp Vec Eq}:
\begin{align}
F_1 &= - \frac{6 A_{0}^2 f_{3 X_{3}}}{a^2} - \frac{6 A_{0}^2 \tilde{f}_{3}}{a^2}, \\
F_2 &= \frac{\vp'^2 f_{2 X_{1} X_{2}}}{2 a^2} + \frac{A_{0} \vp'  f_{2 X_{1} X_{3}}}{a^2} + \frac{1}{2} f_{2 X_{2}} + \frac{A_{0} \vp'  f_{2 X_{2} X_{2}}}{4 a^2} + \frac{A_{0}^2 f_{2 X_{2} X_{3}}}{2 a^2} \nn \\
 &+ 2 f_{3  \varphi } - \frac{2 A_{0}^2 \tilde{f}_{3  \varphi }}{a^2}, \\
F_3 &= - \frac{\vp'^3 f_{2 X_{1} X_{2}}}{2 a^3} - \frac{A_{0} \vp'^2 f_{2 X_{1} X_{3}}}{a^3} - \frac{\vp'  f_{2 X_{2}}}{a} - \frac{A_{0} \vp'^2 f_{2 X_{2} X_{2}}}{2 a^3} - \frac{3 A_{0}^2 \vp'  f_{2 X_{2} X_{3}}}{2 a^3} \nn \\
 &- \frac{2 A_{0} f_{2 X_{3}}}{a} - \frac{A_{0}^3 f_{2 X_{3} X_{3}}}{a^3} - \frac{4 \vp'  f_{3  \varphi }}{a} - \frac{2 A_{0}^2 \vp'  f_{3 \varphi X_{3}}}{a^3} + \frac{8 A_{0}^2 \vp'  \tilde{f}_{3  \varphi }}{a^3} + \frac{2 A_{0}^4 \vp'  \tilde{f}_{3 \varphi X_{3}}}{a^5} \nn \\
 &+ \frac{24 A_{0}^2 f_{3 X_{3}} \mathcal{H}}{a^3} + \frac{6 A_{0}^4 f_{3 X_{3} X_{3}} \mathcal{H}}{a^5} + \frac{24 A_{0}^2 \tilde{f}_{3} \mathcal{H}}{a^3} + \frac{6 A_{0}^4 \tilde{f}_{3 X_{3}} \mathcal{H}}{a^5}, \\
F_4 &= \frac{\vp'  f_{2 \varphi X_{2}}}{2 a} + \frac{A_{0} f_{2 \varphi X_{3}}}{a} + \frac{2 \vp'  f_{3  \varphi \varphi }}{a} - \frac{2 A_{0}^2 \vp'  \tilde{f}_{3  \varphi \varphi }}{a^3} - \frac{6 A_{0}^2 f_{3 \varphi X_{3}} \mathcal{H} }{a^3} - \frac{6 A_{0}^2 \tilde{f}_{3  \varphi } \mathcal{H} }{a^3}, \\
F_5 &= \frac{\vp'^2 f_{2 X_{2} X_{2}}}{4 a^2} + \frac{A_{0} \vp'  f_{2 X_{2} X_{3}}}{a^2} + f_{2 X_{3}} + \frac{A_{0}^2 f_{2 X_{3} X_{3}}}{a^2} + \frac{2 A_{0} \vp'  f_{3 \varphi X_{3}}}{a^2} - \frac{4 A_{0} \vp'  \tilde{f}_{3  \varphi }}{a^2} \nn \\
 &- \frac{2 A_{0}^3 \vp'  \tilde{f}_{3 \varphi X_{3}}}{a^4}- \frac{12 A_{0} f_{3 X_{3}} \mathcal{H} }{a^2} - \frac{6 A_{0}^3 f_{3 X_{3} X_{3}} \mathcal{H} }{a^4} - \frac{12 A_{0} \tilde{f}_{3} \mathcal{H} }{a^2} - \frac{6 A_{0}^3 \tilde{f}_{3 X_{3}} \mathcal{H} }{a^4}, \\
F_6 &= - \frac{2 A_{0} f_{3 X_{3}}}{a} - \frac{2 A_{0} \tilde{f}_{3}}{a}.
\end{align}

\subsection{``Space'' vector field equation of motion}

Coefficients in Eq. \eqref{Pert Space Vec Eq}:
\begin{align}
H_1 &= \frac{2 A_{0}^2 f_{3 X_{3}}}{a^2} + \frac{2 A_{0}^2 \tilde{f}_{3}}{a^2}, \\
H_2 &= - \frac{1}{2} f_{2 X_{2}} - 2 f_{3  \varphi }, \\
H_3 &= - \frac{2 A_{0} f_{3 X_{3}}}{a} - \frac{2 A_{0} \tilde{f}_{3}}{a}, \\
H_4 &= - f_{2 X_{3}} + \frac{2 A'_0 f_{3 X_{3}}}{a^2} + \frac{2 A'_0 \tilde{f}_{3}}{a^2} + \frac{2 A_{0} \vp'  \tilde{f}_{3  \varphi }}{a^2} + \frac{4 A_{0} f_{3 X_{3}} \mathcal{H}}{a^2} + \frac{4 A_{0} \tilde{f}_{3} \mathcal{H}}{a^2}.
\end{align}

\section{Equations with QSA and SHA: coefficients}
\label{App: QSA SHA Coefficients}

Coefficients in Eq. \eqref{QSA SHA Variables}:
\begin{align}
W_1 &= B_{6} (A_{5} -  B_{7}) H_{3}^2, \\
W_2 &= B_6 \[ F_5 H_1 H_2 - F_3 H_2 H_3 + (B_7 - A_5) F_5 H_4 \], \\
W_3 &= B_{6} ( A_{5}^2 - 2 A_{5} B_{7} + B_{6} D_{9}) H_{3}^2, \\
W_4 &= B_{7}^2 H_{1} (F_{5} H_{1} - F_{3} H_{3}) + B_{6} B_{7} ( F_{3} H_{2} H_{3} + F_{4} H_{1} H_{3} -2 F_{5} H_{1} H_{2} + 2 A_{5} F_{5} H_{4}) \nn \\
 &- B_{6} [ A_{5}^2 F_{5} H_{4} - D_{9} F_{3} H_{1} H_{3} + A_{5} ( F_{3} H_{2} H_{3} + F_{4} H_{1} H_{3} - 2 F_{5} H_{1} H_{2}) \nn \\
 &- B_{6} F_{5} H_{2}^2 + D_{9} F_{5} (H_{1}^2 + B_{6} H_{4}) + B_{6} H_{3} (F_{4} H_{2} + H_{3} m_{\varphi }^2)], \\
W_5 &= B_{6} ( F_{5} H_{1}^2 - F_{3} H_{1} H_{3} + B_{6} F_{5} H_{4}) m_{\varphi }^2, \\
W_6 &= - [(B_{7}^2 + B_{6} D_{9}) H_{1} + B_{6} (B_{7} - A_{5}) H_{2}] H_{3}, \\
W_7 &= B_{7}^2 F_{3} H_{4} - B_{6} (F_{4} H_{1} H_{2} - F_{3} H_{2}^2 + D_{9} F_{3} H_{4} + (B_{7} - A_{5}) F_{4} H_{4} + H_{1} H_{3} m_{\varphi }^2), \\
W_8 &= B_{6} F_{3} H_{4} m_{\varphi }^2, \\
W_9 &= (B_{7}^2 - B_{6} D_{9}) F_{5} H_{1} - B_{6} (B_{7} - A_{5}) F_{5} H_{2} - [(B_{7}^2 - B_{6} D_{9}) F_{3} \nn \\
 &- B_{6} (B_{7} - A_{5}) F_{4}] H_{3}, \\
W_{10} &= B_{6} (F_{5} H_{1} - F_{3} H_{3}) m_{\varphi }^2, \\
W_{11} &= (B_{6} D_{9} - A_5 B_7) H_{3}^2, \\
W_{12} &= B_{6} F_{5} (H_{2}^2 - D_{9} H_{4}) + B_7 (A_5 F_5 H_4 + F_3 H_2 H_3 - F_5 H_1 H_2) \nn \\
 &- B_{6} H_{3} (F_{4} H_{2} + H_{3} m_{\varphi }^2), \\
W_{13} &= B_{6} F_{5} H_{4} m_{\varphi }^2, \\
W_{14} &= (B_{6} D_{9} - B_7^2) H_{3}^2, \\
W_{15} &= B_{6} F_{5} (H_{2}^2 - D_{9} H_{4}) - B_{6} H_{3} (F_{4} H_{2} + H_{3} m_{\varphi }^2) + B_7^2 F_5 H_4.
\end{align}

\newpage 

\section{Effective dark energy fluid}
\label{App:4}

Here we use $\kappa = 1$.

\subsection{Theories with non-vanishing anisotropic stress}
\label{App: Ani Stress Coefficients}

Coefficients in Eq. \eqref{Stress Eqs}:
\begin{align}
Z_1 &= G_{4  \varphi } (- A_5^2 B_7 G_{4} - (B_6 - 2) B_6 D_9 G_{4  \varphi } + A_5 B_7 (B_7 G_{4} + (B_6 - 2) G_{4  \varphi })) H_3^2, \\
Z_2 &= - G_{4  \varphi } (- B_6 G_{4  \varphi } (A_5 (- F_5 H_1 H_2 + F_4 H_1 H_3 + A_6 H_2 H_3) \nn \\
 &+ D_9 (H_3 (- A_6 H_1 - F_3 H_1 + A_4 H_3) + F_5 (H_1^2 + (B_6 - 2) H_4)) \nn \\
 &- (B_6 - 2) (F_5 H_2^2 - H_3 (F_4 H_2 + H_3 m_\vp^2))) + B_7 (A_5 F_5 G_{4} H_1 H_2 \nn \\
 &- 2 (B_6 - 1) F_5 G_{4  \varphi } H_1 H_2 + G_{4  \varphi } (B_6 F_4 H_1 -2 F_3 H_2 + B_6 (A_6 + F_3) H_2) H_3 \nn \\
 &- A_5^2 F_5 G_{4} H_4 + A_5 (B_6 - 2) F_5 G_{4  \varphi } H_4 - A_5 G_{4} H_3 (F_3 H_2 + H_3 \mu_\vp)) \nn \\
 &+ B_7^2 (F_5 (G_{4  \varphi } H_1^2 + A_5 G_{4} H_4) + H_3 (A_4 G_{4  \varphi } H_3 - A_6 G_{4  \varphi } H_1 - F_3 G_{4  \varphi } H_1 + G_{4} H_3 \mu_\vp))) \nn \\
 &- \frac{A_2 (A_5 - B_7) B_7 (G_{4  \varphi }^2 - G_{4} G_{4  \varphi \varphi }) H_3^2 \vp'}{a}, \\
Z_3 &=G_{4  \varphi } (B_6 F_3 G_{4  \varphi } H_1 H_3 m_\vp^2 - B_6 F_5 G_{4  \varphi } H_1^2 m_\vp^2 + 2 B_6 F_5 G_{4  \varphi } H_4 m_\vp^2 - B_6^2 F_5 G_{4  \varphi } H_4 m_\vp^2 \nn \\
 &+ A_4 G_{4  \varphi } (B_7^2 F_5 H_4 + B_6 F_5 (H_2^2 - D_9 H_4) - B_6 H_3 (F_4 H_2 + H_3 m_\vp^2)) \nn \\
 &+ A_6 G_{4  \varphi } (- B_7^2 F_3 H_4 + B_6 (F_4 H_1 H_2 - F_3 H_2^2 + D_9 F_3 H_4 + (B_7 - A_5) F_4 H_4 + H_1 H_3 m_\vp^2)) \nn \\
 &+ B_7 F_5 G_{4} H_1 H_2 \mu_\vp - B_7 F_3 G_{4} H_2 H_3 \mu_\vp - A_5 B_7 F_5 G_{4} H_4 \mu_\vp + B_7^2 F_5 G_{4} H_4 \mu_\vp) \nn \\
 &+ \frac{A_2 B_7 (G_{4  \varphi }^2 - G_{4} G_{4  \varphi \varphi }) (F_3 H_2 H_3 - F_5 (H_1 H_2 + (B_7 - A_5) H_4)) \vp'}{a}, \\
Z_4 &= B_6 (A_4 F_5 - A_6 F_3) G_{4  \varphi }^2 H_4 m_\vp^2, \\
Z_5 &= B_6 (A_5^2 -2 A_5 B_7 + B_6 D_9) G_{4  \varphi }^2 H_3^2, \\
Z_6 &= G_{4  \varphi }^2 (B_7^2 H_1 (F_5 H_1 - F_3 H_3) + B_6 B_7 (-2 F_5 H_1 H_2 + F_4 H_1 H_3 + F_3 H_2 H_3 + 2 A_5 F_5 H_4) \nn \\
 &- B_6 (A_5^2 F_5 H_4 - B_6 F_5 H_2^2 - D_9 F_3 H_1 H_3 + A_5 (-2 F_5 H_1 H_2 + F_4 H_1 H_3 + F_3 H_2 H_3) \nn \\
 &+ D_9 F_5 (H_1^2 + B_6 H_4) + B_6 H_3 (F_4 H_2 + H_3 m_\vp^2))), \\
Z_7 &= B_6 G_{4  \varphi }^2 (F_5 (H_1^2 + B_6 H_4) - F_3 H_1 H_3) m_\vp^2, \\
Z_8 &= (A_5 - B_7) B_7 G_{4  \varphi }^2 (B_7 G_{4} - (B_6 - 2) G_{4  \varphi }) H_3^2, \\
Z_9 &= G_{4  \varphi }^2 (- B_6 G_{4  \varphi } H_3 (A_5 B_{11} H_2 - B_{11} D_9 H_1 + B_9 D_9 H_3) + B_7^3 F_5 G_{4} H_4 \nn \\
 &+ B_7^2 ((2 - B_6) F_5 G_{4  \varphi } H_4 + F_5 G_{4} (H_1 H_2 - A_5 H_4) \nn \\
 &+ H_3 (- B_{11} G_{4  \varphi } H_1 - F_3 G_{4} H_2 + B_9 G_{4  \varphi } H_3 -3 G_{4} H_3 \nu_\vp )) \nn \\
 &+ B_7 ((2 - B_6) F_5 G_{4  \varphi } (H_1 H_2 - A_5 H_4) \nn \\
 &+ H_3 (B_{11} B_6 G_{4  \varphi } H_2 + (B_6 - 2) F_3 G_{4  \varphi } H_2 + 3 A_5 G_{4} H_3 \nu_\vp ))) \nn \\
 &+ \frac{B_4 (A_5 - B_7) B_7 G_{4  \varphi } (G_{4  \varphi }^2 - G_{4} G_{4  \varphi \varphi }) H_3^2 \vp'}{a} \nn \\
 &- \frac{B_2 (A_5 - B_7) B_7 H_3^2 (G_{4  \varphi }^2 G_{4  \varphi \varphi } -2 G_{4} G_{4  \varphi \varphi }^2 + G_{4} G_{4  \varphi } G_{4  \varphi \varphi \varphi }) \vp'^2}{a^2} \nn \\
 &+ \frac{B_2 (A_5 - B_7) B_7 H_3^2 (G_{4  \varphi } (G_{4  \varphi }^2 - G_{4} G_{4  \varphi \varphi }) \vp'')}{a^2},
\end{align}

\begin{align}
 Z_{10} &=  G_{4  \varphi }^2 (B_6 B_9 G_{4  \varphi } (- F_5 H_2^2 + F_4 H_2 H_3 + D_9 F_5 H_4 + H_3^2 m_\vp^2) \nn \\
 &+ B_{11} G_{4  \varphi } (B_7^2 F_3 H_4 - B_6 (F_4 H_1 H_2 - F_3 H_2^2 + D_9 F_3 H_4 + (B_7 - A_5) F_4 H_4 + H_1 H_3 m_\vp^2)) \nn \\
 &- B_7 (3 G_{4} (- F_5 H_1 H_2 + F_3 H_2 H_3 + A_5 F_5 H_4) \nu_\vp  + B_7 F_5 H_4 (B_9 G_{4  \varphi } -3 G_{4} \nu_\vp ))) \nn \\
 &+ \frac{B_4 B_7 G_{4  \varphi } (G_{4  \varphi }^2 - G_{4} G_{4  \varphi \varphi }) (F_5 H_1 H_2 - F_3 H_2 H_3 + (B_7 - A_5) F_5 H_4) \vp'}{a} \nn \\
 &+ \frac{B_2 B_7 (F_5 H_1 H_2 - F_3 H_2 H_3 + (B_7 - A_5) F_5 H_4) (2 G_{4} G_{4  \varphi \varphi }^2 - G_{4  \varphi }^2 G_{4  \varphi \varphi }) \vp'^2}{a^2} \nn \\
 &- \frac{B_2 B_7 (F_5 H_1 H_2 - F_3 H_2 H_3 + (B_7 - A_5) F_5 H_4) G_{4} G_{4  \varphi } G_{4  \varphi \varphi \varphi } \vp'^2}{a^2} \nn \\
 &+ \frac{B_2 B_7 (F_5 H_1 H_2 - F_3 H_2 H_3 + (B_7 - A_5) F_5 H_4) G_{4  \varphi } (G_{4  \varphi }^2 - G_{4} G_{4  \varphi \varphi }) \vp''}{a^2}, \\
Z_{11} &= B_6 (B_{11} F_3 - B_9 F_5) G_{4  \varphi }^3 H_4 m_\vp^2, \\
Z_{12} &= G_{4  \varphi }, 
\end{align}

\begin{align}
Z_{13} &= H_3 \Big(G_{4  \varphi } (B_6 G_{4  \varphi } (C_5 D_9 H_1 - A_5 C_5 H_2 - C_3 D_9 H_3) + B_7 (B_6 C_5 G_{4  \varphi } H_2 + A_5 C_4 G_{4} H_3) \nn \\
 &- B_7^2 (C_5 G_{4  \varphi } H_1 + C_4 G_{4} H_3 - C_3 G_{4  \varphi } H_3)) + \frac{(A_5 - B_7) B_7 C_2 (G_{4  \varphi }^2 - G_{4} G_{4  \varphi \varphi }) H_3 \vp'}{a} \Big), \\
Z_{14} &= G_{4  \varphi } (B_7^2 (C_6 G_{4  \varphi } (F_5 H_1 - F_3 H_3) + (C_4 F_5 G_{4} + C_5 F_3 G_{4  \varphi } - C_3 F_5 G_{4  \varphi }) H_4) \nn \\
 &- B_7 (B_6 G_{4  \varphi } (C_6 F_5 H_2 - C_6 F_4 H_3 + C_5 F_4 H_4) \nn \\
 &+ C_4 G_{4} (- F_5 H_1 H_2 + F_3 H_2 H_3 + A_5 F_5 H_4)) \nn \\
 &+ B_6 G_{4  \varphi } (C_6 (- D_9 F_5 H_1 + A_5 F_5 H_2 + D_9 F_3 H_3 - A_5 F_4 H_3) \nn \\
 &- C_5 (F_4 H_1 H_2 - F_3 H_2^2 + D_9 F_3 H_4 - A_5 F_4 H_4 + H_1 H_3 m_\vp^2) \nn \\
 &+ C_3 (- F_5 H_2^2 + F_4 H_2 H_3 + D_9 F_5 H_4 + H_3^2 m_\vp^2))) \nn \\
 &+ \frac{B_7 C_2 (G_{4  \varphi }^2 - G_{4} G_{4  \varphi \varphi }) (F_5 H_1 H_2 - F_3 H_2 H_3 + (B_7 - A_5) F_5 H_4) \vp'}{a}, \\
Z_{15} &= B_6 G_{4  \varphi }^2 (C_6 F_5 H_1 - C_6 F_3 H_3 + C_5 F_3 H_4 - C_3 F_5 H_4) m_\vp^2.
\end{align}

\newpage
\subsection{Theories with vanishing anisotropic stress}
\label{App: No Ani Stress Coefficients}

Coefficients in Eq. \eqref{No Stress Eqs}:

\begin{align}
Y_1 &= - (A_5^2 + (B_6 - 2) D_9) H_3^2, \\
Y_2 &= A_5^2 F_5 H_4 + D_9 (H_3 (A_4 H_3 - A_6 H_1 - F_3 H_1) + F_5 (H_1^2 + (B_6 - 2) H_4)) \nn \\
 &- (B_6 - 2) (F_5 H_2^2 - H_3 (F_4 H_2 + H_3 m_\vp^2)) \nn \\
 &+ A_5 (H_3 (F_4 H_1 + A_6 H_2 + F_3 H_2 + H_3 \mu_\vp) - 2 F_5 H_1 H_2), \\
Y_3 &= A_4 F_5 (H_2^2 - D_9 H_4) - F_5 H_1^2 m_\vp^2 + F_3 H_1 H_3 m_\vp^2 + 2 F_5 H_4 m_\vp^2 - B_6 F_5 H_4 m_\vp^2 \nn \\
 &- A_4 H_3 (F_4 H_2 + H_3 m_\vp^2) + A_6 (F_4 H_1 H_2 - F_3 H_2^2 + D_9 F_3 H_4 - A_5 F_4 H_4  \nn \\
 &+ H_1 H_3 m_\vp^2) + F_5 H_1 H_2 \mu_\vp - F_3 H_2 H_3 \mu_\vp - A_5 F_5 H_4 \mu_\vp, \\
Y_4 &= (A_4 F_5 - A_6 F_3) H_4 m_\vp^2, \\
Y_5 &= (A_5^2 + B_6 D_9) H_3^2, \\
Y_6 &= 2 A_5 F_5 H_1 H_2 + B_6 F_5 H_2^2 + D_9 F_3 H_1 H_3 - A_5 (F_4 H_1 + F_3 H_2) H_3 - A_5^2 F_5 H_4 \nn \\
 &- D_9 F_5 (H_1^2 + B_6 H_4) - B_6 H_3 (F_4 H_2 + H_3 m_\vp^2), \\
Y_7 &= (F_5 (H_1^2 + B_6 H_4) - F_3 H_1 H_3) m_\vp^2, \\
Y_8 &= H_3 (B_{11} D_9 H_1 - A_5 B_{11} H_2 - B_9 D_9 H_3 + 3 A_5 H_3 \nu_\vp ), \\
Y_9 &= - B_{11} (F_4 H_1 H_2 - F_3 H_2^2 + D_9 F_3 H_4 - A_5 F_4 H_4 + H_1 H_3 m_\vp^2) \nn \\
 &+ B_9 (F_4 H_2 H_3 + D_9 F_5 H_4 + H_3^2 m_\vp^2 - F_5 H_2^2) \nn \\
 &+ 3 (F_5 H_1 H_2 - F_3 H_2 H_3 - A_5 F_5 H_4) \nu_\vp, \\
Y_{10} &= (B_{11} F_3 - B_9 F_5) H_4 m_\vp^2, \\
Y_{11} &= H_3 (C_5 D_9 H_1 - A_5 C_5 H_2 + A_5 C_4 H_3 - C_3 D_9 H_3), \\
Y_{12} &= C_4 F_5 H_1 H_2 - C_3 F_5 H_2^2 - C_4 F_3 H_2 H_3 + C_3 F_4 H_2 H_3 \nn \\
 &+ C_6 (A_5 F_5 H_2 + D_9 F_3 H_3 - A_5 F_4 H_3 - D_9 F_5 H_1) - A_5 C_4 F_5 H_4 + C_3 D_9 F_5 H_4 \nn \\
 &+ C_3 H_3^2 m_\vp^2 - C_5 (F_4 H_1 H_2 - F_3 H_2^2 + D_9 F_3 H_4 - A_5 F_4 H_4 + H_1 H_3 m_\vp^2), \\
Y_{13} &= (C_6 F_5 H_1 - C_6 F_3 H_3 + C_5 F_3 H_4 - C_3 F_5 H_4) m_\vp^2,
\end{align}

\bibliographystyle{JHEPmodplain}
\bibliography{paper}

\providecommand{\href}[2]{#2}\begingroup\raggedright\begin{thebibliography}{100}

\bibitem{Abdalla:2022yfr}
E.~Abdalla {\em et~al.}, {\it {Cosmology Intertwined: A Review of the Particle
  Physics, Astrophysics, and Cosmology Associated with the Cosmological
  Tensions and Anomalies}},  in {\em {2022 Snowmass Summer Study}}, 3, 2022.
\newblock \href{http://arxiv.org/abs/2203.06142}{{\sf arXiv:2203.06142}}.

\bibitem{Planck:2018vyg}
{\bf Planck} Collaboration, N.~Aghanim {\em et~al.}, {\it {Planck 2018 results.
  VI. Cosmological parameters}},  {\sl Astron. Astrophys.} {\bf 641} (2020) A6,
  [\href{http://arxiv.org/abs/1807.06209}{{\sf arXiv:1807.06209}}],
  [\href{http://dx.doi.org/10.1051/0004-6361/201833910}{{\sf
  doi:10.1051/0004-6361/201833910}}]. [Erratum: Astron.Astrophys. 652, C4
  (2021)].

\bibitem{Riess:2019cxk}
A.~G. Riess, S.~Casertano, W.~Yuan, L.~M. Macri, and D.~Scolnic, {\it {Large
  Magellanic Cloud} {Cepheid Standards Provide a 1\% Foundation for the
  Determination of the Hubble Constant} and {Stronger Evidence for Physics}
  beyond {$\Lambda$CDM}},  {\sl Astrophys. J.} {\bf 876} (2019), no.~1 85,
  [\href{http://arxiv.org/abs/1903.07603}{{\sf arXiv:1903.07603}}],
  [\href{http://dx.doi.org/10.3847/1538-4357/ab1422}{{\sf
  doi:10.3847/1538-4357/ab1422}}].

\bibitem{Freedman:2021ahq}
W.~L. Freedman, {\it {Measurements of the Hubble Constant: Tensions in
  Perspective}},  {\sl Astrophys. J.} {\bf 919} (2021), no.~1 16,
  [\href{http://arxiv.org/abs/2106.15656}{{\sf arXiv:2106.15656}}],
  [\href{http://dx.doi.org/10.3847/1538-4357/ac0e95}{{\sf
  doi:10.3847/1538-4357/ac0e95}}].

\bibitem{Riess:2021jrx}
A.~G. Riess {\em et~al.}, {\it {A Comprehensive Measurement of the Local Value
  of the Hubble Constant with 1 km/s/Mpc Uncertainty from the Hubble Space
  Telescope and the SH0ES Team}},  \href{http://arxiv.org/abs/2112.04510}{{\sf
  arXiv:2112.04510}}.

\bibitem{DES:2022ign}
{\bf DES, SPT} Collaboration, C.~Chang {\em et~al.}, {\it {Joint analysis of
  DES Year 3 data and CMB lensing from SPT and Planck II: Cross-correlation
  measurements and cosmological constraints}},
  \href{http://arxiv.org/abs/2203.12440}{{\sf arXiv:2203.12440}}.

\bibitem{Gatti:2021uwl}
M.~Gatti {\em et~al.}, {\it {Dark Energy Survey Year 3 results: cosmology with
  moments of weak} {lensing mass maps}},
  \href{http://arxiv.org/abs/2110.10141}{{\sf arXiv:2110.10141}}.

\bibitem{Zurcher:2021bjz}
D.~Z\"urcher {\em et~al.}, {\it {Dark Energy Survey Year 3 results: Cosmology
  with peaks using an} {emulator approach}},
  \href{http://arxiv.org/abs/2110.10135}{{\sf arXiv:2110.10135}}.

\bibitem{Huang:2021tvo}
L.~Huang, Z.~Huang, H.~Zhou, and Z.~Li, {\it {The $S_8$ Tension in Light of
  Updated Redshift-Space Distortion data}},
  \href{http://arxiv.org/abs/2110.08498}{{\sf arXiv:2110.08498}}.

\bibitem{Kobayashi:2021oud}
Y.~Kobayashi, T.~Nishimichi, M.~Takada, and H.~Miyatake, {\it {Full-shape
  cosmology analysis of} {SDSS-III BOSS galaxy power spectrum using
  emulator-based halo model: a $5\%$ determination of $\sigma_8$}},
  \href{http://arxiv.org/abs/2110.06969}{{\sf arXiv:2110.06969}}.

\bibitem{Loureiro:2021ruj}
A.~Loureiro {\em et~al.}, {\it {KiDS \& Euclid: Cosmological implications of a
  pseudo angular power} {spectrum analysis of KiDS-1000 cosmic shear
  tomography}},  \href{http://arxiv.org/abs/2110.06947}{{\sf
  arXiv:2110.06947}}.

\bibitem{PhysRevLett.111.161301}
E.~Macaulay, I.~K. Wehus, and H.~K. Eriksen, {\it Lower growth rate from recent
  redshift space distortion measurements than expected from planck},  {\sl
  Phys. Rev. Lett.} {\bf 111} (Oct, 2013) 161301,
  [\href{http://dx.doi.org/10.1103/PhysRevLett.111.161301}{{\sf
  doi:10.1103/PhysRevLett.111.161301}}].

\bibitem{PhysRevD.91.103508}
R.~A. Battye, T.~Charnock, and A.~Moss, {\it Tension between the power spectrum
  of density perturbations measured on large and small scales},  {\sl Phys.
  Rev. D} {\bf 91} (May, 2015) 103508,
  [\href{http://dx.doi.org/10.1103/PhysRevD.91.103508}{{\sf
  doi:10.1103/PhysRevD.91.103508}}].

\bibitem{Abbott:2017wau}
{\bf DES} Collaboration, T.~M.~C. Abbott {\em et~al.}, {\it {Dark Energy Survey
  year 1 results: Cosmological constraints from galaxy clustering and weak
  lensing}},  {\sl Phys. Rev.} {\bf D98} (2018), no.~4 043526,
  [\href{http://arxiv.org/abs/1708.01530}{{\sf arXiv:1708.01530}}],
  [\href{http://dx.doi.org/10.1103/PhysRevD.98.043526}{{\sf
  doi:10.1103/PhysRevD.98.043526}}].

\bibitem{DES:2021wwk}
{\bf DES} Collaboration, T.~M.~C. Abbott {\em et~al.}, {\it {Dark Energy Survey
  Year 3 results: Cosmological constraints from galaxy clustering and weak
  lensing}},  {\sl Phys. Rev. D} {\bf 105} (2022), no.~2 023520,
  [\href{http://arxiv.org/abs/2105.13549}{{\sf arXiv:2105.13549}}],
  [\href{http://dx.doi.org/10.1103/PhysRevD.105.023520}{{\sf
  doi:10.1103/PhysRevD.105.023520}}].

\bibitem{Blanchard:2021dwr}
A.~Blanchard and S.~Ili\'c, {\it {Closing up the cluster tension?}},
  \href{http://arxiv.org/abs/2104.00756}{{\sf arXiv:2104.00756}}.

\bibitem{Heymans:2020gsg}
C.~Heymans {\em et~al.}, {\it {KiDS-1000 Cosmology: Multi-probe weak
  gravitational lensing and spectroscopic galaxy clustering constraints}},
  {\sl Astron. Astrophys.} {\bf 646} (2021) A140,
  [\href{http://arxiv.org/abs/2007.15632}{{\sf arXiv:2007.15632}}],
  [\href{http://dx.doi.org/10.1051/0004-6361/202039063}{{\sf
  doi:10.1051/0004-6361/202039063}}].

\bibitem{Philcox:2021kcw}
O.~H.~E. Philcox and M.~M. Ivanov, {\it {BOSS DR12 full-shape cosmology:
  \ensuremath{\Lambda}CDM constraints from the large-scale galaxy power
  spectrum and bispectrum monopole}},  {\sl Phys. Rev. D} {\bf 105} (2022),
  no.~4 043517, [\href{http://arxiv.org/abs/2112.04515}{{\sf
  arXiv:2112.04515}}],
  [\href{http://dx.doi.org/10.1103/PhysRevD.105.043517}{{\sf
  doi:10.1103/PhysRevD.105.043517}}].

\bibitem{Joudaki:2016mvz}
S.~Joudaki {\em et~al.}, {\it {CFHTLenS revisited: assessing concordance with
  Planck including astrophysical systematics}},  {\sl Mon. Not. Roy. Astron.
  Soc.} {\bf 465} (2017), no.~2 2033--2052,
  [\href{http://arxiv.org/abs/1601.05786}{{\sf arXiv:1601.05786}}],
  [\href{http://dx.doi.org/10.1093/mnras/stw2665}{{\sf
  doi:10.1093/mnras/stw2665}}].

\bibitem{Fields:2011zzb}
B.~D. Fields, {\it {The primordial lithium problem}},  {\sl Ann. Rev. Nucl.
  Part. Sci.} {\bf 61} (2011) 47--68,
  [\href{http://arxiv.org/abs/1203.3551}{{\sf arXiv:1203.3551}}],
  [\href{http://dx.doi.org/10.1146/annurev-nucl-102010-130445}{{\sf
  doi:10.1146/annurev-nucl-102010-130445}}].

\bibitem{Cyburt:2015mya}
R.~H. Cyburt, B.~D. Fields, K.~A. Olive, and T.-H. Yeh, {\it {Big Bang
  Nucleosynthesis: 2015}},  {\sl Rev. Mod. Phys.} {\bf 88} (2016) 015004,
  [\href{http://arxiv.org/abs/1505.01076}{{\sf arXiv:1505.01076}}],
  [\href{http://dx.doi.org/10.1103/RevModPhys.88.015004}{{\sf
  doi:10.1103/RevModPhys.88.015004}}].

\bibitem{Mathews:2019hbi}
G.~J. Mathews, A.~Kedia, N.~Sasankan, M.~Kusakabe, Y.~Luo, T.~Kajino,
  D.~Yamazaki, T.~Makki, and M.~E. Eid, {\it {Cosmological Solutions to the
  Lithium Problem}},  {\sl JPS Conf. Proc.} {\bf 31} (2020) 011033,
  [\href{http://arxiv.org/abs/1909.01245}{{\sf arXiv:1909.01245}}],
  [\href{http://dx.doi.org/10.7566/JPSCP.31.011033}{{\sf
  doi:10.7566/JPSCP.31.011033}}].

\bibitem{Sbordone:2010zi}
L.~Sbordone {\em et~al.}, {\it {The metal-poor end of the Spite plateau. 1:
  Stellar parameters, metallicities and lithium abundances}},  {\sl Astron.
  Astrophys.} {\bf 522} (2010) A26, [\href{http://arxiv.org/abs/1003.4510}{{\sf
  arXiv:1003.4510}}],
  [\href{http://dx.doi.org/10.1051/0004-6361/200913282}{{\sf
  doi:10.1051/0004-6361/200913282}}].

\bibitem{Pitrou:2020etk}
C.~Pitrou, A.~Coc, J.-P. Uzan, and E.~Vangioni, {\it {A new tension in the
  cosmological model from primordial deuterium?}},  {\sl Mon. Not. Roy. Astron.
  Soc.} {\bf 502} (2021), no.~2 2474--2481,
  [\href{http://arxiv.org/abs/2011.11320}{{\sf arXiv:2011.11320}}],
  [\href{http://dx.doi.org/10.1093/mnras/stab135}{{\sf
  doi:10.1093/mnras/stab135}}].

\bibitem{Iocco:2008va}
F.~Iocco, G.~Mangano, G.~Miele, O.~Pisanti, and P.~D. Serpico, {\it {Primordial
  Nucleosynthesis: from precision cosmology to fundamental physics}},  {\sl
  Phys. Rept.} {\bf 472} (2009) 1--76,
  [\href{http://arxiv.org/abs/0809.0631}{{\sf arXiv:0809.0631}}],
  [\href{http://dx.doi.org/10.1016/j.physrep.2009.02.002}{{\sf
  doi:10.1016/j.physrep.2009.02.002}}].

\bibitem{Bowman:2018yin}
J.~D. Bowman, A.~E.~E. Rogers, R.~A. Monsalve, T.~J. Mozdzen, and N.~Mahesh,
  {\it {An absorption profile centred at 78 megahertz in the sky-averaged
  spectrum}},  {\sl Nature} {\bf 555} (2018), no.~7694 67--70,
  [\href{http://arxiv.org/abs/1810.05912}{{\sf arXiv:1810.05912}}],
  [\href{http://dx.doi.org/10.1038/nature25792}{{\sf
  doi:10.1038/nature25792}}].

\bibitem{PhysRevD.98.083525}
V.~Poulin, T.~L. Smith, D.~Grin, T.~Karwal, and M.~Kamionkowski, {\it
  Cosmological implications of ultralight axionlike fields},  {\sl Phys. Rev.
  D} {\bf 98} (Oct, 2018) 083525,
  [\href{http://dx.doi.org/10.1103/PhysRevD.98.083525}{{\sf
  doi:10.1103/PhysRevD.98.083525}}].

\bibitem{Singh:2022ivh}
S.~Singh, J.~Nambissan~T., R.~Subrahmanyan, N.~Udaya~Shankar, B.~S. Girish,
  A.~Raghunathan, R.~Somashekar, K.~S. Srivani, and M.~Sathyanarayana~Rao, {\it
  {On the detection of a cosmic dawn signal in the radio background}},  {\sl
  Nature Astron.} {\bf 6} (2022), no.~5 607--617,
  [\href{http://dx.doi.org/10.1038/s41550-022-01610-5}{{\sf
  doi:10.1038/s41550-022-01610-5}}].

\bibitem{PhysRevLett.124.161301}
J.~Sakstein and M.~Trodden, {\it Early dark energy from massive neutrinos as a
  natural resolution of the hubble tension},  {\sl Phys. Rev. Lett.} {\bf 124}
  (Apr, 2020) 161301,
  [\href{http://dx.doi.org/10.1103/PhysRevLett.124.161301}{{\sf
  doi:10.1103/PhysRevLett.124.161301}}].

\bibitem{Poulin:2021bjr}
V.~Poulin, T.~L. Smith, and A.~Bartlett, {\it {Dark Energy at early times and
  ACT: a larger Hubble constant without late-time priors}},
  \href{http://arxiv.org/abs/2109.06229}{{\sf arXiv:2109.06229}}.

\bibitem{Hill:2021yec}
J.~C. Hill {\em et~al.}, {\it {The Atacama Cosmology Telescope: Constraints on
  Pre-Recombination Early Dark Energy}},
  \href{http://arxiv.org/abs/2109.04451}{{\sf arXiv:2109.04451}}.

\bibitem{Sabla:2022xzj}
V.~I. Sabla and R.~R. Caldwell, {\it {The Microphysics of Early Dark Energy}},
  \href{http://arxiv.org/abs/2202.08291}{{\sf arXiv:2202.08291}}.

\bibitem{Heavens:2017hkr}
A.~Heavens, Y.~Fantaye, E.~Sellentin, H.~Eggers, Z.~Hosenie, S.~Kroon, and
  A.~Mootoovaloo, {\it {No evidence for extensions to the standard cosmological
  model}},  {\sl Phys. Rev. Lett.} {\bf 119} (2017), no.~10 101301,
  [\href{http://arxiv.org/abs/1704.03467}{{\sf arXiv:1704.03467}}],
  [\href{http://dx.doi.org/10.1103/PhysRevLett.119.101301}{{\sf
  doi:10.1103/PhysRevLett.119.101301}}].

\bibitem{Weinberg:1988cp}
S.~Weinberg, {\it {The Cosmological Constant Problem}},  {\sl Rev. Mod. Phys.}
  {\bf 61} (1989) 1--23, [\href{http://dx.doi.org/10.1103/RevModPhys.61.1}{{\sf
  doi:10.1103/RevModPhys.61.1}}]. [,569(1988)].

\bibitem{Carroll:2000fy}
S.~M. Carroll, {\it {The Cosmological constant}},  {\sl Living Rev. Rel.} {\bf
  4} (2001) 1, [\href{http://arxiv.org/abs/astro-ph/0004075}{{\sf
  arXiv:astro-ph/0004075}}], [\href{http://dx.doi.org/10.12942/lrr-2001-1}{{\sf
  doi:10.12942/lrr-2001-1}}].

\bibitem{Bertone:2016nfn}
G.~Bertone and D.~Hooper, {\it {History of dark matter}},  {\sl Rev. Mod.
  Phys.} {\bf 90} (2018), no.~4 045002,
  [\href{http://arxiv.org/abs/1605.04909}{{\sf arXiv:1605.04909}}],
  [\href{http://dx.doi.org/10.1103/RevModPhys.90.045002}{{\sf
  doi:10.1103/RevModPhys.90.045002}}].

\bibitem{PhysRevLett.125.211101}
K.~Pardo and D.~N. Spergel, {\it What is the price of abandoning dark matter?
  cosmological constraints on alternative gravity theories},  {\sl Phys. Rev.
  Lett.} {\bf 125} (Nov, 2020) 211101,
  [\href{http://dx.doi.org/10.1103/PhysRevLett.125.211101}{{\sf
  doi:10.1103/PhysRevLett.125.211101}}].

\bibitem{PhysRevLett.126.041303}
V.~De~Luca, G.~Franciolini, and A.~Riotto, {\it Nanograv data hints at
  primordial black holes as dark matter},  {\sl Phys. Rev. Lett.} {\bf 126}
  (Jan, 2021) 041303,
  [\href{http://dx.doi.org/10.1103/PhysRevLett.126.041303}{{\sf
  doi:10.1103/PhysRevLett.126.041303}}].

\bibitem{Carr:2020xqk}
B.~Carr and F.~Kuhnel, {\it {Primordial Black Holes as Dark Matter: Recent
  Developments}},  {\sl Ann. Rev. Nucl. Part. Sci.} {\bf 70} (2020) 355--394,
  [\href{http://arxiv.org/abs/2006.02838}{{\sf arXiv:2006.02838}}],
  [\href{http://dx.doi.org/10.1146/annurev-nucl-050520-125911}{{\sf
  doi:10.1146/annurev-nucl-050520-125911}}].

\bibitem{PhysRevD.104.043520}
S.~Ili\ifmmode~\acute{c}\else \'{c}\fi{}, M.~Kopp, C.~Skordis, and D.~B.
  Thomas, {\it Dark matter properties through cosmic history},  {\sl Phys. Rev.
  D} {\bf 104} (Aug, 2021) 043520,
  [\href{http://dx.doi.org/10.1103/PhysRevD.104.043520}{{\sf
  doi:10.1103/PhysRevD.104.043520}}].

\bibitem{PhysRevLett.123.061302}
T.~Tenkanen, {\it Dark matter from scalar field fluctuations},  {\sl Phys. Rev.
  Lett.} {\bf 123} (Aug, 2019) 061302,
  [\href{http://dx.doi.org/10.1103/PhysRevLett.123.061302}{{\sf
  doi:10.1103/PhysRevLett.123.061302}}].

\bibitem{Hertzberg:2019bvt}
M.~P. Hertzberg and M.~Sandora, {\it {Dark Matter and Naturalness}},  {\sl
  JHEP} {\bf 12} (2019) 037, [\href{http://arxiv.org/abs/1908.09841}{{\sf
  arXiv:1908.09841}}], [\href{http://dx.doi.org/10.1007/JHEP12(2019)037}{{\sf
  doi:10.1007/JHEP12(2019)037}}].

\bibitem{doi:10.1146/annurev.aa.32.090194.002531}
B.~Carr, {\it Baryonic dark matter},  {\sl Annual Review of Astronomy and
  Astrophysics} {\bf 32} (1994), no.~1 531--590,
  [\href{http://arxiv.org/abs/https://doi.org/10.1146/annurev.aa.32.090194.002531}{{\sf
  https://doi.org/10.1146/annurev.aa.32.090194.002531}}],
  [\href{http://dx.doi.org/10.1146/annurev.aa.32.090194.002531}{{\sf
  doi:10.1146/annurev.aa.32.090194.002531}}].

\bibitem{Planck:2015sxf}
{\bf Planck} Collaboration, P.~A.~R. Ade {\em et~al.}, {\it {Planck 2015
  results. XX. Constraints on inflation}},  {\sl Astron. Astrophys.} {\bf 594}
  (2016) A20, [\href{http://arxiv.org/abs/1502.02114}{{\sf arXiv:1502.02114}}],
  [\href{http://dx.doi.org/10.1051/0004-6361/201525898}{{\sf
  doi:10.1051/0004-6361/201525898}}].

\bibitem{Steinhardt:2002ih}
P.~J. Steinhardt, N.~Turok, and N.~Turok, {\it {A Cyclic model of the
  universe}},  {\sl Science} {\bf 296} (2002) 1436--1439,
  [\href{http://arxiv.org/abs/hep-th/0111030}{{\sf arXiv:hep-th/0111030}}],
  [\href{http://dx.doi.org/10.1126/science.1070462}{{\sf
  doi:10.1126/science.1070462}}].

\bibitem{Chowdhury:2019otk}
D.~Chowdhury, J.~Martin, C.~Ringeval, and V.~Vennin, {\it {Assessing the
  scientific status of inflation after Planck}},  {\sl Phys. Rev. D} {\bf 100}
  (2019), no.~8 083537, [\href{http://arxiv.org/abs/1902.03951}{{\sf
  arXiv:1902.03951}}],
  [\href{http://dx.doi.org/10.1103/PhysRevD.100.083537}{{\sf
  doi:10.1103/PhysRevD.100.083537}}].

\bibitem{Ijjas:2014nta}
A.~Ijjas, P.~J. Steinhardt, and A.~Loeb, {\it {Inflationary schism}},  {\sl
  Phys. Lett. B} {\bf 736} (2014) 142--146,
  [\href{http://arxiv.org/abs/1402.6980}{{\sf arXiv:1402.6980}}],
  [\href{http://dx.doi.org/10.1016/j.physletb.2014.07.012}{{\sf
  doi:10.1016/j.physletb.2014.07.012}}].

\bibitem{Ijjas:2013vea}
A.~Ijjas, P.~J. Steinhardt, and A.~Loeb, {\it {Inflationary paradigm in trouble
  after Planck2013}},  {\sl Phys. Lett. B} {\bf 723} (2013) 261--266,
  [\href{http://arxiv.org/abs/1304.2785}{{\sf arXiv:1304.2785}}],
  [\href{http://dx.doi.org/10.1016/j.physletb.2013.05.023}{{\sf
  doi:10.1016/j.physletb.2013.05.023}}].

\bibitem{Ratra:1987rm}
B.~Ratra and P.~J.~E. Peebles, {\it {Cosmological Consequences of a Rolling
  Homogeneous Scalar Field}},  {\sl Phys. Rev.} {\bf D37} (1988) 3406,
  [\href{http://dx.doi.org/10.1103/PhysRevD.37.3406}{{\sf
  doi:10.1103/PhysRevD.37.3406}}].

\bibitem{ArmendarizPicon:2000dh}
C.~Armendariz-Picon, V.~F. Mukhanov, and P.~J. Steinhardt, {\it {A Dynamical
  solution to the problem of a small cosmological constant and late time cosmic
  acceleration}},  {\sl Phys. Rev. Lett.} {\bf 85} (2000) 4438--4441,
  [\href{http://arxiv.org/abs/astro-ph/0004134}{{\sf arXiv:astro-ph/0004134}}],
  [\href{http://dx.doi.org/10.1103/PhysRevLett.85.4438}{{\sf
  doi:10.1103/PhysRevLett.85.4438}}].

\bibitem{Copeland:2006wr}
E.~J. Copeland, M.~Sami, and S.~Tsujikawa, {\it {Dynamics of dark energy}},
  {\sl Int. J. Mod. Phys.} {\bf D15} (2006) 1753--1936,
  [\href{http://arxiv.org/abs/hep-th/0603057}{{\sf arXiv:hep-th/0603057}}],
  [\href{http://dx.doi.org/10.1142/S021827180600942X}{{\sf
  doi:10.1142/S021827180600942X}}].

\bibitem{Clifton:2011jh}
T.~Clifton, P.~G. Ferreira, A.~Padilla, and C.~Skordis, {\it {Modified Gravity
  and Cosmology}},  {\sl Phys. Rept.} {\bf 513} (2012) 1--189,
  [\href{http://arxiv.org/abs/1106.2476}{{\sf arXiv:1106.2476}}],
  [\href{http://dx.doi.org/10.1016/j.physrep.2012.01.001}{{\sf
  doi:10.1016/j.physrep.2012.01.001}}].

\bibitem{Li:2018tfg}
J.~Li and G.-B. Zhao, {\it {Cosmological Tests of Gravity with the Latest
  Observations}},  {\sl Astrophys. J.} {\bf 871} (2019), no.~2 196,
  [\href{http://arxiv.org/abs/1806.05022}{{\sf arXiv:1806.05022}}],
  [\href{http://dx.doi.org/10.3847/1538-4357/aaf869}{{\sf
  doi:10.3847/1538-4357/aaf869}}].

\bibitem{Bamba:2012cp}
K.~Bamba, S.~Capozziello, S.~Nojiri, and S.~D. Odintsov, {\it {Dark energy
  cosmology: the equivalent description via different theoretical models and
  cosmography tests}},  {\sl Astrophys. Space Sci.} {\bf 342} (2012) 155--228,
  [\href{http://arxiv.org/abs/1205.3421}{{\sf arXiv:1205.3421}}],
  [\href{http://dx.doi.org/10.1007/s10509-012-1181-8}{{\sf
  doi:10.1007/s10509-012-1181-8}}].

\bibitem{PhysRevLett.116.221101}
{\bf LIGO Scientific and Virgo Collaborations} Collaboration, B.~P. Abbott {\em
  et~al.}, {\it Tests of general relativity with gw150914},  {\sl Phys. Rev.
  Lett.} {\bf 116} (May, 2016) 221101,
  [\href{http://dx.doi.org/10.1103/PhysRevLett.116.221101}{{\sf
  doi:10.1103/PhysRevLett.116.221101}}].

\bibitem{Collett:2018gpf}
T.~E. Collett, L.~J. Oldham, R.~J. Smith, M.~W. Auger, K.~B. Westfall,
  D.~Bacon, R.~C. Nichol, K.~L. Masters, K.~Koyama, and R.~van~den Bosch, {\it
  {A precise extragalactic test of General Relativity}},  {\sl Science} {\bf
  360} (2018) 1342, [\href{http://arxiv.org/abs/1806.08300}{{\sf
  arXiv:1806.08300}}], [\href{http://dx.doi.org/10.1126/science.aao2469}{{\sf
  doi:10.1126/science.aao2469}}].

\bibitem{Ade:2015rim}
{\bf Planck} Collaboration, P.~A.~R. Ade {\em et~al.}, {\it {Planck 2015
  results. XIV. Dark energy and modified gravity}},  {\sl Astron. Astrophys.}
  {\bf 594} (2016) A14, [\href{http://arxiv.org/abs/1502.01590}{{\sf
  arXiv:1502.01590}}],
  [\href{http://dx.doi.org/10.1051/0004-6361/201525814}{{\sf
  doi:10.1051/0004-6361/201525814}}].

\bibitem{Horndeski:1974wa}
G.~W. Horndeski, {\it {Second-order scalar-tensor field equations in a
  four-dimensional space}},  {\sl Int. J. Theor. Phys.} {\bf 10} (1974)
  363--384, [\href{http://dx.doi.org/10.1007/BF01807638}{{\sf
  doi:10.1007/BF01807638}}].

\bibitem{Deffayet:2011gz}
C.~Deffayet, X.~Gao, D.~A. Steer, and G.~Zahariade, {\it {From k-essence to
  generalised Galileons}},  {\sl Phys. Rev. D} {\bf 84} (2011) 064039,
  [\href{http://arxiv.org/abs/1103.3260}{{\sf arXiv:1103.3260}}],
  [\href{http://dx.doi.org/10.1103/PhysRevD.84.064039}{{\sf
  doi:10.1103/PhysRevD.84.064039}}].

\bibitem{Kobayashi:2011nu}
T.~Kobayashi, M.~Yamaguchi, and J.~Yokoyama, {\it {Generalized G-inflation:
  Inflation with the most general second-order field equations}},  {\sl Prog.
  Theor. Phys.} {\bf 126} (2011) 511--529,
  [\href{http://arxiv.org/abs/1105.5723}{{\sf arXiv:1105.5723}}],
  [\href{http://dx.doi.org/10.1143/PTP.126.511}{{\sf
  doi:10.1143/PTP.126.511}}].

\bibitem{Charmousis:2011bf}
C.~Charmousis, E.~J. Copeland, A.~Padilla, and P.~M. Saffin, {\it {General
  second order scalar-tensor theory, self tuning, and the Fab Four}},  {\sl
  Phys. Rev. Lett.} {\bf 108} (2012) 051101,
  [\href{http://arxiv.org/abs/1106.2000}{{\sf arXiv:1106.2000}}],
  [\href{http://dx.doi.org/10.1103/PhysRevLett.108.051101}{{\sf
  doi:10.1103/PhysRevLett.108.051101}}].

\bibitem{Amendola:2017orw}
L.~Amendola, M.~Kunz, I.~D. Saltas, and I.~Sawicki, {\it {Fate of Large-Scale
  Structure in Modified Gravity After GW170817 and GRB170817A}},  {\sl Phys.
  Rev. Lett.} {\bf 120} (2018), no.~13 131101,
  [\href{http://arxiv.org/abs/1711.04825}{{\sf arXiv:1711.04825}}],
  [\href{http://dx.doi.org/10.1103/PhysRevLett.120.131101}{{\sf
  doi:10.1103/PhysRevLett.120.131101}}].

\bibitem{Copeland:2018yuh}
E.~J. Copeland, M.~Kopp, A.~Padilla, P.~M. Saffin, and C.~Skordis, {\it {Dark
  energy after GW170817, revisited}},  {\sl Phys. Rev. Lett.} {\bf 122} (2019),
  no.~6 061301, [\href{http://arxiv.org/abs/1810.08239}{{\sf
  arXiv:1810.08239}}],
  [\href{http://dx.doi.org/10.1103/PhysRevLett.122.061301}{{\sf
  doi:10.1103/PhysRevLett.122.061301}}].

\bibitem{Crisostomi:2017pjs}
M.~Crisostomi and K.~Koyama, {\it {Self-accelerating universe in scalar-tensor
  theories after GW170817}},  {\sl Phys. Rev.} {\bf D97} (2018), no.~8 084004,
  [\href{http://arxiv.org/abs/1712.06556}{{\sf arXiv:1712.06556}}],
  [\href{http://dx.doi.org/10.1103/PhysRevD.97.084004}{{\sf
  doi:10.1103/PhysRevD.97.084004}}].

\bibitem{Creminelli:2017sry}
P.~Creminelli and F.~Vernizzi, {\it {Dark Energy after GW170817 and
  GRB170817A}},  {\sl Phys. Rev. Lett.} {\bf 119} (2017), no.~25 251302,
  [\href{http://arxiv.org/abs/1710.05877}{{\sf arXiv:1710.05877}}],
  [\href{http://dx.doi.org/10.1103/PhysRevLett.119.251302}{{\sf
  doi:10.1103/PhysRevLett.119.251302}}].

\bibitem{Sakstein:2017xjx}
J.~Sakstein and B.~Jain, {\it {Implications of the Neutron Star Merger GW170817
  for Cosmological Scalar-Tensor Theories}},  {\sl Phys. Rev. Lett.} {\bf 119}
  (2017), no.~25 251303, [\href{http://arxiv.org/abs/1710.05893}{{\sf
  arXiv:1710.05893}}],
  [\href{http://dx.doi.org/10.1103/PhysRevLett.119.251303}{{\sf
  doi:10.1103/PhysRevLett.119.251303}}].

\bibitem{Ezquiaga:2017ekz}
J.~M. Ezquiaga and M.~Zumalacarregui, {\it {Dark Energy After GW170817: Dead
  Ends and the Road Ahead}},  {\sl Phys. Rev. Lett.} {\bf 119} (2017), no.~25
  251304, [\href{http://arxiv.org/abs/1710.05901}{{\sf arXiv:1710.05901}}],
  [\href{http://dx.doi.org/10.1103/PhysRevLett.119.251304}{{\sf
  doi:10.1103/PhysRevLett.119.251304}}].

\bibitem{Baker:2017hug}
T.~Baker, E.~Bellini, P.~G. Ferreira, M.~Lagos, J.~Noller, and I.~Sawicki, {\it
  {Strong constraints on cosmological gravity from GW170817 and GRB 170817A}},
  {\sl Phys. Rev. Lett.} {\bf 119} (2017), no.~25 251301,
  [\href{http://arxiv.org/abs/1710.06394}{{\sf arXiv:1710.06394}}],
  [\href{http://dx.doi.org/10.1103/PhysRevLett.119.251301}{{\sf
  doi:10.1103/PhysRevLett.119.251301}}].

\bibitem{TheLIGOScientific:2017qsa}
{\bf Virgo, LIGO Scientific} Collaboration, B.~P. Abbott {\em et~al.}, {\it
  {GW170817: Observation of Gravitational Waves from a Binary Neutron Star
  Inspiral}},  {\sl Phys. Rev. Lett.} {\bf 119} (2017), no.~16 161101,
  [\href{http://arxiv.org/abs/1710.05832}{{\sf arXiv:1710.05832}}],
  [\href{http://dx.doi.org/10.1103/PhysRevLett.119.161101}{{\sf
  doi:10.1103/PhysRevLett.119.161101}}].

\bibitem{Kase:2018aps}
R.~Kase and S.~Tsujikawa, {\it {Dark energy in Horndeski theories after
  GW170817: A review}},  {\sl Int. J. Mod. Phys. D} {\bf 28} (2019), no.~05
  1942005, [\href{http://arxiv.org/abs/1809.08735}{{\sf arXiv:1809.08735}}],
  [\href{http://dx.doi.org/10.1142/S0218271819420057}{{\sf
  doi:10.1142/S0218271819420057}}].

\bibitem{Lewis:1999bs}
A.~Lewis, A.~Challinor, and A.~Lasenby, {\it {Efficient computation of CMB
  anisotropies in closed FRW models}},  {\sl Astrophys. J.} {\bf 538} (2000)
  473--476, [\href{http://arxiv.org/abs/astro-ph/9911177}{{\sf
  arXiv:astro-ph/9911177}}], [\href{http://dx.doi.org/10.1086/309179}{{\sf
  doi:10.1086/309179}}].

\bibitem{Blas:2011rf}
D.~Blas, J.~Lesgourgues, and T.~Tram, {\it {The Cosmic Linear Anisotropy
  Solving System (CLASS) II: Approximation schemes}},  {\sl JCAP} {\bf 1107}
  (2011) 034, [\href{http://arxiv.org/abs/1104.2933}{{\sf arXiv:1104.2933}}],
  [\href{http://dx.doi.org/10.1088/1475-7516/2011/07/034}{{\sf
  doi:10.1088/1475-7516/2011/07/034}}].

\bibitem{Smith:2002dz}
{\bf VIRGO Consortium} Collaboration, R.~E. Smith, J.~A. Peacock, A.~Jenkins,
  S.~D.~M. White, C.~S. Frenk, F.~R. Pearce, P.~A. Thomas, G.~Efstathiou, and
  H.~M.~P. Couchmann, {\it {Stable clustering, the halo model and nonlinear
  cosmological power spectra}},  {\sl Mon. Not. Roy. Astron. Soc.} {\bf 341}
  (2003) 1311, [\href{http://arxiv.org/abs/astro-ph/0207664}{{\sf
  arXiv:astro-ph/0207664}}],
  [\href{http://dx.doi.org/10.1046/j.1365-8711.2003.06503.x}{{\sf
  doi:10.1046/j.1365-8711.2003.06503.x}}].

\bibitem{Zhao:2008bn}
G.-B. Zhao, L.~Pogosian, A.~Silvestri, and J.~Zylberberg, {\it {Searching for
  modified growth patterns with tomographic surveys}},  {\sl Phys. Rev.} {\bf
  D79} (2009) 083513, [\href{http://arxiv.org/abs/0809.3791}{{\sf
  arXiv:0809.3791}}], [\href{http://dx.doi.org/10.1103/PhysRevD.79.083513}{{\sf
  doi:10.1103/PhysRevD.79.083513}}].

\bibitem{Hojjati:2011ix}
A.~Hojjati, L.~Pogosian, and G.-B. Zhao, {\it {Testing gravity with CAMB and
  CosmoMC}},  {\sl JCAP} {\bf 1108} (2011) 005,
  [\href{http://arxiv.org/abs/1106.4543}{{\sf arXiv:1106.4543}}],
  [\href{http://dx.doi.org/10.1088/1475-7516/2011/08/005}{{\sf
  doi:10.1088/1475-7516/2011/08/005}}].

\bibitem{Sakr:2021ylx}
Z.~Sakr and M.~Martinelli, {\it {Cosmological constraints on sub-horizon scales
  modified gravity theories with MGCLASS II}},  {\sl JCAP} {\bf 05} (2022),
  no.~05 030, [\href{http://arxiv.org/abs/2112.14175}{{\sf arXiv:2112.14175}}],
  [\href{http://dx.doi.org/10.1088/1475-7516/2022/05/030}{{\sf
  doi:10.1088/1475-7516/2022/05/030}}].

\bibitem{He:2012wq}
J.-h. He, {\it {Testing $f(R)$ dark energy model with the large scale
  structure}},  {\sl Phys. Rev.} {\bf D86} (2012) 103505,
  [\href{http://arxiv.org/abs/1207.4898}{{\sf arXiv:1207.4898}}],
  [\href{http://dx.doi.org/10.1103/PhysRevD.86.103505}{{\sf
  doi:10.1103/PhysRevD.86.103505}}].

\bibitem{Xu:2015usa}
L.~Xu, {\it {FRCAMB: An $f(R)$ Code for Anisotropies in the Microwave
  Background}},  \href{http://arxiv.org/abs/1506.03232}{{\sf
  arXiv:1506.03232}}.

\bibitem{Gubitosi:2012hu}
G.~Gubitosi, F.~Piazza, and F.~Vernizzi, {\it {The Effective Field Theory of
  Dark Energy}},  {\sl JCAP} {\bf 1302} (2013) 032,
  [\href{http://arxiv.org/abs/1210.0201}{{\sf arXiv:1210.0201}}],
  [\href{http://dx.doi.org/10.1088/1475-7516/2013/02/032}{{\sf
  doi:10.1088/1475-7516/2013/02/032}}]. [JCAP1302,032(2013)].

\bibitem{Hu:2013twa}
B.~Hu, M.~Raveri, N.~Frusciante, and A.~Silvestri, {\it {Effective Field Theory
  of Cosmic Acceleration: an implementation in CAMB}},  {\sl Phys. Rev.} {\bf
  D89} (2014), no.~10 103530, [\href{http://arxiv.org/abs/1312.5742}{{\sf
  arXiv:1312.5742}}], [\href{http://dx.doi.org/10.1103/PhysRevD.89.103530}{{\sf
  doi:10.1103/PhysRevD.89.103530}}].

\bibitem{Battye:2015hza}
R.~A. Battye, B.~Bolliet, and J.~A. Pearson, {\it {$f(R)$ gravity as a dark
  energy fluid}},  {\sl Phys. Rev.} {\bf D93} (2016), no.~4 044026,
  [\href{http://arxiv.org/abs/1508.04569}{{\sf arXiv:1508.04569}}],
  [\href{http://dx.doi.org/10.1103/PhysRevD.93.044026}{{\sf
  doi:10.1103/PhysRevD.93.044026}}].

\bibitem{Battye:2017ysh}
R.~A. Battye, B.~Bolliet, and F.~Pace, {\it {Do cosmological data rule out
  $f(\mathcal{R})$ with $w\neq-1$?}},  {\sl Phys. Rev.} {\bf D97} (2018),
  no.~10 104070, [\href{http://arxiv.org/abs/1712.05976}{{\sf
  arXiv:1712.05976}}],
  [\href{http://dx.doi.org/10.1103/PhysRevD.97.104070}{{\sf
  doi:10.1103/PhysRevD.97.104070}}].

\bibitem{Zumalacarregui:2016pph}
M.~Zumalac\'arregui, E.~Bellini, I.~Sawicki, J.~Lesgourgues, and P.~G.
  Ferreira, {\it {hi\_class: Horndeski in the Cosmic Linear Anisotropy Solving
  System}},  {\sl JCAP} {\bf 08} (2017) 019,
  [\href{http://arxiv.org/abs/1605.06102}{{\sf arXiv:1605.06102}}],
  [\href{http://dx.doi.org/10.1088/1475-7516/2017/08/019}{{\sf
  doi:10.1088/1475-7516/2017/08/019}}].

\bibitem{Capozziello:2006dj}
S.~Capozziello, S.~Nojiri, S.~D. Odintsov, and A.~Troisi, {\it {Cosmological
  viability of f(R)-gravity as an ideal fluid and its compatibility with a
  matter dominated phase}},  {\sl Phys. Lett.} {\bf B639} (2006) 135--143,
  [\href{http://arxiv.org/abs/astro-ph/0604431}{{\sf arXiv:astro-ph/0604431}}],
  [\href{http://dx.doi.org/10.1016/j.physletb.2006.06.034}{{\sf
  doi:10.1016/j.physletb.2006.06.034}}].

\bibitem{Nojiri:2006ri}
S.~Nojiri and S.~D. Odintsov, {\it {Introduction to modified gravity and
  gravitational alternative for dark energy}},  {\sl eConf} {\bf C0602061}
  (2006) 06, [\href{http://arxiv.org/abs/hep-th/0601213}{{\sf
  arXiv:hep-th/0601213}}],
  [\href{http://dx.doi.org/10.1142/S0219887807001928}{{\sf
  doi:10.1142/S0219887807001928}}].

\bibitem{Arjona:2018jhh}
R.~Arjona, W.~Cardona, and S.~Nesseris, {\it {Unraveling the effective fluid
  approach for $f(R)$ models in the subhorizon approximation}},  {\sl Phys.
  Rev.} {\bf D99} (2019), no.~4 043516,
  [\href{http://arxiv.org/abs/1811.02469}{{\sf arXiv:1811.02469}}],
  [\href{http://dx.doi.org/10.1103/PhysRevD.99.043516}{{\sf
  doi:10.1103/PhysRevD.99.043516}}].

\bibitem{Arjona:2019rfn}
R.~Arjona, W.~Cardona, and S.~Nesseris, {\it {Designing Horndeski and the
  effective fluid approach}},  {\sl Phys. Rev. D} {\bf 100} (2019), no.~6
  063526, [\href{http://arxiv.org/abs/1904.06294}{{\sf arXiv:1904.06294}}],
  [\href{http://dx.doi.org/10.1103/PhysRevD.100.063526}{{\sf
  doi:10.1103/PhysRevD.100.063526}}].

\bibitem{Pace:2019uow}
F.~Pace, R.~A. Battye, B.~Bolliet, and D.~Trinh, {\it {Dark sector evolution in
  Horndeski models}},  {\sl JCAP} {\bf 09} (2019) 018,
  [\href{http://arxiv.org/abs/1905.06795}{{\sf arXiv:1905.06795}}],
  [\href{http://dx.doi.org/10.1088/1475-7516/2019/09/018}{{\sf
  doi:10.1088/1475-7516/2019/09/018}}].

\bibitem{Pace:2020qpj}
F.~Pace, R.~Battye, E.~Bellini, L.~Lombriser, F.~Vernizzi, and B.~Bolliet, {\it
  {Comparison of different approaches to the quasi-static approximation in
  Horndeski models}},  {\sl JCAP} {\bf 06} (2021) 017,
  [\href{http://arxiv.org/abs/2011.05713}{{\sf arXiv:2011.05713}}],
  [\href{http://dx.doi.org/10.1088/1475-7516/2021/06/017}{{\sf
  doi:10.1088/1475-7516/2021/06/017}}].

\bibitem{Geng:2021jso}
C.-Q. Geng, Y.-T. Hsu, J.-R. Lu, and L.~Yin, {\it {A Dark Energy model from
  Generalized Proca Theory}},  {\sl Phys. Dark Univ.} {\bf 32} (2021) 100819,
  [\href{http://arxiv.org/abs/2104.06577}{{\sf arXiv:2104.06577}}],
  [\href{http://dx.doi.org/10.1016/j.dark.2021.100819}{{\sf
  doi:10.1016/j.dark.2021.100819}}].

\bibitem{Nakamura:2019phn}
S.~Nakamura, R.~Kase, and S.~Tsujikawa, {\it {Coupled vector dark energy}},
  {\sl JCAP} {\bf 12} (2019) 032, [\href{http://arxiv.org/abs/1907.12216}{{\sf
  arXiv:1907.12216}}],
  [\href{http://dx.doi.org/10.1088/1475-7516/2019/12/032}{{\sf
  doi:10.1088/1475-7516/2019/12/032}}].

\bibitem{Nakamura:2018oyy}
S.~Nakamura, A.~De~Felice, R.~Kase, and S.~Tsujikawa, {\it {Constraints on
  massive vector dark energy models from integrated Sachs-Wolfe-galaxy
  cross-correlations}},  {\sl Phys. Rev. D} {\bf 99} (2019), no.~6 063533,
  [\href{http://arxiv.org/abs/1811.07541}{{\sf arXiv:1811.07541}}],
  [\href{http://dx.doi.org/10.1103/PhysRevD.99.063533}{{\sf
  doi:10.1103/PhysRevD.99.063533}}].

\bibitem{deFelice:2017paw}
A.~de~Felice, L.~Heisenberg, and S.~Tsujikawa, {\it {Observational constraints
  on generalized Proca theories}},  {\sl Phys. Rev. D} {\bf 95} (2017), no.~12
  123540, [\href{http://arxiv.org/abs/1703.09573}{{\sf arXiv:1703.09573}}],
  [\href{http://dx.doi.org/10.1103/PhysRevD.95.123540}{{\sf
  doi:10.1103/PhysRevD.95.123540}}].

\bibitem{PhysRevLett.127.161302}
C.~Skordis and T.~Z\l{}o\ifmmode~\acute{s}\else \'{s}\fi{}nik, {\it New
  relativistic theory for modified newtonian dynamics},  {\sl Phys. Rev. Lett.}
  {\bf 127} (Oct, 2021) 161302,
  [\href{http://dx.doi.org/10.1103/PhysRevLett.127.161302}{{\sf
  doi:10.1103/PhysRevLett.127.161302}}].

\bibitem{PhysRevD.81.104015}
J.~Zuntz, T.~G. Zlosnik, F.~Bourliot, P.~G. Ferreira, and G.~D. Starkman, {\it
  Vector field models of modified gravity and the dark sector},  {\sl Phys.
  Rev. D} {\bf 81} (May, 2010) 104015,
  [\href{http://dx.doi.org/10.1103/PhysRevD.81.104015}{{\sf
  doi:10.1103/PhysRevD.81.104015}}].

\bibitem{PhysRevD.78.063005}
J.~B. Jim\'enez and A.~L. Maroto, {\it Cosmic vector for dark energy},  {\sl
  Phys. Rev. D} {\bf 78} (Sep, 2008) 063005,
  [\href{http://dx.doi.org/10.1103/PhysRevD.78.063005}{{\sf
  doi:10.1103/PhysRevD.78.063005}}].

\bibitem{BeltranJimenez:2008enx}
J.~Beltran~Jimenez and A.~L. Maroto, {\it {Cosmological electromagnetic fields
  and dark energy}},  {\sl JCAP} {\bf 03} (2009) 016,
  [\href{http://arxiv.org/abs/0811.0566}{{\sf arXiv:0811.0566}}],
  [\href{http://dx.doi.org/10.1088/1475-7516/2009/03/016}{{\sf
  doi:10.1088/1475-7516/2009/03/016}}].

\bibitem{BeltranJimenez:2013btb}
J.~Beltran~Jimenez, R.~Durrer, L.~Heisenberg, and M.~Thorsrud, {\it {Stability
  of Horndeski vector-tensor interactions}},  {\sl JCAP} {\bf 10} (2013) 064,
  [\href{http://arxiv.org/abs/1308.1867}{{\sf arXiv:1308.1867}}],
  [\href{http://dx.doi.org/10.1088/1475-7516/2013/10/064}{{\sf
  doi:10.1088/1475-7516/2013/10/064}}].

\bibitem{DeFelice:2016yws}
A.~De~Felice, L.~Heisenberg, R.~Kase, S.~Mukohyama, S.~Tsujikawa, and Y.-l.
  Zhang, {\it {Cosmology in generalized Proca theories}},  {\sl JCAP} {\bf 06}
  (2016) 048, [\href{http://arxiv.org/abs/1603.05806}{{\sf arXiv:1603.05806}}],
  [\href{http://dx.doi.org/10.1088/1475-7516/2016/06/048}{{\sf
  doi:10.1088/1475-7516/2016/06/048}}].

\bibitem{PhysRevD.94.044024}
A.~De~Felice, L.~Heisenberg, R.~Kase, S.~Mukohyama, S.~Tsujikawa, and Y.-l.
  Zhang, {\it Effective gravitational couplings for cosmological perturbations
  in generalized proca theories},  {\sl Phys. Rev. D} {\bf 94} (Aug, 2016)
  044024, [\href{http://dx.doi.org/10.1103/PhysRevD.94.044024}{{\sf
  doi:10.1103/PhysRevD.94.044024}}].

\bibitem{Armendariz-Picon:2004say}
C.~Armendariz-Picon, {\it {Could dark energy be vector-like?}},  {\sl JCAP}
  {\bf 07} (2004) 007, [\href{http://arxiv.org/abs/astro-ph/0405267}{{\sf
  arXiv:astro-ph/0405267}}],
  [\href{http://dx.doi.org/10.1088/1475-7516/2004/07/007}{{\sf
  doi:10.1088/1475-7516/2004/07/007}}].

\bibitem{Koivisto:2008ig}
T.~Koivisto and D.~F. Mota, {\it {Anisotropic Dark Energy: Dynamics of
  Background and Perturbations}},  {\sl JCAP} {\bf 06} (2008) 018,
  [\href{http://arxiv.org/abs/0801.3676}{{\sf arXiv:0801.3676}}],
  [\href{http://dx.doi.org/10.1088/1475-7516/2008/06/018}{{\sf
  doi:10.1088/1475-7516/2008/06/018}}].

\bibitem{Koivisto:2008xf}
T.~Koivisto and D.~F. Mota, {\it {Vector Field Models of Inflation and Dark
  Energy}},  {\sl JCAP} {\bf 08} (2008) 021,
  [\href{http://arxiv.org/abs/0805.4229}{{\sf arXiv:0805.4229}}],
  [\href{http://dx.doi.org/10.1088/1475-7516/2008/08/021}{{\sf
  doi:10.1088/1475-7516/2008/08/021}}].

\bibitem{Thorsrud:2012mu}
M.~Thorsrud, D.~F. Mota, and S.~Hervik, {\it {Cosmology of a Scalar Field
  Coupled to Matter and an Isotropy-Violating Maxwell Field}},  {\sl JHEP} {\bf
  10} (2012) 066, [\href{http://arxiv.org/abs/1205.6261}{{\sf
  arXiv:1205.6261}}], [\href{http://dx.doi.org/10.1007/JHEP10(2012)066}{{\sf
  doi:10.1007/JHEP10(2012)066}}].

\bibitem{Landim:2016dxh}
R.~C.~G. Landim, {\it {Dynamical analysis for a vector-like dark energy}},
  {\sl Eur. Phys. J. C} {\bf 76} (2016), no.~9 480,
  [\href{http://arxiv.org/abs/1605.03550}{{\sf arXiv:1605.03550}}],
  [\href{http://dx.doi.org/10.1140/epjc/s10052-016-4328-x}{{\sf
  doi:10.1140/epjc/s10052-016-4328-x}}].

\bibitem{Gomez:2020sfz}
L.~G. Gomez and Y.~Rodriguez, {\it {Coupled multi-Proca vector dark energy}},
  {\sl Phys. Dark Univ.} {\bf 31} (2021) 100759,
  [\href{http://arxiv.org/abs/2004.06466}{{\sf arXiv:2004.06466}}],
  [\href{http://dx.doi.org/10.1016/j.dark.2020.100759}{{\sf
  doi:10.1016/j.dark.2020.100759}}].

\bibitem{Gomez:2021jbo}
L.~G. Gomez, Y.~Rodriguez, and J.~P.~B. Almeida, {\it {Anisotropic Scalar Field
  Dark Energy with a Disformally Coupled Yang-Mills Field}},  {\sl Int. J. Mod.
  Phys. D} {\bf 31} (2022) 2250060,
  [\href{http://arxiv.org/abs/2103.11826}{{\sf arXiv:2103.11826}}],
  [\href{http://dx.doi.org/10.1142/S0218271822500602}{{\sf
  doi:10.1142/S0218271822500602}}].

\bibitem{Mehrabi:2015lfa}
A.~Mehrabi, A.~Maleknejad, and V.~Kamali, {\it {Gaugessence: a dark energy
  model with early time radiation-like equation of state}},  {\sl Astrophys.
  Space Sci.} {\bf 362} (2017), no.~3 53,
  [\href{http://arxiv.org/abs/1510.00838}{{\sf arXiv:1510.00838}}],
  [\href{http://dx.doi.org/10.1007/s10509-017-3033-z}{{\sf
  doi:10.1007/s10509-017-3033-z}}].

\bibitem{Alvarez:2019ues}
M.~\'Alvarez, J.~B. Orjuela-Quintana, Y.~Rodriguez, and C.~A.
  Valenzuela-Toledo, {\it {Einstein Yang\textendash{}Mills Higgs dark energy
  revisited}},  {\sl Class. Quant. Grav.} {\bf 36} (2019), no.~19 195004,
  [\href{http://arxiv.org/abs/1901.04624}{{\sf arXiv:1901.04624}}],
  [\href{http://dx.doi.org/10.1088/1361-6382/ab3775}{{\sf
  doi:10.1088/1361-6382/ab3775}}].

\bibitem{Orjuela-Quintana:2020klr}
J.~B. Orjuela-Quintana, M.~Alvarez, C.~A. Valenzuela-Toledo, and Y.~Rodriguez,
  {\it {Anisotropic Einstein Yang-Mills Higgs Dark Energy}},  {\sl JCAP} {\bf
  10} (2020) 019, [\href{http://arxiv.org/abs/2006.14016}{{\sf
  arXiv:2006.14016}}],
  [\href{http://dx.doi.org/10.1088/1475-7516/2020/10/019}{{\sf
  doi:10.1088/1475-7516/2020/10/019}}].

\bibitem{Guarnizo:2020pkj}
A.~Guarnizo, J.~B. Orjuela-Quintana, and C.~A. Valenzuela-Toledo, {\it
  {Dynamical analysis of cosmological models with non-Abelian gauge vector
  fields}},  {\sl Phys. Rev. D} {\bf 102} (2020), no.~8 083507,
  [\href{http://arxiv.org/abs/2007.12964}{{\sf arXiv:2007.12964}}],
  [\href{http://dx.doi.org/10.1103/PhysRevD.102.083507}{{\sf
  doi:10.1103/PhysRevD.102.083507}}].

\bibitem{Tasinato:2014eka}
G.~Tasinato, {\it {Cosmic Acceleration from Abelian Symmetry Breaking}},  {\sl
  JHEP} {\bf 04} (2014) 067, [\href{http://arxiv.org/abs/1402.6450}{{\sf
  arXiv:1402.6450}}], [\href{http://dx.doi.org/10.1007/JHEP04(2014)067}{{\sf
  doi:10.1007/JHEP04(2014)067}}].

\bibitem{Heisenberg:2014rta}
L.~Heisenberg, {\it {Generalization of the Proca Action}},  {\sl JCAP} {\bf 05}
  (2014) 015, [\href{http://arxiv.org/abs/1402.7026}{{\sf arXiv:1402.7026}}],
  [\href{http://dx.doi.org/10.1088/1475-7516/2014/05/015}{{\sf
  doi:10.1088/1475-7516/2014/05/015}}].

\bibitem{Allys:2015sht}
E.~Allys, P.~Peter, and Y.~Rodriguez, {\it {Generalized Proca action for an
  Abelian vector field}},  {\sl JCAP} {\bf 02} (2016) 004,
  [\href{http://arxiv.org/abs/1511.03101}{{\sf arXiv:1511.03101}}],
  [\href{http://dx.doi.org/10.1088/1475-7516/2016/02/004}{{\sf
  doi:10.1088/1475-7516/2016/02/004}}].

\bibitem{Allys:2016jaq}
E.~Allys, J.~P. Beltran~Almeida, P.~Peter, and Y.~Rodr\'\i{}guez, {\it {On the
  4D generalized Proca action for an Abelian vector field}},  {\sl JCAP} {\bf
  09} (2016) 026, [\href{http://arxiv.org/abs/1605.08355}{{\sf
  arXiv:1605.08355}}],
  [\href{http://dx.doi.org/10.1088/1475-7516/2016/09/026}{{\sf
  doi:10.1088/1475-7516/2016/09/026}}].

\bibitem{BeltranJimenez:2016rff}
J.~Beltran~Jimenez and L.~Heisenberg, {\it {Derivative self-interactions for a
  massive vector field}},  {\sl Phys. Lett. B} {\bf 757} (2016) 405--411,
  [\href{http://arxiv.org/abs/1602.03410}{{\sf arXiv:1602.03410}}],
  [\href{http://dx.doi.org/10.1016/j.physletb.2016.04.017}{{\sf
  doi:10.1016/j.physletb.2016.04.017}}].

\bibitem{Heisenberg:2018mxx}
L.~Heisenberg, R.~Kase, and S.~Tsujikawa, {\it {Cosmology in
  scalar-vector-tensor theories}},  {\sl Phys. Rev. D} {\bf 98} (2018), no.~2
  024038, [\href{http://arxiv.org/abs/1805.01066}{{\sf arXiv:1805.01066}}],
  [\href{http://dx.doi.org/10.1103/PhysRevD.98.024038}{{\sf
  doi:10.1103/PhysRevD.98.024038}}].

\bibitem{Heisenberg:2018acv}
L.~Heisenberg, {\it {Scalar-Vector-Tensor Gravity Theories}},  {\sl JCAP} {\bf
  10} (2018) 054, [\href{http://arxiv.org/abs/1801.01523}{{\sf
  arXiv:1801.01523}}],
  [\href{http://dx.doi.org/10.1088/1475-7516/2018/10/054}{{\sf
  doi:10.1088/1475-7516/2018/10/054}}].

\bibitem{Kase:2018nwt}
R.~Kase and S.~Tsujikawa, {\it {Dark energy in scalar-vector-tensor theories}},
   {\sl JCAP} {\bf 11} (2018) 024, [\href{http://arxiv.org/abs/1805.11919}{{\sf
  arXiv:1805.11919}}],
  [\href{http://dx.doi.org/10.1088/1475-7516/2018/11/024}{{\sf
  doi:10.1088/1475-7516/2018/11/024}}].

\bibitem{Lagos:2017hdr}
M.~Lagos, E.~Bellini, J.~Noller, P.~G. Ferreira, and T.~Baker, {\it {A general
  theory of linear} {cosmological perturbations: stability conditions, the
  quasistatic limit and dynamics}},  {\sl JCAP} {\bf 03} (2018) 021,
  [\href{http://arxiv.org/abs/1711.09893}{{\sf arXiv:1711.09893}}],
  [\href{http://dx.doi.org/10.1088/1475-7516/2018/03/021}{{\sf
  doi:10.1088/1475-7516/2018/03/021}}].

\bibitem{Oliveros:2022njz}
A.~Oliveros and C.~J. Rodr\'\i{}guez, {\it {Inflation in a
  scalar\textendash{}vector\textendash{}tensor theory}},  {\sl Gen. Rel. Grav.}
  {\bf 54} (2022), no.~1 9, [\href{http://arxiv.org/abs/2201.03629}{{\sf
  arXiv:2201.03629}}],
  [\href{http://dx.doi.org/10.1007/s10714-022-02901-y}{{\sf
  doi:10.1007/s10714-022-02901-y}}].

\bibitem{Hu:1998kj}
W.~Hu, {\it {Structure formation with generalized dark matter}},  {\sl
  Astrophys. J.} {\bf 506} (1998) 485--494,
  [\href{http://arxiv.org/abs/astro-ph/9801234}{{\sf arXiv:astro-ph/9801234}}],
  [\href{http://dx.doi.org/10.1086/306274}{{\sf doi:10.1086/306274}}].

\bibitem{PhysRevLett.122.171301}
{\bf DES Collaboration} Collaboration, T.~M.~C. Abbott {\em et~al.}, {\it
  Cosmological constraints from multiple probes in the dark energy survey},
  {\sl Phys. Rev. Lett.} {\bf 122} (May, 2019) 171301,
  [\href{http://dx.doi.org/10.1103/PhysRevLett.122.171301}{{\sf
  doi:10.1103/PhysRevLett.122.171301}}].

\bibitem{Hogg:2004vw}
D.~W. Hogg, D.~J. Eisenstein, M.~R. Blanton, N.~A. Bahcall, J.~Brinkmann, J.~E.
  Gunn, and D.~P. Schneider, {\it {Cosmic homogeneity demonstrated with
  luminous red galaxies}},  {\sl Astrophys. J.} {\bf 624} (2005) 54--58,
  [\href{http://arxiv.org/abs/astro-ph/0411197}{{\sf arXiv:astro-ph/0411197}}],
  [\href{http://dx.doi.org/10.1086/429084}{{\sf doi:10.1086/429084}}].

\bibitem{Ade:2015hxq}
{\bf Planck} Collaboration, P.~A.~R. Ade {\em et~al.}, {\it {Planck 2015
  results. XVI. Isotropy and statistics of the CMB}},  {\sl Astron. Astrophys.}
  {\bf 594} (2016) A16, [\href{http://arxiv.org/abs/1506.07135}{{\sf
  arXiv:1506.07135}}],
  [\href{http://dx.doi.org/10.1051/0004-6361/201526681}{{\sf
  doi:10.1051/0004-6361/201526681}}].

\bibitem{Marinoni:2012ba}
C.~Marinoni, J.~Bel, and A.~Buzzi, {\it {The Scale of Cosmic Isotropy}},  {\sl
  JCAP} {\bf 10} (2012) 036, [\href{http://arxiv.org/abs/1205.3309}{{\sf
  arXiv:1205.3309}}],
  [\href{http://dx.doi.org/10.1088/1475-7516/2012/10/036}{{\sf
  doi:10.1088/1475-7516/2012/10/036}}].

\bibitem{Abbott:2017oio}
{\bf Virgo, LIGO Scientific} Collaboration, B.~P. Abbott {\em et~al.}, {\it
  {GW170814: A Three-Detector Observation of Gravitational Waves from a Binary
  Black Hole Coalescence}},  {\sl Phys. Rev. Lett.} {\bf 119} (2017), no.~14
  141101, [\href{http://arxiv.org/abs/1709.09660}{{\sf arXiv:1709.09660}}],
  [\href{http://dx.doi.org/10.1103/PhysRevLett.119.141101}{{\sf
  doi:10.1103/PhysRevLett.119.141101}}].

\bibitem{Frusciante:2018jzw}
N.~Frusciante, S.~Peirone, S.~Casas, and N.~A. Lima, {\it {Cosmology of
  surviving Horndeski theory: The road ahead}},  {\sl Phys. Rev. D} {\bf 99}
  (2019), no.~6 063538, [\href{http://arxiv.org/abs/1810.10521}{{\sf
  arXiv:1810.10521}}],
  [\href{http://dx.doi.org/10.1103/PhysRevD.99.063538}{{\sf
  doi:10.1103/PhysRevD.99.063538}}].

\bibitem{McManus:2016kxu}
R.~McManus, L.~Lombriser, and J.~Pe\~narrubia, {\it {Finding Horndeski theories
  with Einstein gravity limits}},  {\sl JCAP} {\bf 1611} (2016), no.~11 006,
  [\href{http://arxiv.org/abs/1606.03282}{{\sf arXiv:1606.03282}}],
  [\href{http://dx.doi.org/10.1088/1475-7516/2016/11/006}{{\sf
  doi:10.1088/1475-7516/2016/11/006}}].

\bibitem{Lombriser:2015sxa}
L.~Lombriser and A.~Taylor, {\it {Breaking a Dark Degeneracy with Gravitational
  Waves}},  {\sl JCAP} {\bf 1603} (2016), no.~03 031,
  [\href{http://arxiv.org/abs/1509.08458}{{\sf arXiv:1509.08458}}],
  [\href{http://dx.doi.org/10.1088/1475-7516/2016/03/031}{{\sf
  doi:10.1088/1475-7516/2016/03/031}}].

\bibitem{Noller:2018wyv}
J.~Noller and A.~Nicola, {\it {Cosmological parameter constraints for Horndeski
  scalar-tensor gravity}},  {\sl Phys. Rev. D} {\bf 99} (2019), no.~10 103502,
  [\href{http://arxiv.org/abs/1811.12928}{{\sf arXiv:1811.12928}}],
  [\href{http://dx.doi.org/10.1103/PhysRevD.99.103502}{{\sf
  doi:10.1103/PhysRevD.99.103502}}].

\bibitem{deRham:2018red}
C.~de~Rham and S.~Melville, {\it {Gravitational Rainbows: LIGO and Dark Energy
  at its Cutoff}},  {\sl Phys. Rev. Lett.} {\bf 121} (2018), no.~22 221101,
  [\href{http://arxiv.org/abs/1806.09417}{{\sf arXiv:1806.09417}}],
  [\href{http://dx.doi.org/10.1103/PhysRevLett.121.221101}{{\sf
  doi:10.1103/PhysRevLett.121.221101}}].

\bibitem{PhysRevLett.117.131302}
D.~Saadeh, S.~M. Feeney, A.~Pontzen, H.~V. Peiris, and J.~D. McEwen, {\it How
  isotropic is the universe?},  {\sl Phys. Rev. Lett.} {\bf 117} (Sep, 2016)
  131302, [\href{http://dx.doi.org/10.1103/PhysRevLett.117.131302}{{\sf
  doi:10.1103/PhysRevLett.117.131302}}].

\bibitem{Mckay2003}
D.~J.~C. MacKay, {\em Information theory, inference, and learning algorithms}.
\newblock Cambridge University Press, 2003.

\bibitem{Romano:2018frb}
A.~E. Romano and S.~A. Vallejo~Pena, {\it {The MESS of cosmological
  perturbations}},  {\sl Phys. Lett. B} {\bf 784} (2018) 367--372,
  [\href{http://arxiv.org/abs/1806.01941}{{\sf arXiv:1806.01941}}],
  [\href{http://dx.doi.org/10.1016/j.physletb.2018.08.016}{{\sf
  doi:10.1016/j.physletb.2018.08.016}}].

\bibitem{Cardona:2014iba}
W.~Cardona, L.~Hollenstein, and M.~Kunz, {\it {The traces of anisotropic dark
  energy in light of Planck}},  {\sl JCAP} {\bf 1407} (2014) 032,
  [\href{http://arxiv.org/abs/1402.5993}{{\sf arXiv:1402.5993}}],
  [\href{http://dx.doi.org/10.1088/1475-7516/2014/07/032}{{\sf
  doi:10.1088/1475-7516/2014/07/032}}].

\bibitem{Tsujikawa:2007gd}
S.~Tsujikawa, {\it {Matter density perturbations and effective gravitational
  constant in modified gravity models of dark energy}},  {\sl Phys. Rev.} {\bf
  D76} (2007) 023514, [\href{http://arxiv.org/abs/0705.1032}{{\sf
  arXiv:0705.1032}}], [\href{http://dx.doi.org/10.1103/PhysRevD.76.023514}{{\sf
  doi:10.1103/PhysRevD.76.023514}}].

\bibitem{Heisenberg:2020xak}
L.~Heisenberg and H.~Villarrubia-Rojo, {\it {Proca in the sky}},  {\sl JCAP}
  {\bf 03} (2021) 032, [\href{http://arxiv.org/abs/2010.00513}{{\sf
  arXiv:2010.00513}}],
  [\href{http://dx.doi.org/10.1088/1475-7516/2021/03/032}{{\sf
  doi:10.1088/1475-7516/2021/03/032}}].

\bibitem{SDSS:2003tbn}
{\bf SDSS} Collaboration, M.~Tegmark {\em et~al.}, {\it {The 3-D power spectrum
  of galaxies from the SDSS}},  {\sl Astrophys. J.} {\bf 606} (2004) 702--740,
  [\href{http://arxiv.org/abs/astro-ph/0310725}{{\sf arXiv:astro-ph/0310725}}],
  [\href{http://dx.doi.org/10.1086/382125}{{\sf doi:10.1086/382125}}].

\bibitem{Sagredo:2018ahx}
B.~Sagredo, S.~Nesseris, and D.~Sapone, {\it {Internal Robustness of Growth
  Rate data}},  {\sl Phys. Rev. D} {\bf 98} (2018), no.~8 083543,
  [\href{http://arxiv.org/abs/1806.10822}{{\sf arXiv:1806.10822}}],
  [\href{http://dx.doi.org/10.1103/PhysRevD.98.083543}{{\sf
  doi:10.1103/PhysRevD.98.083543}}].

\bibitem{Bean:2003fb}
R.~Bean and O.~Dore, {\it {Probing dark energy perturbations: The Dark energy
  equation of state and speed of sound as measured by WMAP}},  {\sl Phys. Rev.}
  {\bf D69} (2004) 083503, [\href{http://arxiv.org/abs/astro-ph/0307100}{{\sf
  arXiv:astro-ph/0307100}}],
  [\href{http://dx.doi.org/10.1103/PhysRevD.69.083503}{{\sf
  doi:10.1103/PhysRevD.69.083503}}].

\end{thebibliography}\endgroup

\end{document}